\documentclass[aps,preprint,nofootinbib]{revtex4}%
\usepackage{amsfonts}
\usepackage{amsmath}
\usepackage{amssymb}
\usepackage{graphicx}%
\providecommand{\U}[1]{\protect\rule{.1in}{.1in}}
\begin{document}
\preprint{ }
\title{The Hamiltonian formulation of General Relativity: myths and reality}
\author{N. Kiriushcheva}
\email{nkiriush@uwo.ca}
\affiliation{Faculty of Arts and Social Science, Huron University College, N6G 1H3}
\affiliation{}
\affiliation{Department of Applied Mathematics, University of Western Ontario, N6A 5B7,
London, Canada}
\author{S.V. Kuzmin}
\email{skuzmin@uwo.ca}
\affiliation{Faculty of Arts and Social Science, Huron University College, N6G 1H3}
\affiliation{}
\affiliation{Department of Applied Mathematics, University of Western Ontario, N6A 5B7,
London, Canada}

\begin{abstract}
A conventional wisdom often perpetuated in the literature states that: (i) a
3+1 decomposition of space-time into space and time is synonymous with the
canonical treatment and this decomposition is essential for any Hamiltonian
formulation of General Relativity (GR); (ii) the canonical treatment
unavoidably breaks the symmetry between space and time in GR and the resulting
algebra of constraints is not the algebra of four-dimensional diffeomorphism;
(iii) according to some authors this algebra allows one to derive only spatial
diffeomorphism or, according to others, a specific field-dependent and
non-covariant four-dimensional diffeomorphism; (iv) the analyses of Dirac
[\textit{Proc. Roy. Soc. A 246 (1958) 333}] and of ADM [\textit{Arnowitt,
Deser and Misner, in \textquotedblleft Gravitation: An Introduction to Current
Research\textquotedblright\ (1962) 227}] of the canonical structure of GR are
equivalent. We provide some general reasons why these statements should be
questioned. Points (i-iii) have been shown to be incorrect in
[\textit{Kiriushcheva et al., Phys. Lett. A 372 (2008) 5101}] and now we
thoroughly re-examine all steps of the Dirac Hamiltonian formulation of GR. By
direct calculation we show that Dirac's references to space-like surfaces are
inessential and that such surfaces do not enter his calculations. In addition,
we show that his assumption $g_{0k}=0$, used to simplify his calculation of
different contributions to the secondary constraints, is unwarranted; yet,
remarkably his total Hamiltonian is equivalent to the one computed without the
assumption $g_{0k}=0$. The secondary constraints resulting from the
conservation of the primary constraints of Dirac are in fact different from
the original constraints that Dirac called secondary (also known as the
\textquotedblleft Hamiltonian\textquotedblright\ and \textquotedblleft
diffeomorphism\textquotedblright\ constraints). The Dirac constraints are
instead particular combinations of the constraints which follow directly from
the primary constraints. Taking this difference into account we found, using
two standard methods, that the generator of the gauge transformation gives
diffeomorphism invariance in four-dimensional space-time; and this shows that
points (i-iii) above cannot be attributed to the Dirac Hamiltonian formulation
of GR. We also demonstrate that ADM and Dirac formulations are related by a
transformation of phase-space variables from the metric $g_{\mu\nu}$ to lapse
and shift functions and the three-metric $g_{km}$, which is not canonical.
This proves that point (iv) is incorrect. Points (i-iii) are mere consequences
of using a non-canonical change of variables and are not an intrinsic property
of either the Hamilton-Dirac approach to constrained systems or Einstein's
theory itself.

\end{abstract}
\volumeyear{year}
\volumenumber{number}
\issuenumber{number}
\eid{ }
\maketitle

%\startpage{101}
%\endpage{102}
%\tableofcontents

\hspace{5cm} \begin{minipage}{10cm}
\textit{\textquotedblleft On ne trouvera point de Figures dans cet Ouvrage.
Les m\'{e}thodes que j'y expose ne demandent ni constructions, ni raisonnemens
g\'{e}om\'{e}triques ou m\'{e}caniques, mais seulement des op\'{e}rations
alg\'{e}briques, assuj\'{e}ties \`{a} une marche r\'{e}guli\`{e}re et
uniforme. Ceux qui aiment \v{l} Analyse, verront avec plaisir la M\'{e}canique
en devenir une nouvelle branche, et me sauront gr\'{e} \v{d} en avoir
\'{e}tendu ainsi le domaine.\textquotedblright}\\
\\
J. L. Lagrange, \textquotedblleft M\'{e}canique Analytique\textquotedblright\ (1788)\\
\\
\textit{ The reader will find no figures in this work. The
methods which I set forth do not require either constructions or geometrical
or mechanical reasonings, but merely algebraic operations subjected to a
regular and uniform rule of procedure. Those who are fond of Mathematical
Analysis will observe with pleasure Mechanics becoming one of its new branches
and they will be grateful to me for having thus extended its
domain.}
\end{minipage}

\section{Introduction}

We begin our paper with words written more than two centuries ago by Lagrange
in the preface to the first edition of the \textquotedblleft M\'{e}canique
Analytique\textquotedblright\ \cite{Lagrange} (the first English translation
appeared in 1997 \cite{Lagrange-eng}) because they express our standpoint in
analyzing of the Hamiltonian formulation of General Relativity (GR). The
results previously obtained by others are reconsidered and classified as
either \textquotedblleft myth\textquotedblright\ or \textquotedblleft
reality\textquotedblright\ depending on whether they were obtained by what
Lagrange called \textit{a regular and uniform rule of procedure, }or by
geometrical or some other reasonings. The results and conclusions constructed
using such reasonings must be checked by explicit calculation without which
they are meaningless and could be misleading, contradicting the rules of
procedure and the essential properties of GR.

Originating more than half a century ago, the Hamiltonian formulation of GR is
not a new subject. It began with advances in the Hamiltonian formulation of
singular Lagrangians due to the pioneering works of Dirac \cite{Dirac-1} and
Bergmann with coauthors \cite{Bergmann-1, Bergmann-2} on generalized
(constrained) Hamiltonian dynamics\footnote{There are references on the
Rosenfeld work\textit{ }\cite{Rosenfeld} but only in a few old papers
\cite{Bergmann-2, Bergmann-3, DeWitt-1, DeWitt-2} and neither in the first
review on this subject \cite{HRT} nor in monographs \cite{Diracbook,Kurt,
Gitmanbook, HTbook} including the newest one \cite{RRbook}. We always wondered
what is the Rosenfeld contribution. Recently, due to an effort of Salisbury
\cite{Preprint}, the paper by Rosenfeld \cite{Rosenfeld} became available for
non-German speaking readers and now anyone can evaluate his contribution and
learn about his views on a Hamiltonian formulation of gauge invariant
Lagrangians. In \cite{Preprint} the original article \cite{Rosenfeld} and its
English translation are accompanied by quite extensive and in many cases
helpful comments, but we want to warn the reader that some of them reflect the
opinion of the commentator, in particular, about the \textquotedblleft crucial
error\textquotedblright\ of Rosenfeld. We will discuss this\ later.}\textit{.
}

We restrict our discussion to the original Einstein metric formulation of GR.
The first-order, affine-metric, form \cite{Einstein, Einstein-rus} (an English
translation of this and a few other Einstein's papers can be found in
\cite{Einstein-eng}) will be just briefly touched; but the analysis presented
here can and must be extended to a affine-metric form and to other
formulations (e.g. Einstein-Cartan formulation of GR).

In chronological order (which is also ranked inversely in popularity) the
Hamiltonian formulation of \textit{metric }GR was considered by Pirani,
Schild, and Skinner (PSS) \cite{Pirani}, Dirac \cite{Dirac}, and Arnowitt,
Deser, and Misner (ADM) \cite{ADM} and references therein. The relationship
among these formulations has not been analyzed; and some authors have adopted
to using the name \textquotedblleft Dirac-ADM\textquotedblright\ or they refer
to Dirac when actually working with the ADM Hamiltonian. This presumes
equivalence of the Dirac and ADM formulations. These two, as we will
demonstrate, are not equivalent.

The Dirac conjecture \cite{Diracbook}, that knowing all the first-class
constraints is sufficient to derive the gauge transformations, was made only
after the appearance of \cite{Pirani, Dirac, ADM}\ and\textit{ }became a well
defined procedure only later \cite{Castref1, Castref2}. In the most general
form (with chains of constraints of any length) and with application to field
theories this procedure was considered for the first time by Castellani
\cite{Castellani} (for alternative approaches see \cite{HTZ, Novel, Novel-1};
problems or, at least, limitations of these approaches are discussed in
\cite{KKaffinemetric}). Deriving the gauge invariance of GR from the complete
set of the first-class constraints should also be viewed as a crucial
consistency condition that must be met by any Hamiltonian formulation of the
theory; yet, this requirement did not attract much attention and it is not
discussed in textbooks on GR, where a Hamiltonian formulation is presented
(e.g. \cite{Gravitation, Waldbook}). In books on constraint dynamics
\cite{Kurt, Gitmanbook, HTbook, RRbook}, even if such a procedure is discussed
\cite{HTbook, RRbook}, it is not applied to the Hamiltonian formulation of GR.
Recently this question was again brought to light by Mukherjee and Saha
\cite{Saha} who applied the method of \cite{Novel} to the ADM Hamiltonian with
the sole emphasis on presenting the method of deriving the gauge invariance,
not on the results themselves. In \cite{Saha} there appears a first complete
derivation of the gauge transformations from the constraint structure of the
ADM Hamiltonian. The expected transformation of the metric tensor is
\cite{Landau}%

\begin{equation}
\delta g_{\mu\nu}=-\xi_{\mu;\nu}-\xi_{\nu;\mu},\label{eqn00}%
\end{equation}
where $\xi_{\mu}$ is the gauge parameter and the semicolon \textquotedblleft%
$;$\textquotedblright\ signifies the covariant derivative. In the literature
on the Hamiltonian formulation of GR, the word \textquotedblleft
diffeomorphism\textquotedblright\ is often used as equivalent to the
transformation (\ref{eqn00}), which is similar to gauge transformations in
ordinary field theories. This meaning is employed in our article\footnote{In
mathematical literature the term diffeomorphism refers to a mapping from one
manifold to another which is differentiable, one-to-one, onto, with a
differentiable inverse.}. The expected transformation (\ref{eqn00}) does not
follow from the constraint structure of ADM Hamiltonian \cite{ADM} and a
\textit{field-dependent} and \textit{non-covariant} redefinition of gauge
parameters is needed\footnote{More detail on the derivation of (\ref{eqn00a})
is given in the last Section where application of Castellani's procedure to
the ADM Hamiltonian is re-examined ($\varepsilon_{ADM}^{\perp}$ and
$\varepsilon_{ADM}^{k}$ are gauge parameters of ADM formulation).} to present
the transformations of \cite{Saha} in the \textit{form} of (\ref{eqn00}), i.e.%

\begin{equation}
\xi^{0}=\left(  -g^{00}\right)  ^{1/2}\varepsilon_{ADM}^{\perp},\left.
{}\right.  \left.  {}\right.  \xi^{k}=\varepsilon_{ADM}^{k}+\frac{g^{0k}%
}{g^{00}}\left(  -g^{00}\right)  ^{1/2}\varepsilon_{ADM}^{\bot}.
\label{eqn00a}%
\end{equation}

The field-dependent redefinition of gauge parameters (\ref{eqn00a}) goes back
to work of Bergmann and Komar \cite{Bergmann} where it was presented for the
first time. The same redefinition of gauge parameters (\ref{eqn00a}), but in a
less transparent form, was obtained for the ADM Hamiltonian by Castellani
\cite{Castellani} for the transformation of the $g_{0\mu}$ components of the
metric tensor to illustrate his procedure for the construction of the gauge
generators. This redefinition of gauge parameters was also discussed from
different points of view in \cite{SalSund, PonsSS, Pons, PonsS} in an attempt
to justify the necessity of such changes or even make them compulsory. We
would like to return to footnote 1 about the Rosenfeld contribution at the
dawn of the Hamiltonian analysis.\textit{ }From the beginning he restricts his
interest to\textit{ field-independent }gauge parameters. This approach differs
from what is advocated by Salisbury with coauthors \cite{SalisburyPS}.\textit{
}As an application of a general but `not-yet' procedure, Rosenfeld considered
a tetrad formulation of gravity and, according to Salisbury, made a
\textquotedblleft crucial error\textquotedblright\ that was corrected by Pons,
Salisbury, and Shepley \cite{SalisburyPS} \textquotedblleft by introducing a
coordinate symmetry transformation group whose elements depend on
metric\textquotedblright\ \cite{Preprint}.\ But the same \textquotedblleft
crucial error\textquotedblright\ emerges from Dirac's treatment of metric
gravity; as we will demonstrate, the Dirac formulation allows one to restore
gauge invariance without the need for field-dependent gauge parameters or the
use of the field-dependent redefinition, i.e. it supports the equivalence of
the Lagrangian and Hamiltonian methods. Such an equivalence is leitmotif and
cornerstone of the Rosenfeld treatment of gauge invariant systems. Exactly the
disappearance of this equivalence in the ADM formulation leads the authors of
\cite{SalisburyPS} to introduce the\textquotedblleft diffeomorphism-induced
gauge symmetry\textquotedblright\ to support ADM variables. So, in such a case
the \textquotedblleft crucial error\textquotedblright\ of Rosenfeld is simply
that he did not use ADM variables\footnote{This point is explicitly stated in
the historical papers of Salisbury and comments, for example, in
\cite{Salisbury-hist}: \textquotedblleft Had he [Rosenfeld] expressed this
orthonormal tetrad in terms of the lapse and shift functions introduced by
Arnowitt, Deser and Misner he would have obtained the triad form of their ADM
Hamiltonian. [11]\textquotedblright\ (The Salisbury's reference [11] is our
\cite{ADM} where neither the tetrad nor the triad form of the ADM Hamiltonian
is discussed.)}. A common feature of different approaches \cite{Saha,
Bergmann, SalSund, PonsSS, Pons, PonsS, SalisburyPS} is that they consider
only the ADM Hamiltonian \cite{ADM} because this is a formulation that does
not lead to (\ref{eqn00}). According to the conclusion of \cite{Bergmann}, the
transformation (\ref{eqn00}) and the one with parameters that depend on the
fields (\ref{eqn00a}) are \textit{distinct}. In \cite{PonsSS} this
transformation is called the \textquotedblleft specific metric-dependent
diffeomorphism\textquotedblright. The authors of \cite{Saha} have a brief and
ambiguous conclusion about (\ref{eqn00a}): \textquotedblleft\lbrack it will]
lead to the \textit{equivalence}\footnote{Here and everywhere in this article
the \textit{Italic} in quotations is ours.} between the diffeomorphism and
gauge transformations\textquotedblright\ and, at the same time,
\textquotedblleft demonstrate the unity of the \textit{different} symmetries
involved\textquotedblright;\ these are contradictory statements.

Soon after appearance of \cite{Saha}, Samanta \cite{Samanta} posed the
question \textquotedblleft whether it is possible to describe the
diffeomorphism symmetries without recourse to the ADM
decomposition\textquotedblright. To answer this question, he derived the
transformation (\ref{eqn00}) starting from the Einstein-Hilbert (EH)
Lagrangian (not the ADM Lagrangian) and applying the Lagrangian method for
recovering gauge symmetries based on the use of certain gauge identities that
appear in \cite{Gitmanbook}. It is important that (\ref{eqn00}) follows
exactly from this procedure without the need of a field-dependent and
non-covariant redefinition of the gauge parameters, which would be necessary
in \cite{Castellani, Saha, Bergmann, SalSund, PonsSS, Pons, PonsS,
SalisburyPS} where the ADM Hamiltonian is used. The question of the
equivalence of (\ref{eqn00}) and (\ref{eqn00a}) does not even arise in the
approach of \cite{Samanta}. In \cite{Samanta} the diffeomorphism
transformations were also derived by applying the same method to the
first-order, affine-metric, formulation \cite{Einstein} of GR. The conclusion
of \cite{Samanta} that \textquotedblleft the ADM splitting, which is
\textit{essential} for discussing diffeomorphism symmetries, is
\textit{bypassed}\textquotedblright\ contradicts the obtained result. Firstly,
any feature that is \textquotedblleft\textit{essential}\textquotedblright%
\ cannot be \textquotedblleft\textit{bypassed}\textquotedblright. Secondly,
the transformations derived from the ADM Hamiltonian in \cite{Saha} are not
those of \cite{Samanta}. It is not a \textquotedblleft
bypass\textquotedblright\ because the \textquotedblleft
destination\textquotedblright,\ of having the invariance of (\ref{eqn00}), is changed.

Comparison of \cite{Saha} and \cite{Samanta} allows us to conclude that ADM
decomposition is not only inessential but incorrect because the EH Lagrangian
leads directly to diffeomorphism invariance without the need of any
field-dependent and non-covariant redefinition of gauge parameters; and a
correct Hamiltonian formulation should give the same result. This discrepancy
between these two recent results vindicates Hawking's old statement
\cite{Hawking} \textquotedblleft the split into three spatial dimensions and
one time dimension seems to be contrary to the whole spirit of
relativity\textquotedblright,\ the more recent statements of Pons \cite{Pons}:
\textquotedblleft Being non-intrinsic, the 3+1 decomposition is somewhat at
odds with a generally covariant formalism, and difficulties arise for this
reason\textquotedblright,\ and Rovelli \cite{Rovelli}: \textquotedblleft The
very foundation of general covariant physics is the idea that the notion of a
simultaneity surface all over the universe is devoid of physical
meaning\textquotedblright.

There is another statement in \cite{Samanta} that can also be found in many
places \textquotedblleft it is well known that this decomposition plays a
central role in \textit{all Hamiltonian formulations} of general
relativity\textquotedblright. This sentence combined with Hawking's
\textquotedblleft spiritual\textquotedblright\ statement forces one to
conclude that the Hamiltonian formulation by itself contradicts the spirit of
GR. This resonates with Pullin's conclusion \cite{Pulin} that
\textquotedblleft Unfortunately, the canonical treatment breaks the symmetry
between space and time in general relativity and the resulting algebra of
constraints is not the algebra of \textit{four diffeomorphism}%
\textquotedblright. We will show in this paper that the canonical formalism is
in fact consistent with the diffeomorphism (\ref{eqn00}) when the Dirac
constraint formalism is applied consistently and that the discrepancies
between the ADM formalism and (\ref{eqn00}) can be explained.

The difference of the results \cite{Saha} and \cite{Samanta} which were
obtained by different methods also implies the non-equivalence of the
Lagrangian and Hamiltonian formulations. In all field theories (e.g., Maxwell
or Yang-Mills) the Hamiltonian and Lagrangian formulations produce the same
result for gauge invariance, so for GR to differ seems unnatural. Could this
be a peculiar property of GR? Is GR a theory in which the Hamiltonian and
Lagrangian formulations lead to different results or was a \textquotedblleft
rule of procedure\textquotedblright\ broken somewhere?

Recently, in collaboration with Racknor and Valluri \cite{KKRV}, we
demonstrated that, by following the most natural first attempt of PSS
\cite{Pirani} and by applying the rules of procedure \cite{Dirac-1, Diracbook,
Kurt, Gitmanbook, HTbook, Castellani}, the Hamiltonian formulation of GR
(without any modifications of the action or change of variables) leads to
consistent results. The gauge transformation of the metric tensor was derived
using the method of \cite{Castellani} and, without any field-dependent
redefinitions of gauge parameters, it gives exactly the same result as the
Lagrangian approach of \cite{Samanta}, as it should. In the Hamiltonian
formulation of GR given in \cite{KKRV} the algebra of constraints is the
algebra of \textquotedblleft\textit{four diffeomorphism}\textquotedblright%
,\ in contradiction to the \textit{general} conclusion of \cite{Pulin}, which
was based on a \textit{particular}, ADM, formulation.

The procedure of passing to a Hamiltonian formulation in field theories based
on the separation of the space and time \textit{components} of the fields and
their \textit{derivatives} (defined on the whole space-time, not on some
hypersurface) is not equivalent to the separation of space-time into space and
time. For example, by rewriting the Einstein equations in components (as was
done before Einstein introduced his condensed notation), we do not abandon
covariance even if it is not manifest. In addition, such explicit separation
of the space and time \textit{components} and the \textit{derivatives} of the
fields does not affect space-time itself and is not to be associated with any
3+1 decomposition, slicing, splitting, foliation, etc. of space-time. The
final result for the gauge transformation of the fields can be presented in
covariant form when using the Hamiltonian formulation of ordinary field
theories (e.g., Yang-Mills, Maxwell), as well as in GR \cite{KKRV}. \ In any
field theory, after rewriting its Lagrangian in components, the Hamiltonian
formulation for singular Lagrangians follows a well defined procedure. Such a
procedure is based on consequent calculations of the Poisson brackets (PB) of
constraints with the Hamiltonian using the fundamental PBs of independent
fields. In the case of field theories they are%

\begin{equation}
\left\{  q\left(  x^{0},\mathbf{x}\right)  ,p\left(  y^{0},\mathbf{y}\right)
\right\}  _{x^{0}=y^{0}}=\delta\left(  \mathbf{x}-\mathbf{y}\right)  .
\label{eqn00b}%
\end{equation}

This is a local relation that does not presume any extended objects or
surfaces. Again, as with separation into components, this locally defined
canonical PB does not affect space-time and is not related to a space-like
surface or any other hypersurface because (\ref{eqn00b}) is zero for
$\mathbf{x}\neq\mathbf{y}$ in the whole space-time; and there is no
information in (\ref{eqn00b}) that, using mathematical language, can allow one
to classify two separate points as points on a particular space-like surface
or on any surface. The canonical procedure does not itself lead to the
appearance of any hypersurfaces; in \cite{KKRV} there are no references to
such surfaces and the result is consistent with the Lagrangian formulation\ of
\cite{Samanta}. Such surfaces are either a phantom of interpretation or
canonical procedure was abandoned by their introduction.

The discussion of an interpretational approach is not on the main road of our
analysis of the Hamiltonian formulations of GR. However, the routes of such an
approach\footnote{We have to confess that we found it hard to understand
approaches which are not analytical and,\ to avoid any misinterpretations, we
will merely quote their advocates. A reader interested in such approaches can
find more details in the articles we cite.} are quite interesting: one
starting from the basic equations of the ADM formulation, according to
\cite{HKT}, \textquotedblleft would like to understand intuitively their
geometrical and physical meaning and derive them from some first principles
rather than by a formal rearrangement of Einstein's law\textquotedblright. By
taking this approach, a formal rearrangement (which is a \textquotedblleft
rule of procedure\textquotedblright) is replaced by some sort of intuitive
understanding. As a result, a new language is created which \textquotedblleft
is much closer to the language of quantum dynamics than the original language
of Einstein's law ever was\textquotedblright\ \cite{HKT}. This language allows
one \textquotedblleft to recover the old comforts of a Hamiltonian-like
scheme: a system of hypersurfaces stacked in a well defined way in space-time,
with the system of dynamic variables distributed over these hypersurfaces and
developing uniquely from one hypersurface to another\textquotedblright%
\ \cite{Kuchar}.\ Such an interpretation, although `reasonable' from the point
of view of classical Laplacian determinism, is hard to justify from the
standpoint of GR \cite{Ann33}. \ In GR, an entire spatial slice can only be
seen by an observer in the infinite future \cite{Ann34} and an observer at any
point on a space-like surface does not have access to information about the
rest of the surface (this is reflected in the local nature of (\ref{eqn00b})
in field theories). It would be non-physical to build any formalism by basing
it on the development in time of data that can be available only in the
infinite future and trying to fit GR into a scheme of classical determinism
and nonrelativistic Quantum Mechanics with its notion of a wave function
defined on a space-like slice. The condition that a space-like surface remains
space-like obviously imposes restrictions on possible coordinate
transformations, thereby destroying four-dimensional symmetry, and, according
to Hawking, \textquotedblleft it restricts the topology of space-time to be
the product of the real line with some three-dimensional manifold, whereas one
would expect that quantum gravity would allow all possible topologies of
space-time including those which are not product\textquotedblright%
\ \cite{Hawking}. This restriction, imposed by the slicing of space-time, must
be lifted at the quantum level \cite{Thiemann}; but, from our point of view,
avoiding it at the outset seems to be the most natural cure for this problem.

The usual interpretation of the ADM variables, constraints, and Hamiltonian
obviously contradicts the spirit of relativity. With restrictions on
coordinate transformations, which are imposed by such an interpretation, it is
quite natural to expect something different from a diffeomorphism
transformation, as was found in \cite{Castellani, Saha}.

Any interpretation, whether or not it contradicts the spirit of GR, cannot
provide a sufficiently strong argument to prove or disprove some particular
result or theory, because an arbitrary interpretation cannot change or affect
the result of formal rearrangements. The transformation, which is different
from the diffeomorphism that follows from the ADM Hamiltonian is the result of
a definite procedure \cite{Castellani, Saha} and is based on calculations
performed with their variables and their algebra of constraints. From the
beginning we will not use the language of 3+1 dimensions, so as to avoid the
necessity of getting ourselves \textquotedblleft out of space and back into
space-time\textquotedblright\ \cite{Smolin} at the end of the calculations. In
any case, it would likely be impossible to do so after we have gone beyond the
point of no return on such a road. We must re-examine the \textit{derivation}
of ADM Hamiltonian right from the start.

It is difficult to compare the results of \cite{KKRV} directly with those of
ADM because some additional modifications of the original GR Lagrangian were
performed by ADM and it is not easy to trace them according to the
\textquotedblleft rules of procedure\textquotedblright. We will start with the
work of Dirac \cite{Dirac}, where all modifications and assumptions are
explicitly stated making it possible for them to be checked and analyzed. In
addition, Dirac's canonical variables are components of the metric tensor
which are the same as those used in \cite{KKRV}, where diffeomorphism
invariance was derived directly from the Hamiltonian and constraints.
Moreover, in \cite{GKK} two Hamiltonian formulations, based on the linearized
Lagrangians of \cite{Pirani, KKRV} and \cite{Dirac}, were considered. Despite
there being different expressions for the primary and secondary constraints,
these two formulations have the same algebra of PBs among the constraints, and
with the Hamiltonian; therefore, they have the same gauge invariance. This is
exactly what one can expect in the case of full GR, provided one makes no
deviation from canonical procedure. In analyzing the ADM formulation we will
follow a different path. We will not start from the GR Lagrangian, but instead
compare the final results of Dirac and ADM and try to determine what deviation
from the canonical procedure leads to the transformations found in
\cite{Castellani, Saha}, which are distinct from those of (\ref{eqn00}).

In the next Section we shall thoroughly re-examine the Dirac derivation of the
GR Hamiltonian \cite{Dirac} with emphasis on the effect of his modifications
of the action and of the other simplifying assumptions he made. In particular,
we will investigate whether space-like surfaces actually play any role in his
derivation, or if they just serve as an illustration, which can be completely
disregarded from the standpoint of the canonical procedure, as in \cite{KKRV}.
In Section III, using Castellani's procedure and the results of Section II
(Dirac's constraints), we derive the gauge transformations of the metric
tensor, diffeomorphism (\ref{eqn00}), without the need of a field-dependent
redefinition of gauge parameters. This result is identical to that found in
\cite{Samanta} for metric GR using the Lagrangian method and in \cite{KKRV}
for the Hamiltonian formulation of PSS. The same result is obtained by
application of the method \cite{Novel} used for the ADM Hamiltonian in
\cite{Saha} when it is applied to the Dirac Hamiltonian. Some peculiarities of
such methods that cannot be seen in ordinary field theories are briefly
discussed and related to the peculiarities of diffeomorphism invariance as it
compares to the gauge invariance in ordinary theories. Finally, we consider
the ADM Hamiltonian formulation of GR. In the last Section IV we demonstrate
that the ADM formulation follows from Dirac's by a change of variables. The
canonicity of this change of variables (the ADM\ lapse and shift functions) is
analyzed. Based on this analysis, the general and more restrictive criteria
for a canonical transformation in the case of singular gauge invariant
theories are discussed.

\section{Analysis of Dirac derivation}

In \cite{KKRV} the GR Hamiltonian, constraints, closure of the Dirac
procedure, and the diffeomorphism transformation of the metric tensor were
derived without any reference to space-like surfaces, the use of any 3+1
decomposition of space-time, or slicing, splitting, foliation, etc., as well
as without modifications of the Lagrangian or the introduction of any new
variables. (The canonical variables of \cite{KKRV} are components of the
metric tensor.) Dirac, when considering the Hamiltonian formulation of GR in
\cite{Dirac}, also used the metric tensor as a canonical variable; but he made
frequent references to space-like surfaces. If such surfaces, which according
to Hawking \cite{Hawking} contradict the whole spirit of General Relativity,
are the part of Dirac's calculations, then one has to expect transformations
different from diffeomorphism and similar to the one found in \cite{Saha} from
the ADM Hamiltonian. Our main interest is to find out, by following all the
steps of Dirac's derivation of the Hamiltonian, the place where (if anywhere)
space-like surfaces enter his \textit{derivation }or where (if anywhere) his
approach \textit{deviates }from a regular and uniform rule of canonical
procedure. If there is no deviation, one should then obtain the diffeomorphism
invariance (\ref{eqn00}), the same as found in \cite{KKRV}. This would
resemble what happens in linearized GR, as discussed in \cite{GKK}.

\subsection{Dirac's modification of Lagrangian and primary constraints}

In \cite{Dirac}, Dirac started the Hamiltonian formulation from the
\textquotedblleft gamma-gamma\textquotedblright\ part of the Einstein-Hilbert
(EH) Lagrangian (Eq. (D8))\footnote{We will refer on Dirac equations quite
often and use the convention, Eq. (D\#), to mean equation \# from\textsf{
\cite{Dirac}.}}\textsf{ }(e.g., see \cite{Landau, Carmeli})%

\begin{equation}
L_{G}=\sqrt{-g}g^{\mu\nu}\left(  \Gamma_{\mu\nu}^{\rho}\Gamma_{\rho\sigma
}^{\sigma}-\Gamma_{\mu\rho}^{\sigma}\Gamma_{\nu\sigma}^{\rho}\right)
=\frac{1}{4}\sqrt{-g}g_{\mu\nu,\rho}g_{\alpha\beta,\sigma}B^{\mu\nu\rho
\alpha\beta\sigma} \label{eqn01}%
\end{equation}
where%

\begin{equation}
B^{\mu\nu\rho\alpha\beta\sigma}=\left(  g^{\mu\alpha}g^{\nu\beta}-g^{\mu\nu
}g^{\alpha\beta}\right)  g^{\rho\sigma}+2\left(  g^{\mu\rho}g^{\alpha\beta
}-g^{\mu\alpha}g^{\beta\rho}\right)  g^{\nu\sigma}. \label{eqn02}%
\end{equation}

The same Lagrangian was used in \cite{Pirani} and \cite{KKRV}. This is a
covariant Lagrangian of a local field theory in four(or any)-dimensional
space-time, and space-like surfaces or any other hypersurfaces are not
intrinsic to such a formulation.

The primary constraints (the $\phi$-equations of \cite{Dirac}) that follow
from (\ref{eqn01}) are%

\begin{equation}
\phi^{\mu0}=p^{\mu0}-\frac{\delta L_{G}}{\delta g_{\mu0,0}}\approx0,
\label{eqn03}%
\end{equation}
where $p^{\mu\nu}$ are momenta conjugate to $g_{\mu\nu}$. The exact form of
$\phi^{\mu0}$ can be found in \cite{Pirani, KKRV} (Greek subscripts run from
$0$ to $d-1$ and Latin ones from $1$ to $d-1$ where $d$ is the dimension of space-time).

In addition to eliminating the second order derivatives of the metric tensor
present in the Ricci scalar in passing from the EH Lagrangian to its
gamma-gamma part (\ref{eqn01}) so that \cite{Carmeli}%

\begin{equation}
L_{EH}=\sqrt{-g}R=L_{G}+\partial_{\mu}V^{\mu}, \label{eqn03.1}%
\end{equation}
Dirac made an additional change to the Lagrangian in order to eliminate the
second term in (\ref{eqn03}). The modified Lagrangian is obtained by adding
two total derivatives which are non-covariant (Eq. (D15))%

\begin{equation}
L^{\ast}=L_{G}+\left[  \left(  \sqrt{-g}g^{00}\right)  _{,v}\frac{g^{v0}%
}{g^{00}}\right]  _{,0}-\left[  \left(  \sqrt{-g}g^{00}\right)  _{,0}%
\frac{g^{v0}}{g^{00}}\right]  _{,v}. \label{eqn04}%
\end{equation}
\ This change does not affect the equations of motion, but leads to simple
primary constraints (Eq. (D14))%
\begin{equation}
\phi^{\mu0}=p^{\mu0}\approx0. \label{eqn05}%
\end{equation}

It was shown in \cite{GKK}, that the linearized version of the modified
(\ref{eqn04}) and unmodified Lagrangians (\ref{eqn01}), despite leading to
different expressions for the constraints and the Hamiltonian, result in the
same constraint structure, the same number of first-class constraints, and the
same gauge invariance, which is the linearized version of diffeomorphism
(\ref{eqn00}). This is what one can also expect in the case of full
GR.\ According to \cite{Dirac}, the simplification (\ref{eqn05})
\textquotedblleft can be achieved only at the expense of abandoning
four-dimensional symmetry\textquotedblright\ which is obviously correct for
this modification of the Lagrangian (\ref{eqn04}); yet Dirac's further
conclusion that \textquotedblleft four-dimensional symmetry is not a
fundamental property of the physical world\textquotedblright\ is too strong
and has to be clarified. Of course, four-dimensional symmetry of the
Lagrangian is destroyed by the modification (\ref{eqn04}); but this change
does not affect the equations of motion, which are the same as the Einstein
equations. Consequently, for the equations of motion, not only
four-dimensional symmetry is preserved, but also general
covariance\footnote{The term \textquotedblleft
four-dimensional\textquotedblright\ symmetry used by Dirac probably reflects
the fact that the gamma-gamma part of the Lagrangian, quadratic in first order
derivatives, is not generally covariant after the elimination of terms with
second order derivatives in the full EH Lagrangian (\ref{eqn03.1}).}.\textsf{
}If four-dimensional symmetry is preserved in the equations of motion, which
are invariant under general coordinate transformations, then diffeomorphism
should be recovered in the course of the Hamiltonian analysis, as in
\cite{KKRV}. \ 

The new Lagrangian $L^{\ast}$ differs from the original one (\ref{eqn01}) only
for terms linear in the time derivatives of a metric (i.e. `velocities'),
which are the parts responsible for the simplification of the primary
constraints. We then have%

\begin{equation}
L^{\ast}=L_{G}\left(  2\right)  +L^{\ast}\left(  1\right)  +L_{G}\left(
0\right)  , \label{eqn06}%
\end{equation}
where the numbers in brackets indicate the order in velocities (for the
Hamiltonian and constraints it will indicate the order in momenta). The exact
form of $L^{\ast}\left(  1\right)  $ is given by Eq. (D18).

This Lagrangian is used to pass to the Hamiltonian%

\begin{equation}
H=g_{\alpha\beta,0}p^{\alpha\beta}-L^{\ast}. \label{eqn06.0}%
\end{equation}
With the modification of (\ref{eqn04}) the part of the Lagrangian
$L_{G}\left(  2\right)  +L^{\ast}\left(  1\right)  $, as was shown by Dirac,
can be written as%

\begin{equation}
L_{G}\left(  2\right)  +L^{\ast}\left(  1\right)  =L_{X}\left(  0\right)
-\sqrt{-g}\frac{1}{g^{00}}E^{rsab}\Gamma_{rs}^{0}\Gamma_{ab}^{0}
\label{eqn06.01}%
\end{equation}
where $\Gamma_{\alpha\beta}^{\mu}$ is the Christoffel symbol%

\begin{equation}
\Gamma_{\alpha\beta}^{\mu}=\frac{1}{2}g^{\mu\nu}\left(  g_{\alpha\nu,\beta
}+g_{\beta\nu,\alpha}-g_{\alpha\beta,\nu}\right)  \label{eqn06.02}%
\end{equation}
and%

\begin{equation}
E^{rsab}=e^{rs}e^{ab}-e^{ra}e^{sb} \label{eqn06.03}%
\end{equation}
with%

\begin{equation}
e^{\alpha\beta}=g^{\alpha\beta}-\frac{g^{0\alpha}g^{0\beta}}{g^{00}}.
\label{eqn06.04}%
\end{equation}
Note, that in the second order formulation, $\Gamma_{\alpha\beta}^{\mu}$,
$E^{\alpha\beta\mu\nu}$, and $e^{\alpha\beta}$ are just short notations and
none of them denote a new and/or independent variable.

Some comments about (\ref{eqn06.01}) are in order. The careful reader will
definitely wonder how the parts of the Lagrangian which are quadratic and
linear in velocities can have contributions without velocities, $L_{X}\left(
0\right)  $; the direct calculation of $L_{G}\left(  2\right)  +L^{\ast
}\left(  1\right)  $ does not have such contributions (see Dirac's unnumbered
equation preceding (D19))%

\begin{equation}
\frac{1}{4}\sqrt{-g}E^{rasb}\left(  g_{rs,0}g_{ab,0}g^{00}+2g_{rs,0}%
g_{ab,v}g^{v0}-4g_{rs,0}g_{a\beta,b}g^{\beta0}\right)  . \label{eqn06.04.1}%
\end{equation}
Dirac completed this expression to the square in velocities,\ leading to the
compact form of (\ref{eqn06.01}). Working with (\ref{eqn06.04.1}) instead of
(\ref{eqn06.01}), will of course not change the results and actually has no
calculational advantage. However, we keep (\ref{eqn06.01}) so as to compare
our calculations with those of Dirac.

The $L_{X}\left(  0\right)  $ in (\ref{eqn06.01}) (explicitly given by
(\ref{eqn08})) is independent of the velocities. The only part of
(\ref{eqn06.0}) that has dependence on $g_{rs,0}$ is%

\begin{equation}
g_{rs,0}p^{rs}+\sqrt{-g}\frac{1}{g^{00}}E^{rsab}\Gamma_{rs}^{0}\Gamma_{ab}%
^{0}. \label{eqn06.05}%
\end{equation}
Performing the variation $\frac{\delta L^{\ast}}{\delta g_{rs,0}}$, we obtain
(see (D18-D21))%

\begin{equation}
p^{rs}=\sqrt{-g}E^{rsab}\Gamma_{ab}^{0}=\frac{1}{2}\sqrt{-g}E^{rsab}\left[
g^{00}\left(  g_{a0,b}+g_{b0,a}-g_{ab,0}\right)  +g^{0k}\left(  g_{ak,b}%
+g_{bk,a}-g_{ab,k}\right)  \right]  . \label{eqn06.1}%
\end{equation}
Equation (\ref{eqn06.1}) is easy to solve for $g_{ab,0}$ due to the
invertability of $E^{rsab}$%

\begin{equation}
E^{rsab}I_{abmn}=\delta_{m}^{r}\delta_{n}^{s}, \label{eqn06.2}%
\end{equation}
where the inverse to $E^{rsab}$ in any space-time dimension $d$ (except $d=2$) is%

\begin{equation}
I_{abmn}=\frac{1}{d-2}g_{ab}g_{mn}-g_{am}g_{bn}. \label{eqn06.3}%
\end{equation}
This result gives%

\begin{equation}
g_{mn,0}=-\frac{2}{\sqrt{-g}g^{00}}p^{rs}I_{rsmn}+g_{m0,n}+g_{n0,m}%
+\frac{g^{0k}}{g^{00}}\left(  g_{mk,n}+g_{nk,m}-g_{mn,k}\right)  .
\label{eqn06.4}%
\end{equation}

After substitution of (\ref{eqn06.4}) into (\ref{eqn06.05}) (note that
(\ref{eqn06.1}) can be solved for $\Gamma_{ab}^{0}$ thus making the
calculations shorter) we obtain the total Hamiltonian%

\begin{equation}
H_{T}=g_{00,0}p^{00}+2g_{0k,0}p^{0k}+H_{G}, \label{eqn06.5}%
\end{equation}
where $H_{G}$ (the canonical part of the Hamiltonian) is given by Eqs. (D33,
D34)) as,%

\begin{equation}
H_{G}=-\frac{1}{g^{00}\sqrt{-g}}I_{rsab}p^{rs}p^{ab}+g_{u0}e^{uv}\left[
p^{rs}g_{rs,v}-2\left(  p^{rs}g_{rv}\right)  _{,s}\right]  -L_{X}\left(
0\right)  -L_{G}\left(  0\right)  , \label{eqn7.1}%
\end{equation}
with $L_{X}\left(  0\right)  $ (Eq. (D19)) and $L_{G}\left(  0\right)  $ (Eq. (D8)):%

\begin{equation}
L_{X}\left(  0\right)  =\frac{1}{4}\frac{\sqrt{-g}}{g^{00}}E^{rsab}\left[
g_{rs,u}g^{u0}-\left(  g_{r\alpha,s}+g_{s\alpha,r}\right)  g^{\alpha0}\right]
\left[  g_{ab,v}g^{v0}-\left(  g_{a\beta,b}+g_{b\beta,a}\right)  g^{\beta
0}\right]  , \label{eqn08}%
\end{equation}

\begin{equation}
L_{G}\left(  0\right)  =\frac{1}{4}\sqrt{-g}g_{\mu\nu,k}g_{\alpha\beta
,t}B^{\mu\nu k\alpha\beta t}.\label{eqn09}%
\end{equation}
Note that the second term of (\ref{eqn7.1}), the part linear in the momenta,
arises only after some rearrangement, and $B^{\mu\nu k\alpha\beta t}$ in
(\ref{eqn09}) was defined in\textit{ }(\ref{eqn02}). The direct substitution
of $g_{rs,0}$ into $g_{rs,0}p^{rs}$ (the only part of (\ref{eqn06.0}) that
leads to terms linear in the momenta) gives%

\begin{equation}
2p^{mn}g_{m0,n}+\frac{g^{0k}}{g^{00}}p^{mn}\left(  2g_{mk,n}-g_{mn,k}\right)
, \label{eqn09.1}%
\end{equation}
which after integration by parts and using $\frac{g^{0k}}{g^{00}}%
=-g_{0m}e^{mk}$ leads to%

\begin{equation}
-2g_{m0}\left[  p_{,n}^{mn}+e^{mk}p^{rn}\left(  g_{rk,n}-\frac{1}{2}%
g_{rn,k}\right)  \right]  +2\left(  p^{mn}g_{m0}\right)  _{,n}.\label{eqn09.2}%
\end{equation}
The first term of (\ref{eqn09.2}) can be written in the form given by Dirac (D41)%

\begin{equation}
-2g_{m0}\left[  p_{,n}^{mn}+e^{mk}p^{rn}\left(  g_{rk,n}-\frac{1}{2}%
g_{rn,k}\right)  \right]  =g_{m0}e^{mv}\mathcal{H}_{v},\label{eqn09.2.1}%
\end{equation}
with%

\begin{equation}
\mathcal{H}_{v}=p^{rs}g_{rs,v}-2\left(  p^{rs}g_{rv}\right)  _{,s}.
\label{eqn09.2.2}%
\end{equation}

We note that in obtaining the expression for the Hamiltonian (\ref{eqn7.1}),
all direct calculations with the initially modified Lagrangian (\ref{eqn06.01}%
) were performed by Dirac without any reference to space-like surfaces or any
additional restrictions or assumptions.

\subsection{Dirac's secondary constraints}

The next step in the canonical procedure is to find the time development of
the primary constraints and see if there are any secondary constraints (or
$\chi$-equations in Dirac's terminology). Poisson brackets (PBs) among the
primary constraints are obviously zero, $\left\{  p^{0\alpha},p^{0\beta
}\right\}  =0$. The PBs of the primary constraints (\ref{eqn05}) with the
total Hamiltonian (\ref{eqn06.5}) are%

\begin{equation}
\left\{  p^{0\sigma},H_{T}\right\}  =\frac{\delta}{\delta g_{0\sigma}}%
H_{G}=\chi^{0\sigma}, \label{eqn09.3}%
\end{equation}
where we keep Dirac's convention for the fundamental PB (Eq. (D11)),%

\begin{equation}
\left\{  p^{\alpha\beta}\left(  x\right)  ,g_{\mu\nu}\left(  x^{\prime
}\right)  \right\}  =\frac{1}{2}\left(  \delta_{\mu}^{\alpha}\delta_{\nu
}^{\beta}+\delta_{\mu}^{\beta}\delta_{\nu}^{\alpha}\right)  \delta_{3}\left(
x-x^{\prime}\right)  . \label{eqn09.4}%
\end{equation}

According to Dirac, \textquotedblleft the second term of (D33) [$H_{G}\left(
0\right)  =-L_{X}\left(  0\right)  -L_{G}\left(  0\right)  $ in our
(\ref{eqn7.1})] is very complicated and a great deal of labour would be needed
to calculate it directly\textquotedblright\ and instead of performing the
variation $\frac{\delta}{\delta g_{0\sigma}}H_{G}\left(  0\right)  $, he uses
some arguments (see (D23-D27)) related to the displacements of surfaces of
constant time, and thus he infers that the Hamiltonian \textquotedblleft must
be of the form\textquotedblright\ (see (D28))%

\[
H=\left(  g^{00}\right)  ^{-\frac{1}{2}}\mathcal{H}_{L}+g_{r0}e^{rs}%
\mathcal{H}_{s}~.
\]
The explicit form of $H_{L}$ and $H_{s}$ was not given at this stage but a
statement about their independence of the $g_{0\mu}$ was made based on some
arguments which are very general and independent of the particular form of the
Lagrangian, i.e. they have no connection with his initial modifications of
$L_{G}$ leading to $L^{\ast}$. And, even in the linearized case \cite{GKK},
without these modifications, the secondary constraints have a dependence on
$g_{0\mu}$; this dependence can be also observed in full GR \cite{KKRV}. In
any case, the explicit form of the constraints cannot be found using such
arguments and explicit calculations are needed; one has to use a well defined
rule of procedure to find them, i.e. we must calculate $\frac{\delta}{\delta
g_{0\sigma}}H_{G}\left(  0\right)  $ . Dirac performed these calculations
using an additional simplifying assumption (see below) and this result has to
be analyzed and compared to what follows from direct calculations.

According to Dirac, there are no contributions from $H_{G}\left(  0\right)  $
to a vector constraint ($\mathcal{H}^{r}=e^{rs}\mathcal{H}_{s}$ in Dirac's
notation) which presumably comes from the time development of the
corresponding primary constraint $\phi^{r0}=p^{r0}$ (\ref{eqn09.3}).
Furthermore, $\mathcal{H}_{L}$, which comes from the time development of the
primary constraint, $\phi^{00}$, can be calculated with the additional
simplifying assumption $g_{r0}=0$, which gives (Eq. (D36)):%

\begin{equation}
g^{r0}=0,\left.  {}\right.  g^{rs}=e^{rs},\left.  {}\right.  g^{00}=\frac
{1}{g_{00}}. \label{eqn010}%
\end{equation}
As a result, all of $L_{X}\left(  0\right)  $, along with the biggest part of
$L_{G}\left(  0\right)  $, is dropped from his calculations. According to
Dirac \cite{Dirac}, the equation for\ $\mathcal{H}_{L}$ \textquotedblleft must
hold also when $g_{r0}$ does not vanish\textquotedblright. It is important to
check this assumption by direct calculation because if the result of
$\frac{\delta}{\delta g_{0\sigma}}H_{G}\left(  0\right)  $ is the same as that
of Dirac's, then the simplifying assumption of (\ref{eqn010}), along with any
references to surfaces of constant time, has nothing to do with his final
result. In such a case, Hawking's criticism of formulations based on the
introduction of space-like surfaces, which is in contradiction with the whole
spirit of General Relativity and restricts topology of
space-time\ \cite{Hawking}, cannot be applied to the Dirac analysis of GR.
This also means that the transformations (\ref{eqn00}) should be derivable in
the Dirac Hamiltonian formulation, as was done in the Lagrangian formulation
\cite{Samanta} or for the Hamiltonian formulation obtained in \cite{KKRV}.

If the results following from the assumption of (\ref{eqn010}) are different
from those where the assumption is not made, then we cannot use (\ref{eqn010})
as an extra condition in the midst of the calculations and we have to go back
to the original Lagrangian to introduce this condition from the outset. This
is the rule followed in ordinary constraint dynamics; all \textit{imposed
}constraints must be solved at the Lagrangian level, or added to the
Lagrangian using Lagrange multipliers, before performing a variation and/or
considering the Hamiltonian formulation.

For example, when Chandrasekhar considers the Hamiltonian for Schwarzschild
space-time he, first of all, writes the Lagrangian using this metric and only
then passes to the Hamiltonian formulation \cite{Chand}. Similarly, the
condition (\ref{eqn010}) corresponds to a particular coordinate system, one
which is \textit{static} \cite{Landau, Carmeli}; and, of course, the momenta
$p^{0k}$, which are conjugate to the eliminated variables $g_{0k}$ cannot
appear in such a formulation. Note that the initial modification of the
Lagrangian (\ref{eqn04}) is irrelevant in a static coordinate system as the
last two terms in (\ref{eqn04}) are zero when $g^{0k}$ is zero. For field
theories, especially generally covariant ones, there is an additional
restriction: the unambiguous canonical formulation must be performed without
explicit reference to ambient space-time by making an \textit{a priori} choice
of a particular coordinate system or subclass of coordinate systems
\cite{Isham}, i.e. without destroying the main feature of a theory from the beginning.

\textit{To find out whether or not Dirac's formulation is correct or any
reference to surfaces of constant time and the simplifications of
(\ref{eqn010}) (or (D36) of \cite{Dirac}\} are relevant to his actual
results}, we perform a \textquotedblleft great deal of
labour\textquotedblright\ to find the functional derivatives $\frac{\delta
}{\delta g_{0\sigma}}$ separately for each contribution of $H_{G}\left(
0\right)  =-L_{X}\left(  0\right)  -L_{G}\left(  0\right)  $ and to compare
the results with those obtained by Dirac.

For $L_{G}\left(  0\right)  $ in (\ref{eqn09}), we find that%

\[
\chi_{G}^{0\sigma}\left(  0\right)  =\left\{  p^{0\sigma},-L_{G}\left(
0\right)  \right\}  =-\frac{1}{2}\sqrt{-g}g_{\alpha\beta,kt}\left(
g^{0\sigma}E^{\alpha\beta kt}-g^{0t}E^{\alpha\beta k\sigma}-g^{0\beta
}E^{tk\sigma\alpha}\right)
\]

\begin{equation}
+\frac{1}{4}\sqrt{-g}g_{\mu\nu,k}g_{\alpha\beta,t}\left[  C^{\sigma\mu\nu
k\alpha\beta t}\left(  eee\right)  +C^{\sigma\mu\nu k\alpha\beta t}\left(
ee\right)  \right]  ,\label{eqn10}%
\end{equation}
where the $C$'s are combinations of terms of different order in $e^{\alpha
\beta}$ (note that the terms of first and zero orders in $e^{\alpha\beta}$ cancel)%

\[
C^{\sigma\mu\nu k\alpha\beta t}\left(  eee\right)  =g^{0\sigma}\left(
-\frac{1}{2}E^{\mu\nu\alpha\beta}e^{kt}+E^{kt\alpha\nu}e^{\mu\beta}%
+2E^{\alpha\beta\nu t}e^{\mu k}\right)
\]

\begin{equation}
+g^{0k}\left(  e^{\beta\mu}E^{\sigma\nu t\alpha}+e^{\sigma\nu}E^{\alpha
t\beta\mu}\right)  +g^{0\alpha}\left(  e^{\mu\nu}E^{\sigma\beta tk}%
+2e^{\nu\beta}E^{\sigma t\mu k}-2e^{\nu t}E^{\sigma\beta\mu k}\right)
\label{eqn11}%
\end{equation}
and%

\[
C^{\sigma\mu\nu k\alpha\beta t}\left(  ee\right)  =\frac{g^{0\alpha}g^{\beta
0}}{g^{00}}\left(  E^{tk\mu\sigma}g^{\nu0}-2E^{kt\mu\sigma}g^{\nu0}-E^{\mu
\nu\sigma t}g^{k0}\right)
\]

\begin{equation}
+\frac{g^{0\sigma}}{g^{00}}\left(  -\frac{1}{2}E^{\mu\nu\alpha\beta}%
g^{k0}g^{t0}+E^{kt\mu\beta}g^{\alpha0}g^{\nu0}+E^{\mu tk\alpha}g^{\beta
0}g^{\nu0}+2E^{\alpha\beta k\mu}g^{\nu0}g^{t0}\right)  . \label{eqn12}%
\end{equation}

When $\sigma=0$, the expression (\ref{eqn10}) is considerably simplified (this
is because $e^{\alpha\beta}$ or $E^{\mu\nu\alpha\beta}$ equal to zero if at
least one index is zero):%

\[
\chi_{G}^{00}\left(  0\right)  =-\frac{1}{2}\sqrt{-g}g_{\alpha\beta,kt}%
g^{00}E^{\alpha\beta kt}+\frac{1}{4}\sqrt{-g}g_{\mu\nu,k}g_{\alpha\beta
,t}\left[  g^{00}\left(  -\frac{1}{2}E^{\mu\nu\alpha\beta}e^{kt}%
+E^{kt\alpha\nu}e^{\mu\beta}+2E^{\alpha\beta\nu t}e^{\mu k}\right)  \right.
\]

\begin{equation}
\left.  -\frac{1}{2}E^{\mu\nu\alpha\beta}g^{k0}g^{t0}+E^{kt\mu\beta}%
g^{\alpha0}g^{\nu0}+E^{\mu tk\alpha}g^{\beta0}g^{\nu0}+2E^{\alpha\beta k\mu
}g^{\nu0}g^{t0}\right]  . \label{eqn13}%
\end{equation}

According to Dirac, this $L_{G}\left(  0\right)  $ is the only source of
contributions to the scalar constraint and he constructed it using the
simplifying assumption of (\ref{eqn010}) and later concluded that it
\textquotedblleft must hold also when $g_{0r}$ does not
vanish\textquotedblright. Let us check this assertion by explicitly separating
all space and time indices in (\ref{eqn13})%

\[
\chi_{G}^{00}\left(  0\right)  =-\frac{1}{2}\sqrt{-g}g_{mn,kt}g^{00}%
E^{mnkt}+\frac{1}{4}\sqrt{-g}g_{mn,k}g_{pq,t}g^{00}\left(  -\frac{1}%
{2}E^{mnpq}e^{kt}+E^{ktpn}e^{mq}+2E^{pqnt}e^{mk}\right)
\]

\[
+\frac{1}{4}\sqrt{-g}g_{m0,k}g_{0q,t}g^{00}g^{00}\left(  E^{ktmq}%
+E^{mtkq}\right)  +\frac{1}{2}\sqrt{-g}g_{m0,k}g_{pq,t}g^{00}\left[
E^{pqkm}g^{t0}+g^{p0}\left(  E^{ktmq}+E^{mtkq}\right)  \right]
\]

\begin{equation}
+\frac{1}{2}\sqrt{-g}g_{mn,k}g_{pq,t}\left(  -\frac{1}{2}E^{mnpq}g^{k0}%
g^{t0}+E^{ktmq}g^{p0}g^{n0}+E^{mtkp}g^{q0}g^{n0}+2E^{pqkm}g^{n0}g^{t0}\right)
. \label{eqn14}%
\end{equation}

Some terms in (\ref{eqn14}) have explicit dependence on the space-time
components of the metric tensor and these components will disappear only if
condition (\ref{eqn010}) is imposed. For $\chi_{G}^{0k}\left(  0\right)  $
there are even more such contributions. Even with condition (\ref{eqn010}),
the result is not zero and this part of the Hamiltonian, $L_{G}\left(
0\right)  $, contributes to the vector constraint.

Now let us find contributions coming from the second part,\ $L_{X}\left(
0\right)  $. After a rearrangement of the terms given in\ (\ref{eqn08}) into a
form which is more suitable for calculation, we obtain%

\[
L_{X}\left(  0\right)  =\frac{1}{2}\frac{\sqrt{-g}}{g^{00}}\left[  \frac{1}%
{2}E^{rskb}g_{rs,u}g_{kb,v}g^{u0}g^{v0}-g_{k\beta,b}g_{rs,u}\left(
E^{rskb}+E^{rsbk}\right)  g^{u0}g^{\beta0}\right.
\]

\begin{equation}
\left.  +g_{r\alpha,s}g_{k\beta,b}\left(  E^{rskb}+E^{rsbk}\right)  g^{\beta
0}g^{\alpha0}\right]  .\label{eqn30}%
\end{equation}
For the $\frac{\delta L_{X}\left(  0\right)  }{\delta g_{0\sigma}}$ part we calculate%

\[
\chi_{X}^{0\sigma}\left(  0\right)  =\left\{  p^{0\sigma},-L_{X}\left(
0\right)  \right\}  =
\]

\begin{equation}
\frac{1}{2}\sqrt{-g}\delta_{k}^{\sigma}\left(  -g_{rs,ub}E^{rskb}%
g^{u0}+g_{r\alpha,sb}E^{rskb}g^{\alpha0}\right)  +C^{\sigma}\left(
eee\right)  +C_{I}^{\sigma}\left(  ee\right)  +C_{II}^{\sigma}\left(
ee\right)  .\label{eqn31}%
\end{equation}

The variation $\frac{\delta L_{X}\left(  0\right)  }{\delta g_{0\sigma}}$
obviously produces contributions which are only third and second order in
$e^{\alpha\beta}$ as in (\ref{eqn10}). For terms of third order we find%

\[
C^{\sigma}\left(  eee\right)  =-\frac{1}{4}\sqrt{-g}g_{\mu\nu,k}g_{\alpha
\beta,t}%
\]

\begin{equation}
\times\left[  g^{0k}\left(  e^{\beta\mu}E^{\sigma\nu t\alpha}+e^{\sigma\nu
}E^{\alpha t\beta\mu}\right)  +g^{0\alpha}\left(  e^{\mu\nu}E^{\sigma\beta
tk}+2e^{\nu\beta}E^{\sigma t\mu k}-2e^{\nu t}E^{\sigma\beta\mu k}\right)
\right]  \label{eqn32}%
\end{equation}
and in second order we have two contributions: the first proportional to
$g^{0\sigma}$%

\begin{equation}
C_{I}^{\sigma}\left(  ee\right)  =\frac{1}{4}\sqrt{-g}\frac{g^{0\sigma}%
}{g^{00}}\left[  \frac{1}{2}g_{rs,u}g_{ab,v}E^{rsab}g^{u0}g^{v0}-g_{a\beta
,b}g^{\beta0}\left(  E^{rsab}+E^{rsba}\right)  \left(  g_{rs,u}g^{u0}%
-g_{r\beta,s}g^{\beta0}\right)  \right]  \label{eqn33}%
\end{equation}
and the second with an index $\sigma$ on $E^{rs\sigma k}$%

\begin{equation}
C_{II}^{\sigma}\left(  ee\right)  =\frac{1}{4}\sqrt{-g}g_{\mu\nu,k}%
\frac{g^{0\nu}g^{0\mu}}{g^{00}}\left[  \frac{1}{2}g_{rs,t}\left(  E^{rs\sigma
k}+E^{rsk\sigma}\right)  g^{t0}-g_{r\beta,t}g^{\beta0}\left(  E^{\sigma
krt}+E^{\sigma ktr}\right)  \right]  . \label{eqn34}%
\end{equation}

Note, that we cannot present the part quadratic in $e^{\alpha\beta}$
(\ref{eqn33}), (\ref{eqn34}) in a compact form, where terms with derivatives
are a common factor, because of the mixture of four and three indices, which
is the result of the original noncovariant modification (\ref{eqn04}) of the
Lagrangian. When performing these calculations we have to consider all
possible combinations separately.

It is not difficult to confirm that $\chi_{X}^{0\sigma}\left(  0\right)  $ is
not zero; even with assumption (\ref{eqn010}), there are contributions to both
constraints $\chi_{X}^{00}\left(  0\right)  $ and $\chi_{X}^{0k}\left(
0\right)  $. Consequently, Dirac's conjecture, if made separately for
$L_{X}(0)$ and $L_{G}(0)$, is not correct; but, when both parts are combined,
the contribution of zeroth order to the secondary constraint is greatly simplified%

\[
\chi^{0\sigma}\left(  0\right)  =\chi_{G}^{0\sigma}\left(  0\right)  +\chi
_{X}^{0\sigma}\left(  0\right)  =
\]

\begin{equation}
\frac{1}{2}\sqrt{-g}g^{0\sigma}\left[  -g_{mn,kt}E^{mnkt}+\frac{1}{4}%
g_{mn,k}g_{pq,t}\left(  -E^{mnpq}e^{kt}+2E^{ktpn}e^{mq}+4E^{pqnt}%
e^{mk}\right)  \right]  . \label{eqn20}%
\end{equation}

The $\chi^{00}\left(  0\right)  $-part is the same with or without condition
(\ref{eqn010}) and $\chi^{0k}\left(  0\right)  $ is given by (\ref{eqn20})
with $\sigma=k$ (it is zero when (\ref{eqn010}) is imposed). Frequently
(\ref{eqn20}) is written in a different form which is based on the following
observation: if in the expression for the four-dimensional Ricci scalar $R$%

\[
R=g^{\alpha\beta}g^{\mu\nu}R_{\alpha\mu\beta\nu}=g_{\alpha\beta,\mu\nu}\left(
g^{\alpha\mu}g^{\beta\nu}-g^{\alpha\beta}g^{\mu\nu}\right)  -\frac{1}%
{4}g_{\alpha\beta,\gamma}g_{\mu\nu,\rho}%
\]

\begin{equation}
\times\left(  g^{\alpha\beta}g^{\mu\nu}g^{\gamma\rho}-3g^{\alpha\mu}%
g^{\beta\nu}g^{\gamma\rho}+2g^{\alpha\rho}g^{\beta\nu}g^{\gamma\mu}%
+4g^{\alpha\gamma}g^{\mu\rho}g^{\beta\nu}-4g^{\alpha\gamma}g^{\beta\rho}%
g^{\mu\nu}\right)  , \label{eqn20.1}%
\end{equation}
we keep only the spatial indices and change the covariant component of
$g^{km}$ to $e^{km}$ or, equivalently, impose the conditions (\ref{eqn010}),
we obtain the expression shown in square brackets of (\ref{eqn20}), which is
often called the three-dimensional Ricci scalar $R_{\left(  3\right)  }$.

Equation (\ref{eqn20}) gives contributions to the secondary constraints of
zeroth order in the momenta $p^{km}$. There are obviously such contributions
in $\chi^{0k}$. Dirac's vector constraint, $\mathcal{H}^{r}$, does not have
such contributions, so it is not directly related to the time development of
the corresponding primary constraint $p^{0k}$ (we will discuss this later).

For $\chi^{00}\left(  0\right)  $, the equation (\ref{eqn20}) has to be
compared to the corresponding expression of Dirac's (D39):%

\begin{equation}
X_{L}\left(  0\right)  =-B+\left\{  \sqrt{-g}g^{00^{1/2}}g_{rs,u}%
E^{rusv}\right\}  _{,v} \label{eqn21}%
\end{equation}
where $B$ (D38)%

\begin{equation}
B=\frac{1}{4}\sqrt{-g}g^{00^{1/2}}g_{rs,u}g_{ab,v}\left\{  E^{rasb}%
e^{uv}+2E^{ruab}e^{sv}\right\}  \label{eqn22}%
\end{equation}
is a part of full expression (\ref{eqn02}) where after passing to
\textquotedblleft$e-$form\textquotedblright\ (substituting $e^{\alpha\beta}$
from (\ref{eqn06.04}) instead of $g^{\alpha\beta}$ in (\ref{eqn02})) only the
terms cubic in $e^{\alpha\beta}$ are present. Terms quadratic and linear in
$e^{\alpha\beta}$ are neglected, which results from the simplifying assumption
because all non-cubic terms have either $g^{0k}$ or the derivatives $g_{0k,m}%
$. In his final expression (\ref{eqn21}) Dirac keeps $e^{km}$, not $g^{km}$,
which is consistent with his statement that this has to be true without the
simplifying assumption which removes the difference between $e^{kn}$ and
$g^{kn}$. In addition, we keep $g=\det\left(  g_{\mu\nu}\right)  $ in all
equations. Dirac used $J^{2}=-\det\left(  g_{\mu\nu}\right)  $ and
$K^{2}=-\det\left(  g_{km}\right)  $ (or now more familiar notation $^{4}g$
for $\det\left(  g_{\mu\nu}\right)  $ and $g$ (or $g_{\left(  3\right)  }$)
for $\det\left(  g_{km}\right)  $ ) which are related by $g^{00}J^{2}=K^{2}$
or $\sqrt{-g}=\sqrt{-\det\left(  g_{km}\right)  /g^{00}}$.

By differentiating the second term of (\ref{eqn21}), it is not difficult to
derive the relation%

\begin{equation}
\chi^{00}\left(  0\right)  =\frac{1}{2}g^{00^{1/2}}X_{L}\left(  0\right)  .
\label{eqn50}%
\end{equation}

Dirac's scalar constraint $\mathcal{H}_{L}$ ($\mathcal{H}_{L}\left(  0\right)
=X_{L}\left(  0\right)  $) is not the result of a direct calculation of
$\left\{  p^{00},H_{G}\right\}  $. This difference is not important for the
proof of closure of the Dirac procedure and one can always consider linear
combinations of \textit{non-primary }constraints. However, the difference
between primary and non-primary constraints becomes important when their
linear combinations are used to construct a gauge generator; this was
discussed in great detail in \cite{KKaffinemetric}.\textit{ }For Castellani's
procedure \cite{Castellani} (or any other procedure) for finding gauge
transformations we have to be careful with the redefinition of constraints, as
we will demonstrate in the next Section (see also \cite{KKaffinemetric}).

Until now, we have been concerned with the most complicated contributions to
the secondary constraints which are zeroth order in the momenta. Let us now
consider the contributions to all orders. In the other two orders we obtain
(using (\ref{eqn7.1}))%

\begin{equation}
\chi^{0\sigma}\left(  2\right)  =\frac{\delta}{\delta g_{0\sigma}}H_{G}\left(
2\right)  =\frac{1}{2}\frac{1}{\sqrt{-g}}\frac{g^{0\sigma}}{g^{00}}\left(
g_{ra}g_{sb}-\frac{1}{2}g_{rs}g_{ab}\right)  p^{rs}p^{ab}, \label{eqn51}%
\end{equation}

\begin{equation}
\chi^{0\sigma}\left(  1\right)  =\frac{\delta}{\delta g_{0\sigma}}H_{G}\left(
1\right)  =-\delta_{u}^{\sigma}\left(  p_{,s}^{us}-\frac{1}{2}e^{uv}%
p^{rs}g_{rs,v}+e^{uv}p^{rs}g_{rv,s}\right)  . \label{eqn52}%
\end{equation}

Note, as $\chi^{00}\left(  1\right)  =0$ there are no contributions linear in
the momenta to the scalar constraint, but $\chi^{0k}\left(  2\right)  \neq0,$
unless we impose (\ref{eqn010}).

$\chi^{0\sigma}\left(  0\right)  $ was already calculated in (\ref{eqn20}).
For the full scalar constraint, $\chi^{00}$, the relation (\ref{eqn50}) is
preserved in all orders%

\begin{equation}
\chi^{00}=\frac{1}{2}g^{00^{1/2}}\mathcal{H}_{L}. \label{eqn53}%
\end{equation}
The vector constraint $\chi^{0k}$ has non-zero contributions in all orders of
the momenta, if (\ref{eqn010}) is not imposed. Before we continue to compare
our direct calculations with those of Dirac, let us try to present the
canonical Hamiltonian as a linear combination of the secondary constraints we
calculated above.

We approach this problem by considering different orders in the momenta. The
highest order is the second and the result is easily obtained from the first
terms of (\ref{eqn7.1}) and (\ref{eqn51})%

\begin{equation}
H_{G}\left(  2\right)  =\frac{1}{g^{00}\sqrt{-g}}\left(  g_{ra}g_{sb}-\frac
{1}{2}g_{rs}g_{ab}\right)  p^{rs}p^{ab}=2g_{0\sigma}\chi^{0\sigma}\left(
2\right)  \label{eqn54}%
\end{equation}
(using $g_{0\sigma}g^{0\sigma}=\delta_{0}^{0}=1$).

By considering (\ref{eqn09.2}), which is equivalent to the second terms of
(\ref{eqn7.1}) and (\ref{eqn52}), we have in first order
\begin{equation}
H_{G}\left(  1\right)  =-2g_{u0}\left(  p_{,s}^{us}-\frac{1}{2}e^{uv}%
p^{rs}g_{rs,v}+e^{uv}p^{rs}g_{rv,s}\right)  =2g_{0\sigma}\chi^{0\sigma}\left(
1\right)  . \label{eqn55}%
\end{equation}
$H_{G}\left(  2\right)  $ and $H_{G}\left(  1\right)  $ are of the same form
and we anticipate $H_{G}\left(  0\right)  $ is also in this form.
Unfortunately, this is not obvious and we have to perform some calculations to
show it. To preserve the structure found in (\ref{eqn54}), (\ref{eqn55}), we
will demonstrate that%

\begin{equation}
H_{G}\left(  0\right)  =-L_{G}\left(  0\right)  -L_{X}\left(  0\right)
=2g_{0\sigma}\chi^{0\sigma}\left(  0\right)  +\left(  ...\right)  _{,k}.
\label{eqn56}%
\end{equation}
Note, that for $H_{G}\left(  1\right)  $ given by (\ref{eqn09.2}) we also
obtain (\ref{eqn55}) only up to a total spatial derivative. Using
(\ref{eqn09}), (\ref{eqn30}), and (\ref{eqn20}) we have%

\[
-L_{G}\left(  0\right)  -L_{X}\left(  0\right)  -2g_{0\sigma}\chi^{0\sigma
}\left(  0\right)  =
\]

\begin{equation}
\left[  \sqrt{-g}E^{mnki}g_{mn,i}-\sqrt{-g}g_{\mu\nu,i}\left(  e^{\nu k}%
\frac{g^{0i}g^{0\mu}}{g^{00}}-e^{\nu i}\frac{g^{0k}g^{0\mu}}{g^{00}}\right)
\right]  _{,k}. \label{eqn74}%
\end{equation}

This equation demonstrates that the relations found for $H_{G}\left(
2\right)  $ and $H_{G}\left(  1\right)  $ are also valid for $H_{G}\left(
0\right)  $ and the canonical Hamiltonian can be written in terms of
$\chi^{0\sigma}$as%

\begin{equation}
H_{G}=2g_{0\sigma}\chi^{0\sigma}. \label{eqn80}%
\end{equation}

Of course, this is correct up to total temporal (see (\ref{eqn04})) and
spatial (see (\ref{eqn04}), (\ref{eqn09.2}), and (\ref{eqn74})) derivatives.
The modification of the initial Lagrangian (\ref{eqn04}) was proposed by Dirac
while (\ref{eqn74}) is obtained in the course of preserving relations found
among contributions of higher order in the momenta to the constraints and the
Hamiltonian. It would be very difficult to guess (\ref{eqn74}) without knowing
the final result. Such an additional integration appearing in (\ref{eqn74}),
is very often performed at the Lagrangian level. For example, in the book by
Gitman and Tyutin \cite{Gitmanbook}, in addition to Dirac's (\ref{eqn04})
(which are $B$ and first term of $C^{i}$ of Eq. (4.4.12) in \cite{Gitmanbook}%
), the integrations of (\ref{eqn74}) were performed at the Lagrangian level
(the second and third terms of $C^{i}$). The integrations of (\ref{eqn74}) can
be derived only in the course of the Hamiltonian procedure, but such
integrations (if they are known) are also correct when applied to the
Lagrangian because (going back to Dirac's derivation) it is clear that
$L_{X}\left(  0\right)  $ was constructed before the elimination of the
velocities (i.e., at the Lagrangian level).

How is this \textquotedblleft covariant\textquotedblright\footnote{We use the
term \textquotedblleft covariant\textquotedblright\ to distinguish constraints
$\chi^{0\sigma}$ that follow directly from time development of the primary
constraints $p^{o\sigma}$ and lead to a compact and symmetric form
(\ref{eqn80}) for the canonical part of the Hamiltonian. This is
\textquotedblleft covariance\textquotedblright\ in the sense of
\cite{Teit-PRL}, i.e. constraints are contracted with components of a true
tensor (components $g_{o\sigma}$), contrary to a similar form used in the ADM
formulation (see, for example, \cite{SalisburyPS}, Eq. (4.12): $H=N^{\mu
}\mathcal{H}_{\mu}$) where $N^{\mu}$ is neither a true vector nor components
of any true tensor.} form of $H_{G}$ (\ref{eqn80}) (which is equivalent to
what was found in \cite{KKRV}) related to Dirac's expression for the
Hamiltonian? Are they equivalent? The relationship between scalar constraints
$\chi^{00}$ and $\mathcal{H}_{L}$ \ was found in (\ref{eqn53}); we now
consider the relation between the vector constraints.

Let us inspect the form of our constraints calculated to different orders
appearing in (\ref{eqn51}), (\ref{eqn52}), and (\ref{eqn20}). There are simple
relations between the contributions of different orders to $\chi^{00}$ and
$\chi^{0k}$:%

\[
\chi^{0k}\left(  2\right)  =\frac{g^{0k}}{g^{00}}\chi^{00}\left(  2\right)
,\left.  {}\right.  \left.  {}\right.  \chi^{00}\left(  1\right)  =0,\left.
{}\right.  \left.  {}\right.  \chi^{0k}\left(  1\right)  =\psi^{0k},\left.
{}\right.  \left.  {}\right.  \chi^{0k}\left(  0\right)  =\frac{g^{0k}}%
{g^{00}}\chi^{00}\left(  0\right)
\]
that allow one to write (to all orders)%

\begin{equation}
\chi^{0k}=\psi^{0k}+\frac{g^{0k}}{g^{00}}\chi^{00} \label{eqn81}%
\end{equation}
with%

\begin{equation}
\psi^{0k}=-p_{,s}^{ks}-e^{kv}p^{rs}\left(  \frac{1}{2}g_{rs,v}-g_{rv,s}%
\right)  . \label{eqn82}%
\end{equation}
Solving (\ref{eqn81}) for $\psi^{0k}$ gives a combination of the constraints
$\chi^{00}$ and $\chi^{0k}$ which were originally calculated from the time
development of the corresponding primary constraints.

In terms of this combination of constraints $\psi^{0k}$ and $\chi^{00}$, we
obtain a different form of the canonical Hamiltonian%

\begin{equation}
H_{G}=2\frac{1}{g^{00}}\chi^{00}+2g_{0k}\psi^{0k}. \label{eqn83}%
\end{equation}

This form of $H_{G}$ is easy to compare with Dirac's, because his vector
constraint is simply related to $\psi^{0k}$%

\begin{equation}
2\psi^{0k}=e^{ks}\mathcal{H}_{s}. \label{eqn84}%
\end{equation}
For $\chi^{0k}$ we find%

\begin{equation}
\chi^{0k}=\frac{1}{2}e^{ks}\mathcal{H}_{s}-\frac{1}{2}g_{0s}e^{sk}g^{00^{1/2}%
}\mathcal{H}_{L}. \label{eqn85}%
\end{equation}

Equation (\ref{eqn84}), together with (\ref{eqn50}), demonstrates the
equivalence of the two different forms of $H_{G}$ given in (\ref{eqn80}) and
(\ref{eqn83}) to Dirac's canonical Hamiltonian%

\begin{equation}
H_{G}=2g_{0\sigma}\chi^{0\sigma}=2\frac{1}{g^{00}}\chi^{00}+2g_{0k}\psi
^{0k}=\left(  g^{00}\right)  ^{-1/2}\mathcal{H}_{L}+g_{r0}e^{rs}%
\mathcal{H}_{s}. \label{eqn86}%
\end{equation}

We would like to emphasize that Dirac's constraints are not a direct result of
the time development of the primary constraints $\phi^{0\sigma}$ which produce
$\chi^{0\sigma}$((\ref{eqn53}) and (\ref{eqn85})). The only place known to us
where this is stated is in the book by Gitman and Tyutin (Eq. (4.4.19) of
\cite{Gitmanbook}); but Dirac's particular combinations of constraints and the
corresponding form of the Hamiltonian are usually used.

The linear approximation of $\chi^{0\sigma}$ gives exactly the constraints of
linearized GR \cite{GKK}. In the linearized case there is no difference
between $\chi^{0k}$ and $\psi^{0k}$; therefore linearized gravity can provide
little \textquotedblleft guidance\textquotedblright\ to full GR, in contrast
to what was emphasized by ADM in \cite{ADM-1}. Any such guidance has to be
taken cautiously.

\subsection{Closure of Dirac procedure for the GR Hamiltonian}

To demonstrate closure of the Dirac procedure, any form of the canonical
Hamiltonian (\ref{eqn86}) is suitable as they are all equivalent; and any
linear combination of \textit{secondary (non-primary) }constraints can be used
for this purpose (e.g., $\chi^{00}=\frac{1}{2}g^{00^{1/2}}\mathcal{H}_{L}$ and
$\psi^{0k}=\frac{1}{2}e^{rs}\mathcal{H}_{s}$). When using Castellani's
procedure to derive the gauge transformations generated by first-class
constraints, we will\textit{ }consider those secondary constraints that
directly follow from the corresponding primary ones and the PBs of secondary
constraints with the total Hamiltonian, not just with its canonical part (this
is also discussed in the next Section).

All of Dirac's secondary constraints have zero PBs with the primary
constraints. In constraint dynamics this means that Lagrange multipliers
cannot be found at this stage. As the PB of the secondary constraints with the
canonical part of the Hamiltonian is zero or proportional to constraints, the
procedure is closed. This is exactly the case here when we are taking into
account the algebra\footnote{This algebra is called \textquotedblleft
hypersurface deformation algebra\textquotedblright\ or \textquotedblleft Dirac
algebra\textquotedblright\ and can be found in slightly different forms in
many places, e.g. \cite{Castellani, Saha, Kuchar, Teit-in-Held, Teit-AOP,
Relativity},\textit{ }but it was not calculated by Dirac for the GR
Hamiltonian.} of PBs among Dirac's combinations of the secondary constraints:%

\[
\left\{  \mathcal{H}_{L}\left(  x\right)  ,\mathcal{H}_{L}\left(  x^{\prime
}\right)  \right\}  =e^{rs}\left(  x\right)  \mathcal{H}_{s}\left(  x\right)
\delta_{,r\left(  x\right)  }\left(  x-x^{\prime}\right)  -e^{rs}\left(
x^{^{\prime}}\right)  \mathcal{H}_{s}\left(  x^{\prime}\right)  \delta
_{,r\left(  x^{\prime}\right)  }\left(  x-x^{\prime}\right)  ,
\]

\begin{equation}
\left\{  \mathcal{H}_{s}\left(  x\right)  ,\mathcal{H}_{L}\left(  x^{\prime
}\right)  \right\}  =\mathcal{H}_{L}\left(  x\right)  \delta_{,s\left(
x\right)  }\left(  x-x^{\prime}\right)  , \label{eqnDA}%
\end{equation}

\[
\left\{  \mathcal{H}_{r}\left(  x\right)  ,\mathcal{H}_{s}\left(  x^{\prime
}\right)  \right\}  =\mathcal{H}_{s}\left(  x\right)  \delta_{,r\left(
x\right)  }\left(  x-x^{\prime}\right)  -\mathcal{H}_{r}\left(  x^{\prime
}\right)  \delta_{,s\left(  x^{\prime}\right)  }\left(  x-x^{\prime}\right)
.
\]

When dealing with the \textquotedblleft covariant\textquotedblright\ secondary
constraints $\chi^{0\sigma}$, the multipliers are also not determined, but now
we have%

\begin{equation}
\left\{  \chi^{0\sigma},p^{0\gamma}\right\}  =\frac{1}{2}g^{\sigma\gamma}%
\chi^{00}. \label{eqn90}%
\end{equation}

The closure of Dirac's procedure is obviously preserved when using the
\textquotedblleft covariant\textquotedblright\ constraints because
$\chi^{0\sigma}$ and Dirac's constraints are simply related by (\ref{eqn53})
and (\ref{eqn85}). This can also be shown by direct calculation of $\left\{
\chi^{0\sigma},H_{G}\right\}  $ without any reference to Dirac's combinations
of constraints and their algebra. We choose the second path because to prove
closure we need the PB with the total Hamiltonian, and in such a calculation
we avoid working with derivatives of delta functions. This is exactly the PB
that we need in implementing Castellani's procedure (see next Section). In
addition, the second equation of (\ref{eqnDA}), as it is usually presented in
the literature, is not symmetric in form, contrary to the rest of the
equations. The symmetric form of this equation was presented in \cite{Faddeev,
Teit-PRD}. Applying Dirac's procedure to the first-order formulation of GR,
the same algebra emerges, but for tertiary constraints where all equations are
symmetric \cite{KKaffinemetric, Gerry-Ramin-2, Gerry-new} (see also the
discussion on this algebra in \cite{KKaffinemetric}). The calculations of PBs
between the secondary constraints and the Hamiltonian are very involved and to
perform them we found it more convenient to work in the intermediate stages
with $\chi^{00}$ and $\psi^{0k}$. This allows us to sort out terms uniquely,
and at the final stage we can express the result, using (\ref{eqn81}), in
terms of the \textquotedblleft covariant\textquotedblright\ constraint
$\chi^{0\sigma}$. The details of such calculations are\textit{ }given in
\cite{KKRV, KKX}. We arrive to the following PBs of $\chi^{0\sigma}$ with the
canonical part of the Hamiltonian%

\begin{equation}
\left\{  \chi^{00},H\right\}  =-\frac{2}{\sqrt{-g}}I_{kmrb}p^{km}g_{0a}%
e^{ab}\chi^{0r}+\chi_{,k}^{0k}+\frac{g^{0\alpha}g^{0\beta}}{g^{00}}%
g_{\alpha\beta,k}\chi^{0k}-\frac{1}{2}g^{0b}g_{00,b}\chi^{00} \label{eqn91}%
\end{equation}
and%

\[
\left\{  \chi^{0k},H\right\}  =\frac{1}{\sqrt{-g}}\frac{1}{g^{00}}\left(
2g_{ra}p^{ak}\chi^{0r}-g_{ab}p^{ab}\chi^{0k}\right)  -\frac{g^{0k}}{g^{00}%
}\frac{2}{\sqrt{-g}}I_{tmrb}p^{tm}g_{0a}e^{ab}\chi^{0r}%
\]

\begin{equation}
+g^{0k}g_{00,t}\chi^{0t}+2g_{0p,t}g^{pk}\chi^{0t}+\frac{g^{0p}}{g^{00}}%
g^{kq}\left(  g_{pq,r}+g_{rp,q}-g_{rq,p}\right)  \chi^{0r}-\frac{1}{2}%
g^{km}g_{00,m}\chi^{00}. \label{eqn92}%
\end{equation}
\bigskip Of course, we can present (\ref{eqn91}) and (\ref{eqn92}) as one
\textquotedblleft covariant\textquotedblright\ equation%

\[
\left\{  \chi^{0\sigma},H\right\}  =-\frac{g^{0\sigma}}{g^{00}}\frac{2}%
{\sqrt{-g}}I_{tmrb}p^{tm}g_{0a}e^{ab}\chi^{0r}+\delta_{m}^{\sigma}\frac
{1}{\sqrt{-g}}\frac{1}{g^{00}}\left(  2g_{ra}p^{am}\chi^{0r}-g_{ab}p^{ab}%
\chi^{0m}\right)  +\delta_{0}^{\sigma}\chi_{,k}^{0k}%
\]

\begin{equation}
-\frac{1}{2}g^{\sigma b}g_{00,b}\chi^{00}+g^{0\sigma}g_{00,t}\chi
^{0t}+2g_{0p,t}g^{p\sigma}\chi^{0t}+\frac{g^{0p}}{g^{00}}g^{\sigma q}\left(
g_{pq,r}+g_{rp,q}-g_{rq,p}\right)  \chi^{0r}. \label{eqn93}%
\end{equation}

These equations, (\ref{eqn91}), (\ref{eqn92}) or (\ref{eqn93}), along with
(\ref{eqn90}), provide proof of the closure of the Dirac procedure: higher
order (tertiary) constraints do not appear and multipliers cannot be found because%

\[
\left\{  \chi^{0\sigma},H_{T}\right\}  \sim\chi^{0\sigma}.
\]

\subsection{Summary}

The Dirac Hamiltonian for GR, which is based on the modified Lagrangian of
(\ref{eqn06}) and the simplifying assumption (\ref{eqn010}), is equivalent to
the result of direct calculations given in (\ref{eqn80}) which are performed
without any reference to surfaces of constant time. All the equivalent forms
of the Hamiltonian of (\ref{eqn86}) are \textit{only} the consequence of an
initial modification that does not affect the equations of motion and
preserves the four-dimensional symmetry. It is natural to expect that Dirac's
Hamiltonian formulation, which is obtained without any \textit{a priori}
assumptions and restrictions (e.g. surfaces of constant time), has to preserve
another manifestation of four-dimensional symmetry: invariance under the
diffeomorphism transformation (\ref{eqn00}). Such a demonstration, given in
the next Section, is an important consistency check of our results. All
constraints are first-class and thus to find the generators of the gauge
transformation, we have to consider \textquotedblleft chains\textquotedblright%
\ of constraints. This means that one has to work, not with some combination
of the constraints $\chi^{0\sigma}$, but with the exact results for $\left\{
\phi^{0\sigma},H_{T}\right\}  $ and $\left\{  \chi^{0\sigma},H_{T}\right\}  $.
These expressions are complicated, especially (\ref{eqn93}), but their
correctness can be verified if they lead to diffeomorphism invariance. A
simple preliminary check of (\ref{eqn93}) is that the linearized version of
this equation gives%

\[
\left\{  \chi^{00},H\right\}  =-p_{,k}^{0k},\left.  {}\right.  \left.
{}\right.  \left\{  \chi^{0k},H\right\}  =0.
\]
This is equivalent to the results of \cite{GKK} (note, that in the linearized
case $\chi_{lin}^{0k}=\psi_{lin}^{0k}$) and the constraint structure leads to
a linearized version of diffeomorphism invariance.

To summarize, the \textit{reality} of Dirac's formulation, that is based on a
modification of the initial Lagrangian which does not affect the equations of
motion, is as follows: notwithstanding Dirac's references to space-like
surfaces, all of his calculations were performed without use of any such
surfaces. Consequently, Hawking's statement \cite{Hawking} about the
contradiction of the Hamiltonian formulation, based on splitting space-time
into three spatial dimensions and one time dimension, is not applicable to
Dirac's Hamiltonian formulation of GR, which does preserve the spirit of GR.
Our own criticism of Dirac's formulation in \cite{KKAnn} was not correct as we
based it only on the `interpretational' aspects of his work. This \textit{faux
pas} is also an illustration of how interpretations or some geometrical (or
any other) reasonings can be dangerous if the \textquotedblleft rule of
procedure\textquotedblright\ referred by Lagrange is neglected.

Dirac's simplifying assumption, $g_{0k}=0$ and (\ref{eqn010}), for
constructing zeroth order in momenta contributions to the secondary
constraints is not correct with respect to the individual parts given in
(\ref{eqn10}), (\ref{eqn31}); but remarkably when these parts are combined
together in (\ref{eqn20}), they are equivalent to his final expression. His
secondary constraints do not follow directly from the time development of the
primary constraints but rather they are particular combinations of the
\textquotedblleft covariant\textquotedblright\ secondary constraints
$\chi^{0\sigma}$ which are the result of direct calculation of PBs of primary
constraints with the canonical Hamiltonian. His secondary constraints cannot
be directly used to find gauge transformations (Dirac himself did not consider
this question). In the next Section we will show that the generator built from
the \textquotedblleft covariant\textquotedblright\ constraints gives the
four-dimensional diffeomorphism\ (\ref{eqn00}), and we can say that the
\textquotedblleft covariant\textquotedblright\ constraints of the Dirac
formulation and their algebra \textit{is} \textquotedblleft the algebra of
four diffeomorphisms\textquotedblright\ \cite{Pulin}.

\section{The gauge generator and transformation of the metric tensor}

The knowledge of the complete set of first-class constraints (primary,
$p^{0\sigma}$,(\ref{eqn05}), and secondary, $\chi^{0\sigma}=\chi^{0\sigma
}\left(  2\right)  +\chi^{0\sigma}\left(  1\right)  +\chi^{0\sigma}\left(
0\right)  $, where contributions of different order in momenta are given by
(\ref{eqn51}), (\ref{eqn52}) and (\ref{eqn20})), as well as the PBs between
the primary and secondary constraints (\ref{eqn90}), and the exact form of the
closure (\ref{eqn93}) are sufficient to find the generators of the gauge
transformations. This possibility is Dirac's old conjecture \cite{Diracbook}
which became a well developed algorithm and exists in a few variations
\cite{Castellani, HTZ, Novel}\footnote{Some methods have to be used with
caution because they are sensitive to the choice of linear combinations of
non-primary constraints (see discussion in \cite{KKaffinemetric}).}. We follow
the work of Castellani \cite{Castellani} where the first application of such a
method to Yang-Mills theory and ADM gravity\footnote{\textsf{I}n
\cite{Castellani} the author referred to the Dirac formulation of GR but in
fact considered the ADM formulation. The non-equivalence of these two
formulations will be discussed in the next Section.} was considered.

\subsection{Castellani's procedure}

Castellani's procedure is based on a derivation of the generator of gauge
transformations which is defined by \textit{chains} of first-class
constraints. One starts with primary first-class constraint(s), $i=1,2,...$,
and construct the chain(s) $\xi_{i}^{\left(  n\right)  }G_{\left(  n\right)
}^{i}$ where $\xi_{i}^{\left(  n\right)  }$ is the $n$th order time derivative
of the gauge parameter $\xi_{i}$ ($n=0,1,...$). The maximum value of $n$
corresponds to the length of the chain (e.g., $n=0,1,2$ for the system with
tertiary constraints). The number of gauge parameters $\xi_{i}$ is equal to
the number of primary first-class constraints. Note, that these chains are an
unambiguous construction once the primary constraints are defined; the
remaining members of the chain are uniquely determined. Of course, one can use
different linear combinations of non-primary constraints, the generator will
have a different form but the gauge transformations will be the same. More
details about the problems arising when linear combinations of primary
constraints are used in constructing of a gauge generator can be found in
\cite{KKaffinemetric}.

From this point, we specialize to the Dirac Hamiltonian formulation of GR with
$n=0,1$ and $i=0,1,...,(d-1)$; if $d=4$ there are four primary and four
secondary constraints\footnote{The following calculations, as well as the
results of the previous Section, are valid in all dimensions, except $d=2$.}.
The functions $G_{\left(  n\right)  }^{i}$ are calculated as follows%

\begin{equation}
G_{\left(  1\right)  }^{\sigma}\left(  x\right)  =p^{0\sigma}\left(  x\right)
, \label{eqn130}%
\end{equation}

\begin{equation}
G_{\left(  0\right)  }^{\sigma}\left(  x\right)  =+\left\{  p^{0\sigma}\left(
x\right)  ,H_{T}\right\}  +\int\alpha_{\gamma}^{\sigma}\left(  x,y\right)
p^{0\gamma}\left(  y\right)  d^{3}y \label{eqn131}%
\end{equation}
where the functions $\alpha_{\gamma}^{\sigma}\left(  x,y\right)  $ have to be
chosen in such a way that the chain beginning with $G_{\left(  1\right)
}^{\sigma}$ in (\ref{eqn130}) ends on the primary constraint surface%

\begin{equation}
\left\{  G_{\left(  0\right)  }^{\sigma},H_{T}\right\}  =primary.
\label{eqn140}%
\end{equation}
The generator $G\left(  \xi_{\sigma}\right)  $ is given by%

\begin{equation}
G\left(  \xi_{\sigma}\right)  =\xi_{\sigma}G_{\left(  0\right)  }^{\sigma}%
+\xi_{\sigma,0}G_{\left(  1\right)  }^{\sigma}. \label{eqn141}%
\end{equation}
There are some peculiarities that arise when applying this algorithm to GR
that cannot be seen in simpler cases like Maxwell, Yang-Mills or linearized GR
theories. We will consider them in detail below.

Firstly, we comment on the use of different linear combinations of
\textit{secondary (non-primary) }constraints. The term $\left\{  p^{0\sigma
}\left(  x\right)  ,H_{T}\right\}  $ in (\ref{eqn131}) is uniquely defined by
primary constraints. After Dirac's modification (\ref{eqn04}) of the original
Lagrangian, these remain just the momenta $p^{0\sigma}$ conjugate to the
$g_{0\sigma}$ components of the metric tensor. Straightforward calculation of
the PBs of these primary constraints with the Hamiltonian gives%

\[
\left\{  p^{00}\left(  x\right)  ,H_{G}\right\}  =\chi^{00}=\frac{1}%
{2}g^{00^{1/2}}\mathcal{H}_{L},
\]

\[
\left\{  p^{0k}\left(  x\right)  ,H_{G}\right\}  =\chi^{0k}=\psi^{0k}%
+\frac{g^{0k}}{g^{00}}\chi^{00}=\frac{1}{2}e^{ks}\mathcal{H}_{s}+\frac{1}%
{2}\left(  g^{00}\right)  ^{-1/2}g^{0k}\mathcal{H}_{L},
\]
and these expressions, $\chi^{00}$ and $\chi^{0k}$, will be used when the
gauge generators are derived. Of course, one can use Dirac's combinations,
$\mathcal{H}_{L}$ and $\mathcal{H}_{s}$, but only with appropriate
coefficients or in appropriate combinations because of the following
inequalities: $\left\{  p^{00},H_{G}\right\}  \neq\mathcal{H}_{L}$, $\left\{
p^{0k},H_{G}\right\}  \neq e^{ks}\mathcal{H}_{s}$. We are not aware of any
other situation where one must consider combinations of secondary constraints
in ordinary field theories, the exception is only the first-order,
affine-metric, GR \cite{KKaffinemetric} where such combinations lead to some
computational advantages but can also be avoided. Such a situation does not
appear, for example, when the gauge generator for Yang-Mills theory is
constructed \cite{Castellani}. In this case, the first term of (\ref{eqn131})
is just a secondary first-class constraint which is the result of direct
calculation of the PB of primary constraints with the Hamiltonian. In the
Hamiltonian formulation of GR it is quite common (if not exclusive) to use the
Dirac combinations of constraints, but one has to be careful when gauge
generators are constructed using $\mathcal{H}_{L}$ and $\mathcal{H}_{s}$.

Secondly, in both the Maxwell and Yang-Mills theories it is possible to choose
the functions $\alpha_{\gamma}^{\sigma}\left(  x,y\right)  $ so that chains
truly end with zero in (\ref{eqn140}) \cite{Castellani}. It does not happen in
GR, and chains end only on the surface of the primary constraints. The effect
of such a difference will be seen in our calculation of the gauge generators
and the associated transformations.

Thirdly, the \textit{total} Hamiltonian should be used in Castellani's
procedure, not just its canonical part (see (\ref{eqn131}), (\ref{eqn140})).
Again, in linearized GR, Yang-Mills and Maxwell theories this difference is
irrelevant because in these theories the PBs of secondary constraints with
primary ones are zero. This is not the case for full GR as can be seen from
(\ref{eqn90}) and similarly from equation (16) of \cite{KKRV}.

Finally, a purely technical comment. There is a change of sign in front of the
first term of (\ref{eqn131}) relative to that used in \cite{Castellani}. This
is the result of Dirac's convention for the fundamental brackets in
(\ref{eqn09.4}) (it is the negative of the fundamental brackets used in
\cite{Castellani}).

\subsection{Castellani's generator}

To construct the generator (\ref{eqn141}) we have to find functions
$\alpha_{\gamma}^{\sigma}\left(  x,y\right)  $ using the condition
(\ref{eqn140})%

\[
\left\{  G_{\left(  0\right)  }^{\sigma},H_{T}\right\}  =\left\{
\chi^{0\sigma}\left(  x\right)  +\int\alpha_{\gamma}^{\sigma}\left(
x,y\right)  p^{0\gamma}\left(  y\right)  d^{3}y,H_{T}\right\}  =
\]

\begin{equation}
\left\{  \chi^{0\sigma}\left(  x\right)  ,H_{T}\right\}  +\int\left\{
\alpha_{\gamma}^{\sigma}\left(  x,y\right)  ,H_{T}\right\}  p^{0\gamma}\left(
y\right)  d^{3}y+\int\alpha_{\gamma}^{\sigma}\left(  x,y\right)  \left\{
p^{0\gamma}\left(  y\right)  ,H_{T}\right\}  d^{3}y. \label{eqn145}%
\end{equation}
Part of the first term has already been calculated and $\left\{  \chi
^{0\sigma}\left(  x\right)  ,H_{G}\right\}  $ is given by (\ref{eqn93}). For
the part involving the primary constraints, $p^{0\gamma}$, we use
(\ref{eqn90}) which gives%

\[
\left\{  \chi^{0\sigma}\left(  x\right)  ,g_{00,0}p^{00}+2g_{0m,0}p^{0m}%
+H_{G}\right\}  =\frac{1}{2}g_{00,0}g^{0\sigma}\chi^{00}+g_{0m,0}g^{\sigma
m}\chi^{00}%
\]

\[
-\frac{g^{0\sigma}}{g^{00}}\frac{2}{\sqrt{-g}}I_{tmrb}p^{tm}g_{0a}e^{ab}%
\chi^{0r}+\frac{1}{\sqrt{-g}}\frac{\delta_{m}^{\sigma}}{g^{00}}\left(
2g_{ra}p^{am}\chi^{0r}-g_{ab}p^{ab}\chi^{0m}\right)  +\delta_{0}^{\sigma}%
\chi_{,k}^{0k}%
\]

\begin{equation}
-\frac{1}{2}g^{\sigma b}g_{00,b}\chi^{00}+g^{0\sigma}g_{00,t}\chi
^{0t}+2g_{0p,t}g^{p\sigma}\chi^{0t}+\frac{g^{0p}}{g^{00}}g^{\sigma q}\left(
g_{pq,r}+g_{rp,q}-g_{rq,p}\right)  \chi^{0r}. \label{eqn146}%
\end{equation}
The second term of (\ref{eqn145}) is irrelevant because it is automatically
zero on the surface of primary constraints, as required by (\ref{eqn140}).

For the last term of (\ref{eqn145}), taking into account the zero value of the
PB among primary constraints, we find that%

\begin{equation}
\left\{  p^{0\gamma},H_{T}\right\}  =\left\{  p^{0\gamma},H_{G}\right\}
=\chi^{0\gamma}. \label{eqn147}%
\end{equation}
This illustrates the advantage of using the \textquotedblleft
covariant\textquotedblright\ constraints $\chi^{0\gamma}$ for deriving the
generator. With (\ref{eqn146})-(\ref{eqn147}), we can now read off the
functions $\alpha_{\gamma}^{\sigma}\left(  x,y\right)  $ from (\ref{eqn145})
that compensates (\ref{eqn146})%

\[
-\alpha_{\gamma}^{\sigma}\left(  x,y\right)  =\frac{1}{2}g_{00,0}\left(
x\right)  g^{0\sigma}\left(  x\right)  \delta_{\gamma}^{0}\left(  x,y\right)
+g_{0m,0}g^{\sigma m}\delta_{\gamma}^{0}%
\]

\[
-\frac{g^{0\sigma}}{g^{00}}\frac{2}{\sqrt{-g}}I_{tmrb}p^{tm}g_{0a}e^{ab}%
\delta_{\gamma}^{r}+\frac{1}{\sqrt{-g}}\frac{\delta_{m}^{\sigma}}{g^{00}%
}\left(  2g_{ra}p^{am}\delta_{\gamma}^{r}-g_{ab}p^{ab}\delta_{\gamma}%
^{m}\right)  +\delta_{0}^{\sigma}\delta_{\gamma,k}^{k}%
\]

\begin{equation}
-\frac{1}{2}g^{\sigma b}g_{00,b}\delta_{\gamma}^{0}+g^{0\sigma}g_{00,t}%
\delta_{\gamma}^{t}+2g_{0p,t}g^{p\sigma}\delta_{\gamma}^{t}+\frac{g^{0p}%
}{g^{00}}g^{\sigma q}\left(  g_{pq,r}+g_{rp,q}-g_{rq,p}\right)  \delta
_{\gamma}^{r}, \label{eqn148}%
\end{equation}
where $\delta_{\gamma}^{\sigma}\left(  x,y\right)  \equiv\delta_{\gamma
}^{\sigma}\delta\left(  x-y\right)  $ and the arguments $x$ and $y$ are
explicitly written only in the first term. Contracting $\alpha_{\gamma
}^{\sigma}\left(  x,y\right)  $ with the primary constraints and performing
integration in the second term of (\ref{eqn131}) we obtain%

\[
G_{\left(  0\right)  }^{\sigma}=\chi^{0\sigma}-\frac{1}{2}g_{00,0}g^{0\sigma
}p^{00}-g_{0m,0}g^{\sigma m}p^{00}%
\]

\[
+\frac{g^{0\sigma}}{g^{00}}\frac{2}{\sqrt{-g}}I_{tmrb}p^{tm}g_{0a}e^{ab}%
p^{0r}-\frac{1}{\sqrt{-g}}\frac{\delta_{m}^{\sigma}}{g^{00}}\left(
2g_{ra}p^{am}p^{0r}-g_{ab}p^{ab}p^{0m}\right)  -\delta_{0}^{\sigma}p_{,k}^{0k}%
\]

\begin{equation}
+\frac{1}{2}g^{\sigma b}g_{00,b}p^{00}-g^{0\sigma}g_{00,t}p^{0t}%
-2g_{0p,t}g^{p\sigma}p^{0t}-\frac{g^{0p}}{g^{00}}g^{\sigma q}\left(
g_{pq,r}+g_{rp,q}-g_{rq,p}\right)  p^{0r}. \label{eqn149}%
\end{equation}
Equation (\ref{eqn149}) completes the calculation of the generator
(\ref{eqn141}). Now the transformation of fields can be found by calculating
their PB with the generator\footnote{Some authors defined transformations as
$\delta\left(  field\right)  =\left\{  G,field\right\}  $ which seems to be
more natural. However, we keep the convention of Castellani that, of course,
affects only an overall sign in the final result, which can always be
incorporated into the gauge parameters (this is not a field-dependent
redefinition, as is used in \cite{Saha}).}%

\begin{equation}
\delta\left(  field\right)  =\left\{  field,G\right\}  . \label{eqn151}%
\end{equation}

For the time-time component of the metric tensor, $g_{00},$ we obtain (using
Dirac's convention of (\ref{eqn09.4}) and keeping only part of the generator
(\ref{eqn141}) with terms proportional to $p^{00}$)%

\[
\delta g_{00}=\left\{  g_{00},G\right\}  =-\frac{\delta}{\delta p^{00}}G=
\]

\[
-\frac{\delta}{\delta p^{00}}\left[  -\xi_{\sigma}\left(  g_{00,0}\frac{1}%
{2}g^{0\sigma}p^{00}+g_{0m,0}g^{\sigma m}p^{00}-\frac{1}{2}g^{\sigma
b}g_{00,b}p^{00}\right)  +\xi_{0,0}p^{00}\right]  =
\]

\begin{equation}
-\xi_{0,0}+\left(  \frac{1}{2}g_{00,0}g^{00}+g_{0m,0}g^{0m}\right)  \xi
_{0}+\left(  \frac{1}{2}g_{00,0}g^{k0}+g_{0m,0}g^{km}\right)  \xi_{k}-\frac
{1}{2}g^{0b}g_{00,b}\xi_{0}-\frac{1}{2}g^{km}g_{00,m}\xi_{k}. \label{eqn166}%
\end{equation}

Let us compare this result with diffeomorphism invariance (\ref{eqn00}) which
can be written in an equivalent form, that is more convenient for comparison
with our calculations:%

\begin{equation}
\delta_{\left(  diff\right)  }g_{\mu\nu}=-\xi_{\mu,\nu}-\xi_{\nu,\mu
}+g^{\alpha\beta}\left(  g_{\mu\beta,\nu}+g_{\nu\beta,\mu}-g_{\mu\nu,\beta
}\right)  \xi_{\alpha}. \label{eqn167}%
\end{equation}
Taking $\mu=\nu=0$ and explicitly separating space and time indices, we have%

\begin{equation}
\delta_{(diff)}g_{00}=-2\xi_{0,0}+\left(  g^{00}g_{00,0}+2g^{0k}%
g_{0k,0}-g^{0k}g_{00,k}\right)  \xi_{0}+\left(  g^{k0}g_{00,0}+2g^{km}%
g_{0m,0}-g^{km}g_{00,m}\right)  \xi_{k}. \label{eqn168}%
\end{equation}
We see that (\ref{eqn168}) is equivalent to (\ref{eqn166}) up to a numerical
factor $2$,%

\begin{equation}
2\left\{  g_{00},G\right\}  =\delta_{(diff)}g_{00}, \label{eqn168.1}%
\end{equation}
that can be incorporated into the gauge parameter by a rescaling $\xi_{\sigma
}\rightarrow2\xi_{\sigma}$.

Similarly, for the space-time components, $g_{0k}$, we have%

\[
\delta g_{0k}=\left\{  g_{0k},G\right\}  =
\]

\[
-\frac{\delta}{\delta p^{0k}}\left[  -\xi_{\sigma}\left(  -\frac{g^{0\sigma}%
}{g^{00}}\frac{2}{\sqrt{-g}}I_{tmrb}p^{tm}g_{0a}e^{ab}p^{0r}+\frac{1}%
{\sqrt{-g}}\frac{\delta_{m}^{\sigma}}{g^{00}}\left(  2g_{ra}p^{am}%
p^{0r}-g_{ab}p^{ab}p^{0m}\right)  +\delta_{0}^{\sigma}p_{,m}^{0m}\right.
\right.
\]

\[
\left.  \left.  +g^{0\sigma}g_{00,t}p^{0t}+2g_{0p,t}g^{p\sigma}p^{0t}%
+\frac{g^{0p}}{g^{00}}g^{\sigma q}\left(  g_{pq,r}+g_{rp,q}-g_{rq,p}\right)
p^{0r}\right)  +\xi_{m,0}p^{0m}\right]  =
\]

\[
=-\frac{1}{2}\xi_{0,k}-\frac{1}{2}\xi_{k,0}%
\]

\[
+\frac{1}{2}\xi_{\sigma}\left[  -\frac{g^{0\sigma}}{g^{00}}\frac{2}{\sqrt{-g}%
}I_{tmkb}p^{tm}g_{0a}e^{ab}+\frac{1}{\sqrt{-g}}\frac{1}{g^{00}}\left(
2g_{ka}p^{am}\delta_{m}^{\sigma}-g_{ab}p^{ab}\delta_{k}^{\sigma}\right)
\right.
\]

\begin{equation}
\left.  +g^{0\sigma}g_{00,k}+2g_{0p,k}g^{p\sigma}+\frac{g^{0p}}{g^{00}%
}g^{\sigma q}\left(  g_{pq,k}+g_{kp,q}-g_{kq,p}\right)  \right]  .
\label{eqn169}%
\end{equation}

There is a difference between the transformation (\ref{eqn169}) and the
transformation of the time-time component (\ref{eqn166}) as the momenta
$p^{ab}$ are present in (\ref{eqn169}). Using the definition of momenta
(\ref{eqn06.1}) and re-expressing $e^{km}$ in terms of $g^{km}$ by
(\ref{eqn06.04}), we obtain%

\[
\delta g_{0k}=-\frac{1}{2}\xi_{0,k}-\frac{1}{2}\xi_{k,0}%
\]

\[
+\frac{1}{2}\left[  g^{00}g_{00,k}+g^{0m}\left(  g_{0m,k}+g_{km,0}%
-g_{0k,m}\right)  \right]  \xi_{0}%
\]

\begin{equation}
+\frac{1}{2}\left[  g^{m0}\left(  g_{00,k}+g_{k0,0}-g_{0k,0}\right)
+g^{mn}\left(  g_{0n,k}+g_{kn,0}-g_{0k,n}\right)  \right]  \xi_{m},
\label{eqn170}%
\end{equation}
which again equals $\delta_{(diff)}g_{\mu\nu}$ as given in (\ref{eqn167}) with
$\mu\nu=0k$; which is true provided we again rescale $\xi_{\sigma}$ by a
factor of $2$.

The last transformation to be checked is the transformation of the space-space components%

\begin{equation}
\delta g_{km}=\left\{  g_{km},G\right\}  =-\frac{\delta}{\delta p^{km}%
}G\left(  p^{pq}\right)  . \label{eqn171}%
\end{equation}
The relevant part of the generator (\ref{eqn141}) which has an explicit
dependence on $p^{pq}$ is%

\begin{equation}
G\left(  p^{pg}\right)  =\xi_{\sigma}\left[  \chi^{0\sigma}+\frac{g^{0\sigma}%
}{g^{00}}\frac{2}{\sqrt{-g}}I_{tmrb}p^{tm}g_{0a}e^{ab}p^{0r}-\delta
_{m}^{\sigma}\frac{1}{\sqrt{-g}}\frac{1}{g^{00}}\left(  2g_{ra}p^{am}%
p^{0r}-g_{ab}p^{ab}p^{0m}\right)  \right]  \label{eqn172}%
\end{equation}
where parts of the secondary constraints ($\chi^{00}\left(  2\right)  $,
$\chi^{0k}\left(  2\right)  $ and $\chi^{0k}\left(  1\right)  $) will also
contribute to the final result. The variation of the last two terms in
(\ref{eqn172}) gives contributions proportional to the primary constraints
(which equal zero on the surface of primary constraints). The only relevant
parts of the generator for $\delta g_{km}$ are given by%

\begin{equation}
G=\xi_{0}\chi^{00}\left(  2\right)  +\xi_{k}\left(  \chi^{0k}\left(  2\right)
+\chi^{0k}\left(  1\right)  \right)  . \label{eqn173}%
\end{equation}
Performing variation of (\ref{eqn173}) with respect to $p^{km}$, using the
expression for momenta given in (\ref{eqn06.1}), and reverting from $e^{km}$
to $g^{km}$ using (\ref{eqn06.04}), we obtain%

\[
\delta g_{km}=-\frac{1}{2}\left(  \xi_{k,m}+\xi_{m,k}\right)
\]

\[
+\frac{1}{2}g^{00}\left(  g_{k0,m}+g_{m0,k}-g_{km,0}\right)  \xi_{0}+\frac
{1}{2}g^{p0}\left(  g_{k0,m}+g_{m0,k}-g_{km,0}\right)  \xi_{p}%
\]

\begin{equation}
+\frac{1}{2}g^{0p}\left(  g_{kp,m}+g_{mp,k}-g_{km,p}\right)  \xi_{0}+\frac
{1}{2}g^{pq}\left(  g_{kq,m}+g_{mq,k}-g_{km,q}\right)  \xi_{p}. \label{eqn174}%
\end{equation}
It is not difficult to check that up to the same numerical factor $2$ (as
occurred in (\ref{eqn168.1})) this is equivalent to $\delta_{(diff)}g_{\mu\nu
}$ (\ref{eqn167}) with $\mu\nu=km$.

We see that transformations of the time-time and space-time components of the
metric tensor are exactly equivalent to a diffeomorphism and the space-space
components give a diffeomorphism only on the surface of primary constraints.
Such a deviation from ordinary field theories like Yang-Mills can be expected
because the derivation of generators is performed (i.e., the functions
$\alpha_{\gamma}^{\sigma}\left(  x,y\right)  $ are found) only on a surface of
primary constraints. This is a consequence of the peculiarities of
diffeomorphism transformations that will be discussed at the end of this Section.

Returning to Dirac's statement \cite{Dirac} about abandoning four--dimensional
symmetry in his approach; we can see that it is restrictive and only related
to his initial modification of the Lagrangian (\ref{eqn04}). This abandoning
of four-dimensional symmetry does not appear, neither in linearized \cite{GKK}
nor in full GR \cite{KKRV}. The Dirac Hamiltonian formulation of GR, as we
demonstrated in this Section, allows one to derive the gauge transformation of
the metric tensor that can be put in covariant form and four-dimensional
symmetry is preserved. The exact meaning of common statements, such as the one
that is found in \cite{Pulin}, \textquotedblleft unfortunately, the canonical
treatment breaks the symmetry between space and time in general
relativity\textquotedblright,\ must also be clarified in light of our results.
Of course, and this is a property of the Hamiltonian approach itself, the
four-dimensional symmetry is not broken, it just is not manifest. For any
generally covariant theory with first-class constraints we can only make a
conclusion about abandoning such a symmetry if the gauge invariance that is
derived from the first-class constraints cannot be presented in covariant
form. In next Section we will show that it happens in the case of the ADM
formulation. If the symmetry possessed by the original Lagrangian disappears
in the Hamiltonian formulation, then it should be regarded as a very strong
indication that there is a mistake in the formulation; this is not a problem
with the initial Lagrangian or with the Hamilton-Dirac method. From this point
of view, if the \textquotedblleft canonical treatment breaks the
symmetry\textquotedblright, then\ such a treatment is \textit{not canonical}.

\subsection{Alternative derivation of a gauge generator}

Let us return to the derivation of the gauge transformations. Our derivation
of the transformation was based on an application of Castellani's method
\cite{Castellani}. There are at least two variations of it: one of them is
based on the extended Hamiltonian \cite{HTZ, HTbook}, where all first-class
constraints are included, and the other, \cite{Novel}, is based only on the
total Hamiltonian (i.e. only primary first-class constraints are included) as
in Castellani's case. The equivalence of the algorithms \cite{HTZ} and
\cite{Novel} was discussed in \cite{Novel} and a comparison of methods
\cite{HTZ} and \cite{Castellani} was made in \cite{HTZ}. Primary constraints
play a special role in all of these methods. The need to include multipliers
associated with the primary constraints was also emphasized in \cite{HTZ,
HTbook} and their importance in gauge transformations was demonstrated by some
simple examples. In \cite{Novel}, the multipliers are also important elements
of this method (see below). Recently, the method of \cite{Novel} was applied
to the ADM Hamiltonian in \cite{Saha}, where the transformation of the metric
tensor was derived and was shown to differ in form from (\ref{eqn00}). For
completeness, we apply the method used in \cite{Saha} to the Dirac Hamiltonian
formulation of GR to demonstrate that the derivation of the diffeomorphism
transformation in the Hamiltonian approach is not an artifact of a particular
procedure for finding a gauge generator. At the same time this demonstration
will illustrate the equivalence between the two different methods described in
\cite{Castellani} and \cite{Novel}\footnote{We would like to note that the
method \cite{Novel} is sensitive to the choice of linear combination of
non-primary constraints and gives equivalent to \cite{Castellani} result only
when using \textquotedblleft covariant\textquotedblright\ constraints of
Dirac's formulation (see more detail in \cite{KKaffinemetric}).}, as well as
the equivalence of both of them to the Lagrangian treatment of this problem in
\cite{Samanta}.

The total Hamiltonian is the starting point of the method outlined in
\cite{Novel}%

\begin{equation}
H_{T}=H_{c}+\lambda_{\mu}\phi^{\mu}. \label{eqn180}%
\end{equation}
For a system with only irreducible secondary first-class constraints and no
tertiary constraints (so that $\left\{  \phi^{\mu},H_{c}\right\}  =\chi^{\mu}%
$) the generator of gauge transformations is simply%

\begin{equation}
G=\eta_{\mu}\phi^{\mu}+\xi_{\mu}\chi^{\mu} \label{eqn181}%
\end{equation}
with two sets of parameters, $\eta_{\mu}$ and $\xi_{\mu}$ (twice the number of
primary constraints), which are related by (see Eq. (17) of \cite{Novel}):%

\begin{equation}
0=\xi_{\mu,0}-\xi_{\nu}\left(  V_{\left(  s\right)  \mu}^{\nu}+\lambda
_{\gamma}B_{\left(  s\right)  \mu}^{\nu\gamma}\right)  -\eta_{\nu}\left(
W_{\left(  s\right)  \mu}^{\nu}+\lambda_{\gamma}C_{\left(  s\right)  \mu}%
^{\nu\gamma}\right)  . \label{eqn182}%
\end{equation}
Here $W_{\mu}^{\nu}$, $V_{\mu}^{\nu}$, $C_{\mu}^{\nu\gamma}$, and $B_{\mu
}^{\nu\gamma}$ are the structure functions of the involutive algebra (see Eqs.
(2) and (3) of \cite{Novel})%

\begin{equation}
\left\{  H_{c},\phi^{\nu}\right\}  =W_{\left(  p\right)  \mu}^{\nu}\phi^{\mu
}+W_{\left(  s\right)  \mu}^{\nu}\chi^{\mu}, \label{eqn183}%
\end{equation}

\begin{equation}
\left\{  H_{c},\chi^{\nu}\right\}  =V_{\left(  p\right)  \mu}^{\nu}\phi^{\mu
}+V_{\left(  s\right)  \mu}^{\nu}\chi^{\mu}, \label{eqn184}%
\end{equation}

\begin{equation}
\left\{  \phi^{\nu},\phi^{\gamma}\right\}  =C_{\left(  p\right)  \mu}%
^{\nu\gamma}\phi^{\mu}+C_{\left(  s\right)  \mu}^{\nu\gamma}\chi^{\mu},
\label{eqn185}%
\end{equation}

\begin{equation}
\left\{  \phi^{\nu},\chi^{\gamma}\right\}  =B_{\left(  p\right)  \mu}%
^{\nu\gamma}\phi^{\mu}+B_{\left(  s\right)  \mu}^{\nu\gamma}\chi^{\mu}.
\label{eqn186}%
\end{equation}
The indices $\left(  p\right)  $ and $\left(  s\right)  $ indicate structure
functions associated with the primary $\phi^{\mu}$ and secondary constraints
$\chi^{\mu}$, respectively. Note, that (\ref{eqn182}) involves only functions
with the subscript $\left(  s\right)  $, i.e. structure functions related to
the primary constraints are not present, so that this equation is valid on the
primary constraint surface. This is similar to what happens in Castellani's procedure.

To find the generator (\ref{eqn181}), one has to solve (\ref{eqn182}) for
$\eta_{\nu}$ in terms of $\xi_{\nu}$; and as in Castellani's procedure the
number of independent parameters becomes equal to the number of primary
constraints. For the Dirac Hamiltonian of GR, which is obtained after
modification of the gamma-gamma part (with no effect on the equations of
motion, canonical variables $g_{\mu\nu}$, and conjugate momenta $p^{\mu\nu}$)
we have the simple primary constraints (\ref{eqn05}) $\phi^{0\mu}=p^{0\mu}$
and the secondary constraints (for explicit expressions see (\ref{eqn51}),
(\ref{eqn52}) and (\ref{eqn20}))%

\begin{equation}
\left\{  \phi^{0\mu},H_{c}\right\}  =\chi^{0\mu}. \label{eqn188}%
\end{equation}
This allows us to write the Hamiltonian in a compact and symmetric form
$H_{c}=2g_{0\mu}\chi^{0\mu}.$

The possibility of solving (\ref{eqn182}) for $\eta_{\nu}$, which is an
ordinary algebraic equation, depends on the structure functions $W_{\left(
s\right)  \mu}^{\nu}$ and $C_{\left(  s\right)  \mu}^{\nu\gamma}$. In the case
of the Dirac Hamiltonian formulation of GR (see (\ref{eqn188})) they are%

\begin{equation}
W_{\left(  s\right)  \mu}^{\nu}=\delta_{\mu}^{\nu},\left.  {}\right.  \left.
{}\right.  C_{\left(  s\right)  \mu}^{\nu\gamma}=0 \label{eqn190}%
\end{equation}
that give for solution of (\ref{eqn182}) for $\eta_{\mu}$:%

\begin{equation}
\eta_{\mu}=\xi_{\mu,0}-\xi_{\nu}\left(  V_{\left(  s\right)  \mu}^{\nu
}+\lambda_{\gamma}B_{\left(  s\right)  \mu}^{\nu\gamma}\right)  .
\label{eqn191}%
\end{equation}
The structure functions $V_{\left(  s\right)  \mu}^{\nu}$ for GR are
complicated (see (\ref{eqn93})), for $B_{\left(  s\right)  \mu}^{\nu\gamma}$
we have (see (\ref{eqn90}))%

\[
B_{\left(  s\right)  \mu}^{\nu\gamma}=\frac{1}{2}g^{\nu\gamma}\delta_{\mu}%
^{0},
\]
and (\ref{eqn191}) becomes%

\begin{equation}
\eta_{\mu}=\xi_{\mu,0}-\xi_{\nu}V_{\left(  s\right)  \mu}^{\nu}-\xi_{\nu
}g_{0\gamma,0}\frac{1}{2}g^{\nu\gamma}\delta_{\mu}^{0}. \label{eqn195}%
\end{equation}

Note that the structure functions $B_{\left(  s\right)  \mu}^{\nu\gamma}$ do
not equal zero and the Lagrange multipliers, which are the velocities
$g_{0\sigma,0}$ that cannot be expressed in term of $p^{0\sigma}$ enter
(\ref{eqn195}) explicitly ($\lambda_{0}=g_{00,0}$ and $\lambda_{k}=2g_{0k,0}%
$).\footnote{Expressions for the multipliers came from the Legendre
transformation and from the fact that the modified Lagrangian is independent
of $g_{0\mu.0}$, so that $H=g_{\nu\mu,0}p^{\nu\mu}-L=g_{00,0}p^{00}%
+2g_{0\mu,0}p^{0\mu}+...$} After the substitution of (\ref{eqn195}) into
(\ref{eqn181}) we obtain a one-parameter $\xi_{\nu}$ (the number of components
equals to the number of primary constraints) generator%

\begin{equation}
G\left(  \xi_{\nu}\right)  =-\xi_{\nu}g_{00,0}\frac{1}{2}g^{\nu0}\phi^{0}%
-\xi_{\nu}g_{0m,0}g^{\nu m}\phi^{0}+\xi_{\nu,0}\phi^{\nu}-\xi_{\nu}V_{\left(
s\right)  \mu}^{\nu}\phi^{\mu}+\xi_{\nu}\chi^{\nu}. \label{eqn196}%
\end{equation}
This expression has to be compared with the generator found using Castellani's procedure%

\begin{equation}
G=\xi_{\nu,0}G_{\left(  1\right)  }^{\nu}+\xi_{\nu}G_{\left(  0\right)  }%
^{\nu}. \label{eqn197}%
\end{equation}
The third term of (\ref{eqn196}) is exactly the same as the first term of
(\ref{eqn197}); the first, second, and last terms of (\ref{eqn196}) are also
the same as in Castellani's approach (see the first line of (\ref{eqn149})).
The fourth term of (\ref{eqn196}), to be compared with our result obtained
using Castellani's approach, is $-\xi_{\nu}V_{\left(  s\right)  \mu}^{\nu}%
\phi^{\mu}$. The structure function, $V_{\left(  s\right)  \mu}^{\nu}$,
originates from the calculation of the PB of (\ref{eqn184}), which in the case
of a field theory is%

\begin{equation}
\left\{  H_{c},\chi^{\nu}\left(  x\right)  \right\}  =\int V_{\left(
s\right)  \mu}^{\nu}\left(  x,y\right)  \chi^{\mu}\left(  y\right)  d^{3}y.
\label{eqn198}%
\end{equation}
The direct calculation of $\left\{  H_{c},\chi^{\nu}\right\}  $ gives terms
proportional to $\chi^{\mu}$ and its derivatives (for its explicit form see
(\ref{eqn93}))%

\begin{equation}
\left\{  H_{c},\chi^{\nu}\right\}  =K_{\mu}^{\nu}\chi^{\mu}+M_{\mu}^{\nu
k}\chi_{,k}^{\mu}. \label{eqn199}%
\end{equation}
This equation is usually presented in the following form in order to find the
structure functions%

\begin{equation}
\left\{  H_{c},\chi^{\nu}\right\}  =\int\left[  K_{\mu}^{\nu}\left(  x\right)
\delta\left(  x-y\right)  -M_{\mu}^{\nu k}\left(  x\right)  \frac{\partial
}{\partial y_{k}}\delta\left(  x-y\right)  \right]  \chi^{\mu}\left(
y\right)  d^{3}y. \label{eqn210}%
\end{equation}
This is the standard form for an intermediate result in such calculations
(e.g. see \cite{Castellani}). To find the generator (\ref{eqn196}), it is not
necessary to rewrite (\ref{eqn199}) in the form of (\ref{eqn210}); and the
direct substitution $\chi^{\mu}\rightarrow\phi^{\mu}$ into (\ref{eqn199})
gives the corresponding part, $-\xi_{\nu}V_{\left(  s\right)  \mu}^{\nu}%
\phi^{\mu}$, of the generator (\ref{eqn196}). This is equivalent to the second
and third lines of (\ref{eqn149}).

If the \textit{novel} method of \cite{Novel}\ is applied to the Dirac
Hamiltonian of GR and not to the ADM Hamiltonian, as was done in \cite{Saha},
then the generator (\ref{eqn196}) is equivalent to the one obtained using the
old procedure of \cite{Castellani}; and so they generate the same gauge
transformations when these two methods are applied to the same Hamiltonian.

The peculiarities of Castellani's procedure, when it is applied to the
Hamiltonian of GR, were discussed at the beginning of this Section and
illustrated by a derivation of (\ref{eqn00}) using the methods
\cite{Castellani} and \cite{Novel}. They are originated from the algebra of
constraints either in Dirac's formulation, given in this Section, or in the
`covariant' formulation of \cite{KKRV}. This algebra is different from that of
ordinary gauge theories. It reflects the peculiarities of diffeomorphism
invariance if compared to the gauge invariance of ordinary gauge theories. We
now briefly consider this topic.

\subsection{General coordinate transformations and diffeomorphism}

In the Introduction, we restrict our discussion to a particular meaning of
diffeomorphism given in (\ref{eqn00}) that is generally accepted in literature
on GR and which is similar to the usual gauge transformations. It is in
exactly this sense that diffeomorphism invariance can be derived from the
Hamiltonian approach. Now we would like to describe this transformation
without recourse to the Hamiltonian formulation as it is usually presented in
textbooks (e.g. see \cite{Landau}); and we wish to reveal how a difference
between these two views of diffeomorphism invariance manifests itself. This
exercise will also demonstrate the connection between (\ref{eqn00}) and
general coordinate transformations.

The principle of general covariance, the cornerstone of GR, puts severe
restrictions on the possible forms of the Lagrangian. The simplest is the
Einstein-Hilbert Lagrangian \cite{Landau, Carmeli}\footnote{Of course, it is
not unique and there are many posibilities: such as Lovelock gravity
\cite{Lovelock} or $f\left(  R\right)  $.}%

\[
L_{EH}=\sqrt{-g}R=\sqrt{-g}g^{\mu\nu}R_{\mu\nu}=\sqrt{-g}g^{\mu\nu}\left(
\Gamma_{\mu\nu,\alpha}^{\alpha}-\Gamma_{\mu\alpha,\nu}^{\alpha}+\Gamma_{\mu
\nu}^{\alpha}\Gamma_{\alpha\beta}^{\beta}-\Gamma_{\mu\beta}^{\alpha}%
\Gamma_{\alpha\nu}^{\beta}\right)  .
\]

The EH action and the Einstein equations are invariant under a general
coordinate transformation%

\begin{equation}
x^{\prime\mu}=f^{\mu}\left(  x^{\nu}\right)  \label{eqnD00}%
\end{equation}
and the corresponding transformation of the metric tensor%

\begin{equation}
g^{\prime\mu\nu}\left(  x^{\prime}\right)  =\frac{\partial x^{\prime\mu}%
}{\partial x^{\alpha}}\frac{\partial x^{\prime\nu}}{\partial x^{\beta}%
}g^{\alpha\beta}\left(  x\right)  . \label{eqnD1}%
\end{equation}
For infinitesimal transformations%

\begin{equation}
x^{\mu}\rightarrow x^{\prime\mu}=x^{\mu}+\xi^{\mu}\left(  x\right)
\label{eqnD0}%
\end{equation}
(\ref{eqnD1}) can be written as%

\begin{equation}
g^{\prime\mu\nu}\left(  x^{\prime}\right)  =g^{\mu\nu}\left(  x\right)
+\xi_{,\alpha}^{\nu}g^{\mu\alpha}\left(  x\right)  +\xi_{,\alpha}^{\mu
}g^{\alpha\nu}\left(  x\right)  +O\left(  \xi^{2}\right)  . \label{eqnD3}%
\end{equation}
Note that the components $\xi^{\mu}$ form a true vector \cite{Carmeli, Torre},
in contrast to the parameters $\varepsilon^{\perp}$ and $\varepsilon^{k}$
which appear in the ADM formulation of \cite{Pons} (see (\ref{eqn00a}) and its
derivation in the next Section).

If we consider $\xi^{\mu}\left(  x\right)  $ as being a small parameter and
restrict our interest to the first-order contributions in $\xi^{\mu}$, then
the exact invariance\footnote{We use the phrase \textquotedblleft exact
invariance\textquotedblright\ when a variation of the Lagrangian and equations
of motion under the gauge transformation gives zero exactly. We will see that
for GR the invariance is not \textquotedblleft exact\textquotedblright\ as the
variation of the Lagrangian equals to the total derivative, while the
variation of the Einstein equations is proportional to the equations
themselves.} of the EH action and the Einstein equations of motion is lost as
only the inclusion of all terms in the expansion of (\ref{eqnD3}) will
preserve it. In addition, if we want to present (\ref{eqnD3}) in a form
similar to a gauge transformation, in which the invariance with respect to
replacement of field variables is written in the same coordinate frame of
reference \cite{Grishchuk}, we should write both sides of (\ref{eqnD3}) in the
same coordinate system. This can be done by an additional approximation, using
the Taylor expansion of $g_{\mu\nu}^{\prime}\left(  x^{\prime}\right)  $:%

\begin{equation}
g^{\prime\mu\nu}\left(  x^{\prime}\right)  =g^{\prime\mu\nu}\left(  x^{\alpha
}+\xi^{\alpha}\left(  x\right)  \right)  =g^{\prime\mu\nu}\left(  x\right)
+g_{,\alpha}^{\mu\nu}\xi^{\alpha}+O\left(  \xi^{2}\right)  \label{eqnD4}%
\end{equation}
where in the second term (which is already linear in $\xi^{\alpha}$) we used
$\left.  \frac{\partial g^{\prime\mu\nu}\left(  x^{\prime}\right)  }{\partial
x^{\prime\alpha}}\right\vert _{x^{\prime}=x}=$ $g_{,\alpha}^{\mu\nu}+O\left(
\xi\right)  $.

Combining (\ref{eqnD3}) and (\ref{eqnD4}) and keeping only terms linear in
$\xi^{\alpha}$, we obtain%

\begin{equation}
\delta g^{\mu\nu}=g^{\prime\mu\nu}\left(  x\right)  -g^{\mu\nu}\left(
x\right)  =\xi_{,\alpha}^{\nu}g^{\mu\alpha}\left(  x\right)  +\xi_{,\alpha
}^{\mu}g^{\alpha\nu}\left(  x\right)  -g_{,\alpha}^{\mu\nu}\xi^{\alpha}
\label{eqnD5}%
\end{equation}
which is equivalent to%

\begin{equation}
\delta g^{\mu\nu}=\xi^{\mu;\nu}+\xi^{\mu;\nu}. \label{eqnD7}%
\end{equation}

By repeating similar calculations, or by using $\delta\left(  g_{\mu\alpha
}g^{\alpha\nu}\right)  =0$, one can find transformations for the covariant
components of a metric tensor%

\begin{equation}
\delta g_{\mu\nu}=-\xi_{\mu;\nu}-\xi_{\nu;\mu}=-\xi_{\mu,\nu}-\xi_{\nu,\mu
}+2\Gamma_{\mu\nu}^{\alpha}\xi_{\alpha}. \label{eqnD6}%
\end{equation}
This equation is just (\ref{eqn00}) (or its equivalent form (\ref{eqn167})
that was more convenient for comparison with our calculations). With $\xi
^{\mu}$, being a true vector, the transformations (\ref{eqnD7}) and
(\ref{eqnD6}) are generally covariant (these are covariant derivatives of a
true vector), so they are independent of the choice of coordinate system;
these are the transformations we derived from the Dirac Hamiltonian of GR and
in \cite{KKRV}.

Similarly, we can obtain the transformation of the Christoffel symbols. Using
the relation between $\Gamma_{\mu\nu}^{\alpha}$ and $g_{\mu\nu}$ in
(\ref{eqn06.02}) and the transformation $\delta g_{\mu\nu}$ in (\ref{eqnD6}),
we obtain%

\begin{equation}
\delta\Gamma_{\mu\nu}^{\alpha}=-\xi^{\beta}\Gamma_{\mu\nu,\beta}^{\alpha
}+\Gamma_{\mu\nu}^{\beta}\xi_{,\beta}^{\alpha}-\Gamma_{\mu\beta}^{\alpha}%
\xi_{,\nu}^{\beta}-\Gamma_{\nu\beta}^{\alpha}\xi_{,\mu}^{\beta}-\xi_{,\mu\nu
}^{\alpha}. \label{eqnD10}%
\end{equation}
Note that the presence of second-order derivatives of the parameters
($\xi_{,\mu\nu}^{\alpha}=\partial_{\mu}\partial_{\nu}\xi^{\alpha}$)
immediately allows one to come to a general conclusion about the constraint
structure of the Hamiltonian formulation of GR in the first-order form, the
Einstein affine-metric formulation of \cite{Einstein}, where $g^{\mu\nu}$ and
$\Gamma_{\mu\nu}^{\alpha}$ are treated as independent fields.\footnote{This
formulation was originally introduced by Einstein \cite{Einstein} (for English
translation see \cite{Einstein-eng}), but mistakenly attributed to Palatini
\cite{Franc} (see also Palatini's original paper \cite{Palatini} and its
English translation \cite{Palatini-eng}).} The presence of the second-order
derivatives of the parameters in the transformation $\delta\Gamma_{\mu\nu
}^{\alpha}$ implies that the generators must have the same order of
derivatives, i.e. the tertiary constraints must appear in such a formulation.
Of course, direct calculations confirmed this simple observation \cite{Kummer,
KKAnn, KKM, GM, Gerry-Ramin-2, Gerry-new, KKaffinemetric}.\ In the first
discussion of the Hamiltonian formulation of the first-order form of GR which
was presented in \cite{ADM-1}, the tertiary constraints did not appear because
some first-class constraints were solved \textit{before} closure of Dirac's
procedure was reached; this procedure is not a consistent implementation of
Dirac's procedure for treating constrained systems. This fact was clearly
demonstrated in the Appendix of \cite{Faddeev} and was discussed in
\cite{KKAnn}. The loss of tertiary constraints is also due to a misleading
analogy between the first-order formulations of Electrodynamics and linearized
GR appearing in \cite{ADM-1, ADM-2}. In the first-order formulation of
Electrodynamics, where the field strength is treated as an independent
variable, there is no increase in the order of the derivatives of the gauge
parameters in the generator of gauge transformations because the variation of
the field strength is zero. In contrast, the variation of $\Gamma_{\mu\nu
}^{\alpha}$ under diffeomorphism transformations, (\ref{eqnD10}), is not zero
and cannot even be written in covariant form as $\delta g_{\mu\nu}$ was in
(\ref{eqnD6}). This characteristic is consistent with $\Gamma_{\mu\nu}%
^{\alpha}$ not being a true tensor \cite{Eisenhart, Landau}\footnote{The
$\Gamma_{\mu\nu}^{\alpha}$ behaves like a tensor only with respect to linear
coordinate trasformations \cite{Landau}. Probably, this was the origin of the
analogy between Electrodynamics and linearized GR and of the conjecture that
this analogy should be extended to full GR \cite{ADM-1}.}. Also there is no
linear combination of the first-order derivatives of the metric tensor that is
exactly invariant under diffeomorphism transformations, as well as under
general coordinate transformations\footnote{Actually, such a combination
exists which is a true tensor: this is the covariant derivative of the metric
tensor, $g_{\mu\nu;\gamma}$, but it identically equals zero in GR.}. This
invariance is also related to the fact that a generally covariant action for
GR cannot be built from terms only quadratic in the first-order derivatives of
the metric tensor; and the simplest generally covariant, EH Lagrangian, is
proportional to a Ricci scalar and this involves second-order derivatives of
the metric tensor. This is why the affine-metric formulation of GR leads
unavoidably to tertiary constraints and consequently increases the length of
the chain of constraints and the order of the derivatives in the gauge
generator. This general arguments were recently confirmed by direct
calculations in \cite{KKaffinemetric, Gerry-Ramin-2, Gerry-new}.

Let us compare (\ref{eqnD10}) with the gauge invariance of Yang-Mills theory.
We have the field strength $F_{\mu\nu}^{a}$ whose variation does not vanish,
in contrast to Electrodynamics,%

\[
\delta F_{\mu\nu}^{a}=f^{abc}\theta^{b}F_{\mu\nu}^{c},
\]
with $f^{abc}$ a totally antisymmetric structure constant\ and $\theta^{b}$ a
gauge parameter. We do not have exact invariance for $F_{\mu\nu}^{a}$ (i.e.
$\delta F_{\mu\nu}^{a}\neq0$), but the gauge parameters enter only linearly
(without derivatives). Thus in the first-order formulation of Yang-Mills
theory, if we consider the field strength as an independent variable, there is
no increase in the length of the chains of constraints needed to accommodate
these transformations. In GR it is not possible to build any combination of
first-order derivatives which is exactly invariant under diffeomorphism; and
it is also impossible to find a combination whose variation is proportional to
the gauge parameter or its first-order derivatives.

The transformation (\ref{eqn00}) (or other equivalent forms) is written in the
same coordinate system and, because this combination is a true tensor, it is
independent of the coordinate transformations. In this sense it is
\textquotedblleft analogous to the gauge transformation\textquotedblright%
\ \cite{Muller, Weinberg}; but the absence of combinations of derivatives that
do not lead to an increase of the order of the derivatives of gauge parameters
in the generator is a distinct property of GR.

In addition, the Lagrangian of GR is not exactly invariant ($\delta L\neq0$)
under a diffeomorphism transformation, in contrast to Maxwell and Yang-Mills
theories. From (\ref{eqnD5}) and (\ref{eqnD10}) and by using%

\[
R_{\mu\nu}=\Gamma_{\mu\nu,\alpha}^{\alpha}-\Gamma_{\mu\alpha,\nu}^{\alpha
}+\Gamma_{\mu\nu}^{\alpha}\Gamma_{\alpha\beta}^{\beta}-\Gamma_{\mu\beta
}^{\alpha}\Gamma_{\alpha\nu}^{\beta}%
\]
we can find the transformations of $R_{\mu\nu}$ and $R=g^{\alpha\beta
}R_{\alpha\beta}$:
\begin{equation}
\delta R_{\mu\nu}=-\xi^{\rho}R_{\mu\nu,\rho}-\xi_{,\mu}^{\rho}R_{\nu\rho}%
-\xi_{,\nu}^{\rho}R_{\mu\rho},\left.  {}\right.  \left.  {}\right.  \delta
R=-\xi^{\rho}R_{,\rho}~. \label{eqnD15}%
\end{equation}
From the above relations it is not difficult to confirm that the
transformation of the EH Lagrangian under a diffeomorphism gives the total divergence%

\begin{equation}
\delta\left(  \sqrt{-g}R\right)  =\left(  -\xi^{\mu}\sqrt{-g}R\right)  _{,\mu
}. \label{eqnD16}%
\end{equation}

This lack of exact invariance is distinct from what occurs in the Maxwell and
Yang-Mills theories, but other models exist with Lagrangians which are also
invariant up to a total divergence, e.g. Topologically Massive Electrodynamics
(TME) of \cite{Notexact}. (See \cite{Notexact1, Notexact2, Notexact3} for a
discussion of its first-order formulation.) Despite differences in the
invariance property of Lagrangians, which can be either exact as in ordinary
Electrodynamics or up to a total divergence as in TME \cite{Notexact}, the
equations of motion are \textit{exactly} invariant under gauge
transformations. In GR the transformation of the equations of motion is
proportional to the equations themselves \cite{Grishchuk}. Using
(\ref{eqnD15}) we can easily find transformations of the Einstein field equations%

\begin{equation}
\delta G_{\mu\nu}=-\xi^{\rho}G_{\mu\nu,\rho}-\xi_{,\mu}^{\rho}G_{\nu\rho}%
-\xi_{,\nu}^{\rho}G_{\mu\rho} \label{eqnD17}%
\end{equation}
where%

\begin{equation}
G_{\mu\nu}=R_{\mu\nu}-\frac{1}{2}g_{\mu\nu}R \label{eqnD17a}%
\end{equation}
is the Einstein tensor. So, in GR under diffeomorphism transformations
(\ref{eqn00}), the equations of motion are invariant only `on-shell' which
does not contradict the principle of gauge invariance: a solution to the
equations of motion maps into a solution. This last result, the `on-shell'
invariance of the equations of motion might cause some confusion because
(\ref{eqn00}) was obtained from infinitesimal coordinate transformations
(\ref{eqnD0}), which are a particular case of general coordinate
transformations (\ref{eqnD00}). The EH action and the Einstein equations are
exactly invariant under (\ref{eqnD00}) and therefore, under (\ref{eqnD0}) (in
fact, the Einstein equations were originally postulated so as to satisfy
(\ref{eqnD00})). After writing (\ref{eqnD0}) in \textit{the same coordinate
system} and using the approximations (\ref{eqnD3}) and (\ref{eqnD4}), exact
invariance is lost. Such a mapping, (\ref{eqnD17}), by itself, is not peculiar
to GR because a similar property is present in the Yang-Mills theory where the
transformation of the equations of motion is%

\begin{equation}
\delta\left(  D_{\mu}F_{a}^{\mu\nu}\right)  =f_{abc}\theta_{b}D_{\mu}%
F_{c}^{\mu\nu}. \label{ym}%
\end{equation}
\qquad As in GR, the Yang-Mills field equation is only invariant `on-shell';
but this transformation is proportional to the gauge parameter, whereas the
transformation of $\delta G_{\mu\nu}$ in (\ref{eqnD17}) also contains
derivatives of the gauge parameter. The peculiar algebra of constraints in GR
is related to this increase in the order of the derivatives of the gauge
parameters in the transformations of the equations of motion, as well as the
impossibility of finding a combination of derivatives of the metric tensor
which is either exactly invariant or whose variation does not require
derivatives of the gauge parameters.

\subsection{Noether's differential identities and diffeomorphism}

In ordinary field theories gauge invariance is derivable from the
corresponding actions without any reference to coordinate transformations; and
GR being a field theory is no exception. All properties of a system are
encoded in its action and the presence of gauge symmetries reflects the fact
that some relations exist between equations of motion. This is stated in
Noether's theorems \cite{Noether} (see English translation of \cite{Noether}
in \cite{Noether-eng}) that allows us to find differential identities if a
symmetry of an action is known. The converse of both theorems also holds; that
is why we can try to build such identities without \textit{a priori} knowledge
of the gauge invariance(s) of a particular action. Such a construction for
covariant theories is a very simple iterative procedure: starting from Euler
derivatives (variation of the Lagrangian with respect to basic field(s)) of an
action one can try to build a basic differential identity by differentiating
the Euler derivative (ED) and then finding combinations of EDs that
identically give zero. This procedure works for all field theories and we
demonstrated it in detail by deriving the gauge invariance of the first-order
Einstein-Cartan action in \cite{trans}, where the basic differential
identities lead to translational and rotational invariances in the tangent
space \cite{trans}.

We can also apply this method to GR. Using the variational principle (without
any references to a coordinate system) we find a variation of (\ref{eqn03.1})
with respect to only the independent field variable $g_{\alpha\beta}$. That
gives us the Euler derivative:%

\begin{equation}
E^{\alpha\beta}=\frac{\delta L_{EH}}{\delta g_{\alpha\beta}}=\sqrt{-g}\left(
\frac{1}{2}g^{\alpha\beta}R-R^{\alpha\beta}\right)  =-\sqrt{-g}G^{\alpha\beta
}, \label{diff1}%
\end{equation}
where $G^{\alpha\beta}$ is the contravariant Einstein tensor. Note that the
equations of motion of GR, which are in fact the Euler derivative,
$E^{\alpha\beta}$, differ from Einstein's equation $G^{\alpha\beta}=0$ by
$\sqrt{-g}$. This difference is not important if we want to obtain solutions
of Einstein's equations (or equations of motion) as we always assume that
$\det\left(  g_{\mu\nu}\right)  \neq0$; but in deriving the identities and the
corresponding transformations of fields, the Euler derivative must be used.
The combination $\sqrt{-g}G^{\alpha\beta}$ in (\ref{diff1}) appears naturally,
in contrast to the method employed in \cite{Samanta}, where the gauge identity
has to be imposed from the outset.

We start to build the basic differential identity from the derivative of
(\ref{diff1})%

\begin{equation}
\partial_{\alpha}E^{\alpha\beta}=\partial_{\alpha}\left(  -\sqrt{-g}%
G^{\alpha\beta}\right)  =-\frac{1}{2}\sqrt{-g}g^{\mu\nu}g_{\mu\nu,\alpha
}G^{\alpha\beta}-\sqrt{-g}\partial_{\alpha}G^{\alpha\beta}. \label{diff2}%
\end{equation}
Compare (\ref{diff2}) with (\ref{diff1}) as in \cite{trans} (where details can
be found), one obtains%

\begin{equation}
\partial_{\alpha}E^{\alpha\beta}=-\sqrt{-g}\Gamma_{\alpha\lambda}^{\beta
}G^{\alpha\lambda}=-\Gamma_{\alpha\lambda}^{\beta}E^{\alpha\lambda
}.\label{diff3}%
\end{equation}
Equally well (and in a much shorter way) the identity (\ref{diff3}) can be
obtained using the Bianci identity: $G_{~;\alpha}^{\alpha\beta}=\partial
_{\alpha}G^{\alpha\beta}+\Gamma_{\alpha\lambda}^{\alpha}G^{\beta\lambda
}+\Gamma_{\alpha\lambda}^{\beta}G^{\alpha\lambda}=0$. Note that the equations
of motion were not imposed in deriving the Bianci identity \cite{Landau}. This
is important as a differential identity should be satisfied `off-shell',
without imposing equations of motion. Finally, the differential identity for
the EH action is:
\begin{equation}
I^{\beta}=\partial_{\alpha}E^{\alpha\beta}+\Gamma_{\alpha\lambda}^{\beta
}E^{\alpha\lambda}=0.\label{diff4}%
\end{equation}

Now, forming a scalar density by contracting $I^{\beta}$ (which is a
contravariant vector density) with the gauge parameter of the corresponding
tensorial dimension (a covariant vector in this case)%
\begin{equation}%
%TCIMACRO{\dint }%
%BeginExpansion
{\displaystyle\int}
%EndExpansion
I^{\beta}\varepsilon_{\beta}d^{4}x=0 \label{diff5}%
\end{equation}
we find (using integration by parts and eliminating a surface term)%

\begin{equation}%
%TCIMACRO{\dint }%
%BeginExpansion
{\displaystyle\int}
%EndExpansion
I^{\beta}\varepsilon_{\beta}d^{4}x=%
%TCIMACRO{\dint }%
%BeginExpansion
{\displaystyle\int}
%EndExpansion
\left(  \partial_{\alpha}E^{\alpha\beta}+\Gamma_{\alpha\lambda}^{\beta
}E^{\alpha\lambda}\right)  \varepsilon_{\beta}d^{4}x=%
%TCIMACRO{\dint }%
%BeginExpansion
{\displaystyle\int}
%EndExpansion
\left(  -\frac{1}{2}\left(  \partial_{\alpha}\varepsilon_{\beta}%
+\partial_{\beta}\varepsilon_{\alpha}\right)  +\Gamma_{\alpha\beta}^{\gamma
}\varepsilon_{\gamma}\right)  E^{\alpha\beta}d^{4}x \label{diff6}%
\end{equation}
(as $E^{\alpha\beta}$ is symmetric in $\alpha$ and $\beta$).

Comparing (\ref{diff5}) with the variation of the action $\delta S=%
%TCIMACRO{\dint }%
%BeginExpansion
{\displaystyle\int}
%EndExpansion
E^{\alpha\beta}\delta g_{\alpha\beta}d^{4}x=0$ and using (\ref{diff6}) we can
read off the transformation of the metric tensor:%

\begin{equation}
\delta g_{\alpha\beta}=-\frac{1}{2}\left(  \partial_{\alpha}\varepsilon
_{\beta}+\partial_{\beta}\varepsilon_{\alpha}\right)  +\Gamma_{\alpha\beta
}^{\gamma}\varepsilon_{\gamma}\label{diff7}%
\end{equation}
where $\Gamma_{\alpha\beta}^{\gamma}$ is just a short-hand notation for
(\ref{eqn06.02}). The transformation (\ref{diff7}) coincides with the
diffeomorphism (\ref{eqnD6}) if we redefine $\varepsilon_{\alpha}=2\xi
_{\alpha}$. This result follows from the action without reference to any
properties of the coordinate transformations or the transformations of fields.
Note that in (\ref{diff7}) the numerical factor $2$ appears, which is the same
as we obtained in the gauge transformations when applying Castellani's
procedure. This consideration, based on Noether's second theorem, is an
illustration of how diffeomorphism invariance appears in GR in exactly the
same sense (in terms of differential identities) as in other field theories.

\subsection{Summary}

To summarize: firstly, the diffeomorphism transformation (\ref{eqn00}) is
related to coordinate transformations and as a consequence to general
covariance; secondly, the Hamiltonian formulation of GR is performed using the
same \textquotedblleft rule of procedure\textquotedblright\ as in ordinary
gauge theories, allowing one to obtain the same transformation (\ref{eqn00});
thirdly, the same transformation can be derived using purely Lagrangian
methods, which are valid for all field theories, without any reference to a
coordinate transformations or relying to the Hamiltonian methods. The
resulting invariance makes these transformations distinct from ordinary gauge
theories because of the presence of the derivatives of the gauge parameters
(compare (\ref{ym}) and (\ref{eqnD17}) for transformations of the equations of
motion). Such distinct transformations should manifest themselves in all steps
of the procedure and this is exactly what we have found. For example, the
non-zero PB among primary and secondary constraints in (\ref{eqn90}) and the
`on-shell' of primary constraints result of (\ref{eqn172}) are properties that
are not observed in ordinary gauge theories. These peculiarities are present
in Dirac's formulation as considered in this article, as well as in
\cite{KKRV}; and both of these formulations allow one to derive the
diffeomorphism invariance (\ref{eqn00}). At least two of the above mentioned
properties, which are related to (\ref{eqn90}) and (\ref{eqn172}), are absent
in the ADM formulation; and the transformations derived from the ADM
Hamiltonian are different from a diffeomorphism. Diffeomorphism, as a gauge
transformation according to Noether's theorems, reflects a very special
connection between the equations of motion; and such a relationship specific
to Einstein's GR cannot disappear in the Hamiltonian analysis. A comparison
between the Dirac and ADM formulations will be made in the next Section.

\section{The Dirac Hamiltonian of GR versus the ADM Hamiltonian of
geometrodynamics}

\subsection{Prelude}

In the previous Section we demonstrated that, following the \textquotedblleft
rule of procedure\textquotedblright\ as applied to Dirac's Hamiltonian of GR
(\ref{eqn86}) (with the covariant metric $g_{\mu\nu}$ and corresponding
conjugate momentum, $p^{\mu\nu}$, as the fundamental canonical variables) the
diffeomorphism invariance (\ref{eqn00}) can be derived using Castellani's
procedure \cite{Castellani} or the method of \cite{Novel}. The same result was
recently obtained in the Hamiltonian formulation of GR \cite{KKRV} without
using Dirac's modifications (\ref{eqn04}) and in the Lagrangian approach of
\cite{Gitmanbook} by Samanta \cite{Samanta}. This result is exactly what one
would expect for the invariance of GR, as well as the equivalence of the
results in the Hamiltonian and Lagrangian approaches.

In addition, because Dirac's non-covariant modification of the Lagrangian does
not change the equations of motion, four-dimensional symmetry\ is preserved
and it is reflected in the covariant form of the transformations
(\ref{eqn00}). We have also demonstrated that Dirac's references to space-like
surfaces are not part of his actual calculations. As a result, Hawking's
statement \cite{Hawking} that\ \textquotedblleft the split into three spatial
dimensions and one time dimension seems to be contrary to the whole spirit of
relativity\textquotedblright\ is not related to Dirac's formulation, where
only \textit{manifest }invariance (but not the invariance itself) is broken by
explicitly considering different components of the metric tensor. We have seen
that working with the original Einstein variable, the metric tensor, we have
the Hamiltonian formulation of GR that is consistent with diffeomorphism
symmetry, and the spirit of GR is \textquotedblleft alive\textquotedblright.

There exists a more popular Hamiltonian than Dirac's Hamiltonian which is
based on a different set of variables: the \textit{lapse} and \textit{shift}
functions and the space-space components of the metric tensor. It is
frequently, but mistakenly called the Dirac Hamiltonian (e.g.
\cite{Castellani}) and even its variables are called \textquotedblleft Dirac's
lapse and shift\textquotedblright\ \cite{SalSund}\footnote{The names
\textquotedblleft lapse\textquotedblright and \textquotedblleft
shift\textquotedblright\ were introduced neither by ADM nor by Dirac and
appeared only later, for example, in \cite{Gravitation, Kuchar}. To the best
of our knowledge, it was Wheeler who coined the names of these variables,
\textquotedblleft lapse\textquotedblright and \textquotedblleft
shift\textquotedblright\ in \cite{Relativity}.}. This formulation (with the
lapse and shift functions) is due to Arnowitt, Deser and Misner (see
\cite{ADM} and references therein). The name \textquotedblleft
Dirac-ADM\textquotedblright\ is also not correct; moreover, despite the
apparent similarities between the Dirac and ADM Hamiltonians, they are
different (see below). The appropriate and best known names for the ADM
formulation are \textquotedblleft ADM gravity\textquotedblright\ and
geometrodynamics, as opposed to Einstein gravity. The gauge transformation
derived from the ADM Hamiltonian by the methods of \cite{Castellani, Novel},
using ADM variables and constraints, is not diffeomorphism invariance
(\ref{eqn00}). This fact was recently demonstrated very clearly using the
method of \cite{Novel} in \cite{Saha} and was discussed in our Introduction.
The field-dependent redefinition of the parameters (\ref{eqn00a}) for the ADM
Hamiltonian distinct from the exact result that was obtained in Section III
from the Dirac formulation, where no such a redefinition was used, and in
\cite{Samanta} and \cite{KKRV} where the question of equivalence with
diffeomorphism does not arise.

We feel that it is insufficient to say that the formulations of \cite{Pirani,
KKRV} and that of Dirac \cite{Dirac}, both of which use the metric tensor as
the canonical variable, correctly describe the Hamiltonian of GR and one is
obliged to work with the original Einstein variables. One should not try to
recast GR into a description of the motion of surfaces; more precisely, one
should not change variables to make such an interpretation plausible. It is
necessary to understand why two such closely related Hamiltonians, those of
Dirac and ADM, which are mistakenly said to be equivalent, produce different
results. This is a general question related to the Hamiltonian formulation of
singular Lagrangians. The understanding of the peculiarities of the
Hamiltonian method for constrained systems can prevent from repeating some
mistakes that have been made when considering formulations of such theories.

The analysis of the ADM formulation is also interesting from another point of
view, as it provides a very instructive example of what might be called an
\textquotedblleft interpretational\textquotedblright\ approach to physics. In
the original work of ADM (e.g., \cite{ADM}), an interpretation of the
variables they introduced was proposed and in later work, this interpretation
became the cornerstone of that formulation. Attempts were made to
\textquotedblleft derive\textquotedblright\ results starting from that
interpretation, i.e. by elevating the interpretation to the level of first
principles \cite{Kuchar}. Such an approach exhibits lack of rigour and relies
on some geometrical reasoning. The essence of this approach is perfectly
expressed in the following quotation from \cite{Pictorial}: \textquotedblleft
I capture as much of the classical theory as I can by \textit{pictorial
visualization}\textquotedblright\footnote{This becomes a well-known
pedagogical approach in teaching conceptional (not mathematical) physics for
non-science students.} and \textquotedblleft The reader is encouraged to
follow the broad outlines and not worry about technical
details.\textquotedblright\ The \textquotedblleft advantage\textquotedblright%
\ of such an approach is that it cannot be disproved; yet it prevents one from
obtaining any reliable prediction or result. For example, if we accept Dirac's
references to space-like surfaces as a part of his formalism, then Hawking's
statement that introduction of a family of space-like surfaces
\textquotedblleft seems to be contrary to the whole spirit of
relativity\textquotedblright\ \cite{Hawking} forces us to reject this
formulation as an inappropriate formulation of GR. And yet, as we have
demonstrated, Dirac's formulation leads to a direct restoration of
diffeomorphism invariance and, because of this, it is consistent with the
spirit of GR and is the correct Hamiltonian formulation of GR. This
demonstration shows how a purely interpretational consideration can lead to a
wrong conclusion and that the interpretational approach without having to
\textquotedblleft worry about technical details\textquotedblright\ is
meaningless. Dirac's derivation follows the \textquotedblleft rule of
procedure\textquotedblright\ and allows us to check any interpretation by
explicit calculations. If something is constructed only using pure
interpretation, then the final result cannot be checked by calculations and
can only be analyzed by comparing its consistency with general principles. An
\textquotedblleft interpretation\textquotedblright\ cannot serve as a ground
to disprove a result and in fact could be wrong, as in the case of Dirac's
formulation in which he makes references to space-like surfaces that were not
used in his calculations. A general understanding of the limitations of the
interpretational approach probably provides an explanation of why Hawking's
words \cite{Hawking}, spoken almost thirty years ago on the occasion of the
centenary of Einstein's birth, were not enough to cause people to immediately
abandon the ADM formulation.

Yet more disturbing, is that this \textquotedblleft
interpretational\textquotedblright\ language has completely prevailed in the
Hamiltonian formulation of GR. As an example, consider the work of Samanta
\cite{Samanta} in which he used the Lagrangian formulation, and where there
are no surfaces of constant time, space-like, slicing, etc.; then his position
is abruptly altered when he refers to the Hamiltonian formulation of the same
theory and states that \textquotedblleft slicing is essential for Hamiltonian
formulation\textquotedblright. This assertion is obviously wrong as slicing is
\textit{not essential} and the Hamiltonian formulations of GR obtained without
any reference to slicing gives a consistent result, as we have demonstrated in
the previous Section and in \cite{KKRV}.

The general trend in Physics and the main goal of many physicists is the
unification of theories and methods on all possible levels; but even now, when
we are a few years away from the 100th jubilee of the discovery of GR
\cite{EinsteinGR} there remains an inconsistency when discussing different
formulations (Lagrange and Hamilton-Dirac) of the same theory, Einstein's GR!
This leads to erroneous observations, such as \textquotedblleft It is worth
noting that generalized Hamiltonian dynamics is not completely equivalent to
Lagrangian formulation of the original theory. In the Hamiltonian formalism
the constraints generate transformations of phase space variables; however,
the group of these transformations \textit{does not have to be equivalent} to
the group of gauge transformations of Lagrangian theory\textquotedblright%
\ \cite{Shestakova}.

We consider it important to understand and find explicitly where and why the
ADM Hamiltonian formulation contradicts the spirit of GR, and why it cannot be
associated with Einstein's theory (i.e. it is not the Hamiltonian formulation
of GR, but rather the Hamiltonian formulation of distinct formulation:
\textquotedblleft geometrodynamics\textquotedblright). One can talk about
abandoning the spirit of GR if one is discussing a different formulation built
on different principles (such as \cite{AndBar}) but the ADM formulation has
been given the appearance of having a formal basis on GR (according to the ADM
papers and the presentation in many textbooks); indeed the summary of ADM's
work in \cite{ADM} has the title \textquotedblleft Dynamics of General
Relativity\textquotedblright. The mathematical manifestation of the spirit of
Einstein's GR is the general covariance of Einstein's equations of motion for
the metric tensor. Einstein's GR is a field theory and the methods used in
ordinary field theories, those of Lagrange and Hamilton, if applied correctly,
must not destroy its spirit. This is exactly what we want to investigate:
where was \textquotedblleft a regular and uniform rule of
procedure\textquotedblright\ broken in the ADM approach? In Hamiltonian
language, we want to see where the canonical procedure was destroyed by
passing from the Dirac Hamiltonian to the ADM Hamiltonian and why the ADM
formulation with their variables is not equivalent to GR or, in other words,
is not a canonical formulation of GR.

\textsf{ }It does not seem possible to start from the Lagrangian of GR, where
surfaces are not present, and then after introducing new variables have such
surfaces. We will follow a path distinct from the interpretational approach
and pay attention to technical details by using the \textquotedblleft uniform
rule of procedure\textquotedblright\ in analyzing the ADM formulation. In the
previous Section we demonstrated that, when we are using the rule of
procedure, surfaces do not appear in either Dirac's calculations or in
\cite{KKRV}. We will not repeat the calculations of the previous Section or of
\cite{KKRV} for the ADM formulation, but instead we will analyze how the ADM
formulation is related to that of Dirac.

\subsection{Relationship between Dirac and ADM Hamiltonians}

Let us compare the two Hamiltonians of Dirac and ADM. Castellani himself
considered the GR and Yang-Mills theories as illustrative examples of his
algorithm for finding gauge transformations. Referring to Dirac's book
\cite{Diracbook}, Castellani \cite{Castellani} started with the statement
\textquotedblleft from the Hilbert action one \textit{derives} the
Hamiltonian\textquotedblright\footnote{Such a Hamiltonian can be
\textquotedblleft derived\textquotedblright\ without recourse to the EH
Lagrangian. We refer the reader interested in \textquotedblleft
visualization\textquotedblright\ or in the geometrical reasoning behind a
derivation or geometrical meaning of this equation to the numerous figures in
\cite{Kuchar}. We disregard such approaches as inadequate for any proofs.}%

\begin{equation}
H^{ADM}=N\mathcal{H}_{\bot}^{ADM}+N^{i}\mathcal{H}_{i}^{ADM}+N_{,0}\Pi
+N_{,0}^{i}\Pi_{i}, \label{eqn400}%
\end{equation}
where $\Pi$ and $\Pi_{i}$ are momenta conjugate to $N$ and $N^{i}$, respectively.

This Hamiltonian never appears in Dirac's book or in any of his papers.
Dirac's \textit{derivation} of the Hamiltonian of GR is in the article
\cite{Dirac} that we discussed in previous Sections. It is different from
(\ref{eqn400}) and given by (\ref{eqn86}) which is the canonical part of the
Hamiltonian, and his primary constraints are the momenta $p^{0\mu}$ conjugate
to $g_{0\mu}$ (\ref{eqn05}). Equation (\ref{eqn400}) is, in fact, the ADM
result \cite{ADM} and we have indicated so by using the superscript `$ADM$'.
In order to compare the results of the Dirac and ADM formulations we use a
slightly different convention for the Dirac Hamiltonian which appeared in a
subsequent article of Dirac (see Eq. (7) of \cite{Dirac-2} for the canonical
part of Hamiltonian)%

\begin{equation}
H_{D}=\left(  -g^{00}\right)  ^{-1/2}\mathcal{H}_{L}+g_{r0}e^{rs}%
\mathcal{H}_{s}+g_{00,0}p^{00}+2g_{0k,0}p^{0k}. \label{eqn401}%
\end{equation}
In this convention, $g_{00}$ is negative \cite{Dirac-2}.

Let us compare the secondary constraints of the two formulations.
$\mathcal{H}_{i}^{ADM}$, the \textquotedblleft
diffeomorphism\textquotedblright\ constraint, is given in many sources (e.g.
see Eq. (3.14b) of \cite{ADM}) as%

\[
\mathcal{H}_{i}^{ADM}=g_{ik}\mathcal{H}_{ADM}^{k}=-2g_{ik}\Pi_{\mid j}^{kj}%
\]
where $\Pi^{kj}$ is a momentum conjugate to the spatial metric $g_{kj}$. The
symbol \textquotedblleft$\mid$\textquotedblright\ seems to indicate the
covariant derivative with respect to $g_{kj}$ \cite{ADM}; but the definition
of the particular covariant derivative used in ADM is non-standard and is not
easily found (see Eq. (5.1) of \cite{ADM-6} and Eq. (2.3b) of \cite{ADM-12}
which are consequently papers six and twelve in their series)%

\[
\Pi_{\mid j}^{kj}\equiv\Pi_{,j}^{kj}+\Pi^{lm}\Gamma_{lm}^{k}%
\]
which gives%

\[
\mathcal{H}_{i}^{ADM}=-2g_{ik}\Pi_{,j}^{kj}-2\Pi^{kj}g_{ik,j}+\Pi^{kj}%
g_{jk,i}.
\]
This is exactly the Dirac constraint (see Eq. (D41) of \cite{Dirac}) or our
(\ref{eqn09.2.2})). We note that the ADM definition of a covariant derivative
of a contravariant second rank tensor mimics a covariant derivative of a
contravariant vector (the first rank tensor) or a covariant derivative of the
tensor density \cite{Carmeli}. If we use the standard definition of a
covariant derivative of the second rank tensor \cite{Landau, Eisenhart,
Carmeli} we will obtain a different result. This result can only be presented
as a standard covariant derivative (but with respect to the spatial metric
$g_{ik}$ only) if we write it as%

\[
\mathcal{H}_{i}^{Dirac}=\mathcal{H}_{i}^{ADM}=-2g_{is}\sqrt{\det g_{km}%
}\left(  \frac{\Pi^{ls}}{\sqrt{\det g_{km}}}\right)  _{;l}%
\]
(see Eq. (37) of \cite{Franke} or Eq. (E.2.34) of \cite{Waldbook}).

The scalar, \textquotedblleft Hamiltonian\textquotedblright, constraint
$\mathcal{H}_{\bot}^{ADM}$ is given by Eq. (3.14b) of \cite{ADM}%

\[
\mathcal{H}_{\bot}^{ADM}=-\sqrt{g}\left[  ^{3}R+g^{-1}\left(  \frac{1}{2}%
\Pi^{2}-\Pi^{ij}\Pi_{ij}\right)  \right]  =
\]

\begin{equation}
-\sqrt{g}~^{3}R+\frac{1}{\sqrt{g}}\left(  g_{ik}g_{jm}-\frac{1}{2}g_{ij}%
g_{km}\right)  \Pi^{ij}\Pi^{km}. \label{eqnADMs}%
\end{equation}

Using (50) (employing the convention of \cite{Dirac-2} where $g^{00}$ is
negative), from (48) we obtain the second term of (\ref{eqnADMs}). Also from
(43), and taking into account (44), the first term of (\ref{eqnADMs}) follows.
Thus, this constraint is also equivalent to Dirac's $\mathcal{H}_{L}$.

Dirac's combinations $\mathcal{H}_{L}$ and $\mathcal{H}_{s}$ of the
\textquotedblleft covariant\textquotedblright\ secondary constraints
$\chi^{0\sigma}$ (given in (\ref{eqn53}), (\ref{eqn85})) are \textit{exactly}
the same as the ADM secondary\footnote{Actually, in the ADM formulation they
appear as primary \cite{ADM-7, ADM}. We will return to this later.}
constraints $\mathcal{H}_{\bot}^{ADM}$ and $\mathcal{H}_{i}^{ADM}$%

\begin{equation}
\mathcal{H}_{\bot}^{ADM}=\mathcal{H}_{L},\left.  {}\right.  \left.  {}\right.
\mathcal{H}_{i}^{ADM}=\mathcal{H}_{i}. \label{eqn402}%
\end{equation}
The only difference between the first two terms of (\ref{eqn400}) and
(\ref{eqn401}) is that the field-dependent coefficients in front of Dirac
combinations of constraints are called new variables by ADM \cite{ADM}. These
are the \textit{lapse} and \textit{shift} functions%

\begin{equation}
N\equiv\left(  -g^{00}\right)  ^{-1/2}, \label{eqn407}%
\end{equation}

\begin{equation}
N^{i}\equiv g_{j0}e^{ji}=-\frac{g^{0i}}{g^{00}}. \label{eqn408n}%
\end{equation}

In addition, Dirac's $e^{ji}$ (\ref{eqn06.04}) (the reciprocal of $g_{ji}$) is
called the \textquotedblleft three-dimensional\textquotedblright\ metric
$g^{ji}$ \cite{ADM}. (In fact, $g^{ji}$ is the space-space component of the
full four-dimensional metric in four-dimensional space-time.) To distinguish
$e^{ji}$ from the space-space components of Einstein's four-dimensional
contravariant metric tensor, the latter is defined to be $^{4}g^{ji}$
\cite{ADM}. Dirac's notation is more transparent as it arises in his
derivation of the Hamiltonian that was analyzed in Section II.\footnote{The
ADM renaming was probably introduced to underline the geometrical
interpretation of their variables (see \cite{Relativity}).}

The confused notion (in the literature) that these two formulations are
equivalent, is understandable, especially in light of relation (\ref{eqn402}).
Another reason is that in many presentations of the ADM Hamiltonian
(\ref{eqn400}), such as in \cite{Pulin}, the primary constraints are ignored
by imposing the idea that the lapse and shift variables are merely the
Lagrange multipliers for the constraints $\mathcal{H}_{\bot}$ and
$\mathcal{H}_{i}$ and that they can be treated as `primary' rather than
`secondary' constraints \cite{SalisburyPS}. Firstly, in such an approach even
derivation of the gauge transformations of all components of the metric tensor
becomes impossible as we are no longer dealing with the full phase space of
the Hamiltonian (e.g., see the remark on p. 3288 of \cite{Pons}). The methods
of derivation of Castellani \cite{Castellani} and of \cite{Novel} (which is
used in \cite{Saha}) employ the complete phase space and so all fields and
their conjugate momenta must be included. Secondly, dropping primary
constraints contradicts the methods of constraint dynamics: primary
constraints are first-class and must not be solved as this does not allow to
restore the gauge symmetry present in the Lagrangian. Only second-class
constraints can be solved and then some of the phase-space variables can be
eliminated provided PBs are replaced by Dirac brackets. All this means that if
we derive generators of gauge transformations, following any procedure using
$\mathcal{H}_{\bot}^{ADM}$ and $\mathcal{H}_{i}^{ADM}$ as the first-class
primary constraints, we will have generators independent of the momenta
conjugate to the \textquotedblleft multipliers\textquotedblright, so that\ the
gauge transformation of, for example $N$, would be zero%

\begin{equation}
\delta N=\left\{  N,G\right\}  =0. \label{eqnN}%
\end{equation}

This result means that $\delta g^{00}=0$, which is different from a
diffeomorphism transformation at all. Without the primary constraints the time
derivatives of the lapse function, for example, would vanish according to the
Hamiltonian formulation,%

\[
\left\{  N,H\right\}  =0.
\]
Thus, $N$ is constant in time, yet if we use the total Hamiltonian
(\ref{eqn401}), we obtain%

\[
\left\{  N,H_{T}\right\}  =N_{,0}.
\]

Finally, in the Hamiltonian approach, the primary constraints come from
varying the Lagrangian with respect to velocities, and if we follow this rule,
then $\mathcal{H}_{\bot}^{ADM}$ and $\mathcal{H}_{i}^{ADM}$ are not primary
constraints. We will consequently work with the total Hamiltonian.

Let us continue to compare these two formulations. The relation between their
primary constraints is as yet unclear and we shall return to this later. Form
(\ref{eqn407}) and (\ref{eqn408n}) the metric and its inverse are (e.g., see
\cite{Castellani, Gravitation})%

\begin{equation}
g_{\mu\nu}=\left(
\begin{array}
[c]{cc}%
g_{ij}N^{i}N^{j}-N^{2} & g_{ij}N^{j}\\
g_{ij}N^{j} & g_{ij}%
\end{array}
\right)  ,\left.  {}\right.  \left.  {}\right.  g^{\mu\nu}=\left(
\begin{array}
[c]{cc}%
-\frac{1}{N^{2}} & \frac{N^{i}}{N^{2}}\\
\frac{N^{i}}{N^{2}} & ^{3}g^{ij}-\frac{N^{i}N^{j}}{N^{2}}%
\end{array}
\right)  .\label{eqn409}%
\end{equation}
The fundamental PBs of the canonical variables, the components of the
covariant metric tensor and their corresponding conjugate momenta, for Dirac's
Hamiltonian are \cite{Dirac}%

\begin{equation}
\left\{  p^{\alpha\beta}\left(  x\right)  ,g_{\mu\nu}\left(  x^{\prime
}\right)  \right\}  =\frac{1}{2}\left(  \delta_{\mu}^{\alpha}\delta_{\nu
}^{\beta}+\delta_{\nu}^{\alpha}\delta_{\mu}^{\beta}\right)  \delta_{3}\left(
x-x^{\prime}\right)  , \label{eqn410}%
\end{equation}
whereas for the ADM approach the fundamental PBs are (e.g., see
\cite{Castellani, Saha, deWitt})%

\begin{equation}
\left\{  g_{ij}\left(  x\right)  ,\Pi^{kl}\left(  x^{\prime}\right)  \right\}
=\frac{1}{2}\left(  \delta_{i}^{k}\delta_{j}^{l}+\delta_{i}^{l}\delta_{j}%
^{k}\right)  \delta_{3}\left(  x-x^{\prime}\right)  =\Delta_{ij}^{kl}%
\delta_{3}\left(  x-x^{\prime}\right)  , \label{eqn411}%
\end{equation}

\begin{equation}
\left\{  N^{i}\left(  x\right)  ,\Pi_{j}\left(  x^{\prime}\right)  \right\}
=\delta_{j}^{i}\delta_{3}\left(  x-x^{\prime}\right)  , \label{eqn412}%
\end{equation}

\begin{equation}
\left\{  N\left(  x\right)  ,\Pi\left(  x^{\prime}\right)  \right\}
=\delta_{3}\left(  x-x^{\prime}\right)  . \label{eqn413}%
\end{equation}
Other PBs presumably equal zero (i.e. $\left\{  N\left(  x\right)  ,\Pi
^{kl}\left(  x^{\prime}\right)  \right\}  =0$, etc.) if these variables are to
be canonical.

We now investigate why the Dirac and ADM approaches to the canonical treatment
of GR lead to diffeomorphism transformations in the former case and to
transformations that correspond to a diffeomorphism only after a non-covariant
field-dependent redefinition of gauge parameters in the latter case.

\subsection{Complete treatment of the ADM formulation using Castellani
algorithm}

In Section III we have demonstrated that the gauge transformations that follow
from Dirac's Hamiltonian can be derived using both the methods of
\cite{Castellani} and \cite{Novel}. The method \cite{Novel} was applied to ADM
gravity in \cite{Saha}. The result of \cite{Saha} is not new, but it is
probably the first complete consideration of how one can derive the gauge
transformations from the constraint structure of the ADM formulation. The
application of Castellani's method to the ADM Hamiltonian given in Appendix of
\cite{Castellani} is opaque and incomplete as the relation between the
diffeomorphism and the ADM parameters is not explicitly given and only the
transformations of $g_{0\mu}$ are found. The calculations themselves were
performed in an unnatural way - in order to find the transformations of the
metric tensor, the ADM variables were expressed in terms of the metric in the
generator. For completeness, we shall use Castellani's approach to find the
gauge generator with the ADM Hamiltonian and compare it with the result of
\cite{Saha}. In addition, we will show some of the peculiarities in this
calculation which are related to the somewhat confusing notation used by ADM.

According to Castellani's procedure, the generators of a gauge transformation
can be constructed for the Hamiltonian using the so-called algebra of the
Dirac constraints (\ref{eqnDA}) (unnumbered equation preceding Eq. (29) of
\cite{Castellani})\footnote{Starting from here we are considering the ADM
formulation and do not use the superscript $ADM$.}%

\begin{equation}
\left\{  \mathcal{H}_{\bot},H\right\}  =N_{,r}e^{rs}\mathcal{H}_{s}+\left(
Ne^{rs}\mathcal{H}_{s}\right)  _{,r}+\left(  N^{r}\mathcal{H}_{\bot}\right)
_{,r}, \label{eqn415}%
\end{equation}

\begin{equation}
\left\{  \mathcal{H}_{i},H\right\}  =N_{,i}\mathcal{H}_{\bot}+N_{,i}%
^{j}\mathcal{H}_{j}+\left(  N^{j}\mathcal{H}_{i}\right)  _{,j}. \label{eqn416}%
\end{equation}
These lead to the generator (see Eq. (29) of \cite{Castellani})%

\[
G=-\int\left\{  \left[  \varepsilon^{\bot}\left(  \mathcal{H}_{\bot}%
+N_{,i}e^{ij}\Pi_{j}+\left(  N\Pi_{i}e^{ij}\right)  _{,j}+\left(  \Pi
N^{j}\right)  _{,j}\right)  +\varepsilon_{,0}^{\bot}\Pi\right]  \right.
\]

\begin{equation}
+\left.  \left[  \varepsilon^{i}\left(  \mathcal{H}_{i}+N_{,i}^{j}\Pi
_{j}+\left(  N^{j}\Pi_{i}\right)  _{,j}+N_{,i}\Pi\right)  +\varepsilon
_{,0}^{i}\Pi_{i}\right]  \right\}  d^{3}x. \label{eqn417}%
\end{equation}

There are some differences between these equations and the corresponding ones
of \cite{Castellani} as we use Dirac's notation, $e^{rs}$, which in ADM is
called the three-dimensional metric. Actually, this renaming is sloppy and
confusing because (see (\ref{eqn409})) $^{4}g^{km}=\left.  {}\right.
^{3}g^{km}-\frac{N^{k}N^{m}}{N^{2}}=\left.  {}\right.  ^{3}g^{km}+\frac
{g^{0k}g^{0m}}{g^{00}}$ which, when solved for $^{3}g^{km}$, is equivalent to
Dirac's $e^{km}$ (\ref{eqn06.04}). We keep $e^{rs}$ to avoid the temptation to
raise spatial indices with this tensor. We can do this for the spatial metric
(as they are inverses) but we cannot do this for derivatives, as was done in
\cite{Castellani} in the second term of (\ref{eqn417}) where $N^{,j}\Pi_{j}$
is correct \textit{only} if we consider\ $N^{,j}$ as short for $N_{,i}e^{ij}$.
This is because, according to the standard rules of raising indices in GR,%

\begin{equation}
\partial^{j}=g^{j\mu}\partial_{\mu}=g^{ji}\partial_{i}+g^{j0}\partial
_{0}=e^{ji}\partial_{i}+\frac{g^{0i}g^{0j}}{g^{00}}\partial_{i}+g^{j0}%
\partial_{0}, \label{eqn418}%
\end{equation}
and%

\begin{equation}
\partial^{j}=e^{ji}\partial_{i} \label{eqn419}%
\end{equation}
only if $g^{j0}=0$ which is Dirac's simplifying assumption (\ref{eqn010}). If
we use (\ref{eqn419}) instead of (\ref{eqn418}) we would obtain a different result.

The generator (\ref{eqn417}) allows one to find the transformations of the ADM
fields ($N$, $N^{i}$ and $g_{km}$) and then by using the definition of these
variables (\ref{eqn407}), (\ref{eqn408n}), we can \textit{formally} find the
transformations of $g_{\mu\nu}$. We do this in natural order - we first find
the transformations of the ADM variables using the generator in terms of the
ADM variables and then revert to the metric tensor. Transformations of the ADM
fields are calculated using $\delta\left(  field\right)  =\left\{
field,G\right\}  $.

Starting with the simplest variable, $N$, we obtain%

\begin{equation}
\delta_{ADM}N=\left\{  N,G\right\}  =\varepsilon_{,j}^{\bot}N^{j}%
-\varepsilon_{,0}^{\bot}-\varepsilon^{i}N_{,i} \label{eqn420a}%
\end{equation}
which using $N=\left(  -g^{00}\right)  ^{-1/2}$ gives%

\[
\delta_{ADM}N=\frac{1}{2}\left(  -g^{00}\right)  ^{-3/2}\delta_{ADM}g^{00}%
\]
and so we find%

\begin{equation}
\delta_{ADM}g^{00}=2\left(  -g^{00}\right)  ^{+3/2}\delta N=2\left(
-g^{00}\right)  ^{+3/2}\left[  \varepsilon_{,j}^{\bot}\left(  -g^{0j}%
/g^{00}\right)  -\varepsilon_{,0}^{\bot}-\varepsilon^{i}\left(  \left(
-g^{00}\right)  ^{-1/2}\right)  _{,i}\right]  . \label{eqn420}%
\end{equation}
This differs from the diffeomorphism transformation (\ref{eqnD7}):%

\[
\delta_{\left(  diff\right)  }g^{00}=2\xi^{0,0}-\xi^{0}g_{,0}^{00}-\xi
^{k}g_{,k}^{00}.
\]

In ADM variables we cannot restore diffeomorphism invariance; the most that
can be done is to present (\ref{eqn420}) in a \textit{form} similar to a diffeomorphism:%

\[
\delta_{ADM}g^{00}=2\left(  -g^{00}\right)  ^{+1/2}\left[  g^{0j}%
\varepsilon_{,j}^{\bot}+g^{00}\varepsilon_{,0}^{\bot}\right]  -\varepsilon
^{i}g_{,i}^{00}%
\]
and using $\partial^{0}=g^{0\mu}\partial_{\mu}=g^{00}\partial_{0}%
+g^{0k}\partial_{k}$ this becomes%

\[
\delta_{ADM}g^{00}=2\left(  -g^{00}\right)  ^{+1/2}\varepsilon^{\bot
,0}-\varepsilon^{i}g_{,i}^{00}=
\]

\[
2\left[  \varepsilon^{\bot}\left(  -g^{00}\right)  ^{+1/2}\right]
^{,0}-\left[  \varepsilon^{\bot}\left(  -g^{00}\right)  ^{+1/2}\right]
g_{,0}^{00}-\left[  \varepsilon^{k}+\frac{g^{0k}}{g^{00}}\varepsilon^{\bot
}\left(  -g^{00}\right)  ^{+1/2}\right]  g_{,k}^{00}.
\]
The combinations in square brackets \textquotedblleft
correspond\textquotedblright\ to the diffeomorphism parameters%

\begin{equation}
\xi^{0}=\varepsilon^{\bot}\left(  -g^{00}\right)  ^{+1/2}=\frac{1}%
{N}\varepsilon^{\bot}, \label{eqn425}%
\end{equation}

\begin{equation}
\xi^{k}=\varepsilon^{k}+\frac{g^{k}}{g^{00}}\varepsilon^{\bot}\left(
-g^{00}\right)  ^{+1/2}=\varepsilon^{k}-\frac{N^{k}}{N}\varepsilon^{\bot
}\label{eqn426}%
\end{equation}
or%

\begin{equation}
\varepsilon^{\bot}=N\xi^{0}=\left(  -g^{00}\right)  ^{-1/2}\xi^{0},
\label{eqn427}%
\end{equation}

\begin{equation}
\varepsilon^{k}=\xi^{k}+N^{k}\xi^{0}=\xi^{k}-\frac{g^{0k}}{g^{00}}\xi
^{0}.\label{eqn428}%
\end{equation}

Equations (\ref{eqn425})-(\ref{eqn428}) are equivalent to the result of
\cite{Saha} (where the methods of \cite{Novel} were used) and also to what is
presented in \cite{Bergmann, Pons}. The relations (\ref{eqn425}),
(\ref{eqn427}) can be found in the Appendix of Castellani's article
\cite{Castellani}, but (\ref{eqn426}), (\ref{eqn428}) were not given there
explicitly. This field-dependent redefinition of gauge parameters provides a
\textquotedblleft correspondence\textquotedblright\ \cite{Saha}, but not an
equivalence with the diffeomorphism, that follows directly from consideration
of the Dirac Hamiltonian.

For the next variable, $N^{k}$, we obtain%

\begin{equation}
\delta_{ADM}N^{k}=\left\{  N^{k},G\right\}  =-\left[  \varepsilon^{\bot}%
N_{,j}e^{jk}-\varepsilon_{,j}^{\bot}Ne^{k^{j}}+\varepsilon^{j}N_{,j}%
^{k}-\varepsilon_{,j}^{k}N^{j}+\varepsilon_{,0}^{k}\right]  \label{eqn429}%
\end{equation}
and using $N^{k}=-\frac{g^{0k}}{g^{00}}$ we find%

\[
\delta_{ADM}g^{0k}=\frac{g^{0k}}{g^{00}}\delta_{ADM}g^{00}+g^{00}\left[
\varepsilon^{\bot}N_{,j}e^{jk}-\varepsilon_{,j}^{\bot}Ne^{jk}+\varepsilon
^{j}N_{,j}^{k}-\varepsilon_{,j}^{k}N^{j}+\varepsilon_{,0}^{k}\right]  .
\]

After expressing $N$ and $N^{k\text{ }}$in terms of the metric%

\[
\delta_{ADM}g^{0k}=\frac{g^{0k}}{g^{00}}\delta_{ADM}g^{00}%
\]

\[
+g^{00}\left[  \varepsilon^{\bot}\left[  \left(  -g^{00}\right)
^{-1/2}\right]  _{,j}e^{jk}-\varepsilon_{,j}^{\bot}\left(  -g^{00}\right)
^{-1/2}e^{kj}+\varepsilon^{j}\left(  -\frac{g^{0k}}{g^{00}}\right)
_{,j}-\varepsilon_{,j}^{k}\left(  -\frac{g^{0j}}{g^{00}}\right)
+\varepsilon_{,0}^{k}\right]  ,
\]
we again have an invariance that is not a diffeomorphism (\ref{eqnD7})%

\[
\delta_{\left(  diff\right)  }g^{0k}=\xi^{0,k}+\xi^{k,0}-\xi^{0}g_{,0}%
^{0k}-\xi^{m}g_{,m}^{0k}.
\]
If we perform the field-dependent change of parameters (\ref{eqn427}),
(\ref{eqn428}) we again can present $\delta_{ADM}g^{0k}$ in the \textit{form}
of a diffeomorphism transformation.

The transformation of the space-space components $g_{km}$ was not considered
in \cite{Castellani} because, according to the author, it is well-known that
(\ref{eqn417}) generates a diffeomorphism transformation of $g_{km}$. Let us
check this statement;%

\begin{equation}
\delta g_{km}=\left\{  g_{km},G\right\}  =-\frac{\delta}{\delta\Pi^{km}%
}\left[  \varepsilon^{\bot}\mathcal{H}_{\bot}+\varepsilon^{i}\mathcal{H}%
_{i}\right]  \label{eqn431}%
\end{equation}
which, keeping only the $\Pi^{pq}$-dependent part of secondary constraints, gives%

\begin{equation}
-\varepsilon^{\bot}\frac{1}{\sqrt{-g}}\left(  -g^{00}\right)  ^{-1/2}\left(
g_{ip}g_{jq}-\frac{1}{2}g_{ij}g_{pq}\right)  2\Delta_{km}^{ij}\Pi
^{pq}-2\left(  \varepsilon^{i}g_{ip}\right)  _{,q}\Delta_{km}^{pq}%
+\varepsilon^{i}\left(  2g_{pi,q}-g_{pq,i}\right)  \Delta_{km}^{pq}.
\label{eqn431a}%
\end{equation}
We must express $\Pi^{pq}$ in terms of $g_{ij}$ and its derivatives; using Eq.
(7-3.9b) of \cite{ADM}%

\[
\Pi^{ij}=\sqrt{-^{4}g}\left(  ^{4}\Gamma_{pq}^{0}-g_{pq}\left.  {}\right.
^{4}\Gamma_{rs}^{0}g^{rs}\right)  g^{ip}g^{jq}%
\]
which is (taking into account that the \textquotedblleft three-dimensional
quantity\textquotedblright\ $g^{ip}$ in ADM is Dirac's $e^{ip}$)%

\begin{equation}
\Pi^{ij}=\sqrt{-g}\left(  \Gamma_{pq}^{0}e^{ip}e^{jq}-\Gamma_{rs}^{0}%
e^{rs}e^{ij}\right)  =-\sqrt{-g}E^{ijab}\Gamma_{ab}^{0}.\label{eqn430}%
\end{equation}

This expression is equivalent to Dirac's $p^{ij}$ (\ref{eqn06.1}).
Substituting (\ref{eqn430}) into (\ref{eqn431a}) and using (\ref{eqn06.03}),
(\ref{eqn06.2}) and (\ref{eqn06.3}), we obtain%

\[
\delta_{ADM}g_{km}=-\varepsilon^{\bot}\left(  -g^{00}\right)  ^{-1/2}%
2\Gamma_{km}^{0}-2\left(  \varepsilon^{i}g_{ip}\right)  _{,q}\Delta_{km}%
^{pq}+\varepsilon^{i}\left(  2g_{pi,q}-g_{pq,i}\right)  \Delta_{km}^{pq},
\]
or in the explicit form, using (\ref{eqn06.1}), for the ADM formulation%

\[
\delta_{ADM}g_{km}=\varepsilon^{\bot}\frac{1}{N}\left[  N_{k,m}+N_{m,k}%
-g_{km,0}-N^{i}\left(  g_{ki,m}+g_{mi,k}-g_{km,i}\right)  \right]
\]

\begin{equation}
-\varepsilon_{,m}^{i}g_{ik}-\varepsilon_{,k}^{i}g_{im}-\varepsilon^{i}g_{km,i}
\label{eqn440a}%
\end{equation}
and for Dirac's variables%

\[
\delta_{ADM}g_{km}=-\varepsilon^{\bot}\left(  -g^{00}\right)  ^{-1/2}\left[
g^{00}\left(  g_{k0,m}+g_{m0,k}-g_{km,0}\right)  +g^{0i}\left(  g_{ki,m}%
+g_{mi,k}-g_{km,i}\right)  \right]
\]

\begin{equation}
-\varepsilon_{,m}^{i}g_{ik}-\varepsilon_{,k}^{i}g_{im}-\varepsilon^{i}%
g_{km,i}. \label{eqn440}%
\end{equation}
This is again different from the transformation of the spatial components of
the metric tensor under a diffeomorphism (\ref{eqnD6})%

\begin{equation}
\delta_{\left(  diff\right)  }g_{km}=-g_{km,0}\xi^{0}-g_{k0}\xi_{,m}%
^{0}-g_{m0}\xi_{,k}^{0}-g_{km,i}\xi^{i}-g_{ki}\xi_{,m}^{i}-g_{mi}\xi_{,k}^{i}.
\label{eqn441}%
\end{equation}

Again, only after the field-dependent redefinition of parameters
(\ref{eqn427}), (\ref{eqn428}) we can obtain a \textquotedblleft
correspondence\textquotedblright\ between $\delta_{ADM}g_{km}$ and
$\delta_{\left(  diff\right)  }g_{km}$. (Note that both parameters have to be
redefined despite the apparent equivalence of the last three terms in both
equations (\ref{eqn440}) and (\ref{eqn441}).) So, as in the case of Dirac's
Hamiltonian, both methods \cite{Castellani, Novel} produce the same result
(\ref{eqn00a}) for the ADM Hamiltonian.

We would like to note that even a spatial diffeomorphism does not follow
directly from the ADM formulation despite what is often stated in the
literature (e.g. \cite{SalisburyPS, Pulin}). If the lapse and shift functions
are treated as `multipliers'\footnote{The authors of \cite{Gravitation} in the
\textquotedblleft Historical remark" on p. 486 stated that "The great payoff
of this work [ADM] was recognition of the lapse and shift functions of
equation (21.40) [the same as in (21.42) or our (\ref{eqn409})] as Lagrange
multipliers, the coefficients of which gave directly and simply Dirac's
constraints.\textquotedblright\ As we have shown, \textquotedblleft the great
payoff\textquotedblright\ of this recognition is that the ADM formulation lost
the connection with GR and the gauge transformations derived from it are
different from diffeomorphism. We would also like to mention that in the same
\textquotedblleft Historical remark\textquotedblright\ the authors wrote:
\textquotedblleft Dirac paid no particular attention to any variational
principle\textquotedblright. The interested reader is encouraged to look at
Dirac's papers, especially at \cite{Dirac}, to recognize that this is not
correct.} and the secondary constraints are considered as being primary (the
contradictions that result from such manipulations have already been
discussed), this would not correct this problem because a spatial
diffeomorphism does not follow. In this case, according to Castellani's
procedure, the generator is simply
\begin{equation}
G=\varepsilon^{\perp}H_{\perp}+\varepsilon^{i}H_{i} \label{eqnG}%
\end{equation}
which is equivalent to (\ref{eqn431}). Even in such a \textquotedblleft
formulation\textquotedblright\ the gauge parameters have to be redefined and,
in addition, the transformations of lapse and shift functions equal zero (see,
e.g. (\ref{eqnN})).

There is only one way to \textquotedblleft derive\textquotedblright\ spatial
diffeomorphism invariance and it explains the origin of the term
\textquotedblleft diffeomorphism constraint\textquotedblright. It behooves us
to warn the reader that such a \textquotedblleft derivation\textquotedblright%
\ has nothing to do with any procedure. If, in addition to eliminating primary
constraints and promoting secondary to being primary (which leads to the
generator (\ref{eqnG})), we also consider only the second term of this
generator, then%

\begin{equation}
\delta g_{km}=\left\{  g_{km},\varepsilon^{i}H_{i}\right\}  \label{eqnGa}%
\end{equation}
will give the spatial diffeomorphism (see the second line of (\ref{eqn440})).
The only possible explanation of why such manipulations were accepted is that
it seems to follow the \textquotedblleft guidance\textquotedblright\ which
comes from linearized gravity. From the derivation of the gauge
transformations for linearized gravity in \cite{GKK}, it is clear that only
one part of the generator which is proportional to $\chi^{0n}$ contributes to
the transformation of the space-space components of the metric tensor, but
this is not the case for full GR.

Ironically, the result (\ref{eqnGa}), which is just the consequence of a
series of manipulations that contradict any consistent procedure, is often
presented as being the \textquotedblleft problem\textquotedblright\ of the
Hamiltonian formulation of GR: \textquotedblleft Hamiltonian\textquotedblright%
\ and \textquotedblleft diffeomorphism\textquotedblright\ constraints are
treated in a different manner \cite{Nicolai, Pulin}. Furthermore, the
conclusion is drawn, based on (\ref{eqnGa}), that \textquotedblleft the
diffeomorphism constraint can be shown to be associated with the invariance of
general relativity under spatial diffeomorphism\textquotedblright%
\ \cite{Pulin} (see also \cite{deWitt}). Finally, because (\ref{eqnGa}) leads
to a spatial diffeomorphism, this result is interpreted as \textquotedblleft
disappearance of Diff$\mathcal{M}$\textquotedblright\ and considered as
\textquotedblleft the problem that has worried many people working in
geometrodynamics for so long\textquotedblright\ \cite{Isham-diff-II}.

Statements similar to some of the quotations above can be found in many
articles. They are based on questionable manipulations; but, at the same time,
they clearly demonstrate that some authors correctly consider that the
restoration of diffeomorphism invariance for all components of the metric
tensor is to be expected (however, \textquotedblleft this expectation has
never been fully realized...\textquotedblright\ \cite{Isham-diff-II}%
)\footnote{This expectation is fully realized in \cite{KKRV} and in the
previous Section.} and its absence is taken to be a deficiency or a
contradiction arising in the Hamiltonian formulation. To have the possibility
of restoring transformations (whatever it may be) of all variables one must
work in the full phase space and use all of the first-class constraints. We
disregard \textquotedblleft approaches\textquotedblright\ leading to
(\ref{eqnG}) and (\ref{eqnGa}) and return to the analysis of the total
Hamiltonians (\ref{eqn400}) and (\ref{eqn401}).

\subsection{ADM change of variables and canonicity}

We now ask why, in the ADM case, we must redefine the gauge parameters
(\ref{eqn425})-(\ref{eqn428}) to have a \textquotedblleft
correspondence\textquotedblright\ with diffeomorphism, whereas from the Dirac
Hamiltonian (and also from the formulation without Dirac's modifications
\cite{KKRV}) diffeomorphism arises directly. Dirac's Hamiltonian was obtained
from the Lagrangian of GR, after some integrations that do not affect the
equations of motion; and the ADM Hamiltonian apparently follows from the same
Lagrangian. The two Hamiltonian formulations of the same Lagrangian ought to
be equivalent and should give the same gauge transformation, which does not
happen in the case of the Dirac and ADM approaches.

It is well-known that different sets of phase-space variables can be used to
describe Hamiltonian systems. In the ordinary Classical Mechanics of
non-singular systems (e.g., \cite{Lanczos, Whittaker}), for a given
Hamiltonian $H\left(  q_{i},p_{i}\right)  $ and the Hamilton equations,%

\begin{equation}
q_{i}=\left\{  q_{i},H\right\}  =\frac{\partial H}{\partial p_{i}},\left.
{}\right.  p_{i}=\left\{  p_{i},H\right\}  =-\frac{\partial H}{\partial q_{i}%
}, \label{eqnA}%
\end{equation}
we can pass to another set of phase-space variables $\left(  Q_{i}%
,P_{i}\right)  $:%

\begin{equation}
q_{i}=q_{i}\left(  Q_{k},P_{k}\right)  ,\left.  {}\right.  p_{i}=p_{i}\left(
Q_{k},P_{k}\right)  ,\left.  {}\right.  K\left(  Q_{k},P_{k}\right)  =H\left(
q_{i}\left(  Q_{k},P_{k}\right)  ,~p_{i}\left(  Q_{k},P_{k}\right)  \right)  ,
\label{eqnB}%
\end{equation}
such that%

\begin{equation}
Q_{i}=\left\{  Q_{i},K\right\}  =\frac{\partial K}{\partial P_{i}},\left.
{}\right.  P_{i}=\left\{  P_{i},K\right\}  =-\frac{\partial K}{\partial Q_{i}%
}. \label{eqnC}%
\end{equation}

If this system of Hamilton equations can be solved, then one can return to the
original variables by using the inverse of (\ref{eqnB})%

\begin{equation}
Q_{i}=Q_{i}\left(  q_{k},p_{k}\right)  ,\left.  {}\right.  P_{i}=P_{i}\left(
q_{k},p_{k}\right)  . \label{eqnD}%
\end{equation}

In Classical Mechanics, even for non-singular systems, it is well-known, that
\cite{Lanczos} \textquotedblleft...at first sight we might think that
arbitrary point transformations of the phase space are now at our disposal.
This would mean that the 2$n$ coordinates $q_{i}$ and $p_{i}$ can be
transformed into some new $Q_{i\text{ }}$ and $P_{i}$ by any functional
relations we please. This, however, is not the case.\textquotedblright\ For
the non-singular systems, the necessary and sufficient condition that the
transformation (\ref{eqnB}) to the new set of variables $\left(  Q_{i}%
,P_{i}\right)  $ is a canonical transformation (and keeps the two formulation
equivalent), is%

\[
\left\{  Q_{i},Q_{k}\right\}  _{Q,P}=\left\{  Q_{i}\left(  p,q\right)
,Q_{k}\left(  p,q\right)  \right\}  _{p,q}=0,
\]

\begin{equation}
\left\{  P_{i},P_{k}\right\}  _{Q,P}=\left\{  P_{i}\left(  p,q\right)
,P_{k}\left(  p,q\right)  \right\}  _{p,q}=0, \label{eqnE}%
\end{equation}

\[
\left\{  Q_{i},P_{k}\right\}  _{Q,P}=\left\{  Q_{i}\left(  p,q\right)
,P_{k}\left(  p,q\right)  \right\}  _{p,q}=\delta_{ik}.
\]

The change of phase-space variables (canonical transformations) for
unconstrained Hamiltonians is an old and well established topic that can be
found in many textbooks on Classical Mechanics. For constrained Hamiltonians
the situation is different and even the number of papers (e.g. see
\cite{Gomis, Gomis2}) that discuss the general questions or particular
examples (which are mainly final dimensional and artificial) of such changes
is minuscule compared to the number of articles in which such changes are used
without any analysis of their consequences. Such changes are especially common
in Hamiltonian formulations of GR (in both from Einstein to ADM \cite{ADM} and
Einstein-Cartan to ADM-inspired formulations \cite{DeserIsham, Stefano}, with
a rare exception like \cite{CastNieu}). For the Hamiltonian and Lagrangian
treatments of the Einstein-Cartan action without any change of variables see
\cite{3D, Report, trans}.

The Dirac's generalization of the Hamiltonian formulation to constrained
systems leads to the system of equations which is similar to (\ref{eqnA}):%

\[
q_{i}=\left\{  q_{i},H_{T}\right\}  ,\left.  {}\right.  p_{i}=\left\{
p_{i},H_{T}\right\}  ,
\]
where $H_{T}$ also includes the primary constraints. We restrict our
discussion to gauge invariant systems with only first-class constraints. For
such systems, the conditions that should be imposed on possible changes of
variables in phase space seems to be more restrictive. The reason for this is
that gauge invariance depends on all of the first-class constraints and their
PB algebra \cite{Castellani, HTZ, Novel}. In addition, it is related to the
total Hamiltonian of a particular model, in contrast to unconstrained systems
where the conditions (\ref{eqnE}) are, in fact, independent of the
Hamiltonian. For gauge invariant systems the change of variables (\ref{eqnD})
must preserve gauge invariance, i.e. gauge invariance derived in the $\left(
Q_{i},P_{i}\right)  $ variables, after using the inverse transformations, must
produce the same result as with the original $\left(  q_{i},p_{i}\right)  $
variables. The set of phase-space transformations (\ref{eqnB}) that preserves
this property is the \textit{equivalent set }or, as in non-singular case, we
can call such changes canonical transformations. In \cite{KKaffinemetric} the
special role of primary constraints is emphasized and discussed in detail in
both derivation of a gauge generator and in canonical transformations as, for
example, in first-order formulation of GR the primary constraints are momenta
which are phase-space variables and unjustified manipulations with them are
not allowed.

In linearized versions of the Dirac formulation (see the previous Sections)
and that of \cite{Pirani} which were considered in \cite{GKK}, both
formulations, despite having different constraints and Hamiltonians, lead to
the same gauge invariance. The relation between the two formulations was
discussed and it was shown that they are related by canonical transformations
(\ref{eqnE}) \cite{KKX}, which is exactly the condition known to be needed for
non-singular Hamiltonians \cite{Lanczos, Whittaker}. Despite there being far
more complicated expressions for constraints and transformations, a similar
relation (\ref{eqnE}) exists between the non-linearized Hamiltonian
formulation of Dirac considered in the present paper and \cite{KKRV}; and
these, as we have shown, also have the same gauge invariance. These examples
demonstrate that the ordinary condition for the transformation to be
canonical, which is known for unconstrained Hamiltonians is also valid for the
general (constrained) case; i.e. it is a necessary condition as before. Yet,
it is not sufficient and we shall demonstrate this fact by way of example.

Surprisingly, we have been unable to find in the literature, either a
discussion of the relations between the ADM and Dirac phase-space variables or
the transformations between them, which is strange as many authors presume
their equivalence by calling this formulation \textquotedblleft
Dirac-ADM\textquotedblright. In \cite{ADM-footnote} the authors called the
variables $N$, $N^{i}$ and $g_{km}$ \textquotedblleft an equivalent
set\textquotedblright\ which \textquotedblleft is analytically convenient and
geometrically more significant\textquotedblright. In footnote 5 of
\cite{ADM-footnote} (appeared in 1959) it is stated that \textquotedblleft The
properties of these variables are discussed in detail in a forthcoming paper
by C. Misner\textquotedblright\ that we have not been able to find. The
convenience and significance are not our present concern, but the question of
equivalence of the ADM and Dirac sets of variables is important. We are
interested in the equivalence of gauge transformations in the two approaches,
i.e. we must work in the full phase space. The complete relation between these
two sets of variables is not known, but there is one PB, $\left\{  N\left(
x\right)  ,\Pi^{kl}\left(  x^{\prime}\right)  \right\}  $, that can be easily
checked. The space-space components of the metric tensor and corresponding
momenta are the same in both formulations, i.e. $\Pi^{kl}=\Pi^{kl}\left(
p^{kl}\right)  =p^{kl}$. The $\Pi^{kl}$ of ADM given by (\ref{eqn430}) is
equivalent to $p^{kl}$ of Dirac given by (\ref{eqn06.1}). It is sufficient to
check using the Dirac variables $g_{\mu\nu}$, $p^{\mu\nu}$ the PB $\left\{
N\left(  x\right)  ,\Pi^{kl}\left(  x^{\prime}\right)  \right\}  $ which, if
ADM variables are canonical, must give zero \cite{Lanczos, Whittaker}. Using
the corresponding fundamental PBs (\ref{eqn411})-(\ref{eqn413}) we obtain:%

\begin{equation}
\left\{  N\left(  x\right)  ,\Pi^{kl}\left(  x^{\prime}\right)  \right\}
=\left\{  \left(  -g^{00}\right)  ^{-1/2},p^{kl}\right\}  _{g,p}=-\frac
{\delta}{\delta g_{kl}}\left(  -g^{00}\right)  ^{-1/2}=\frac{1}{2}\left(
-g^{00}\right)  ^{-3/2}g^{0k}g^{0l}\neq0. \label{eqn445}%
\end{equation}

Once again, we have the result that depends on Dirac's simplifying assumption
(\ref{eqn010}): if we impose $g^{0k}=0$, then the PB of (\ref{eqn445}) gives
zero. In general this PB is not zero and the transformation from $\left(
g_{\mu\nu},p^{\mu\nu}\right)  $ to $\left(  N,\Pi\right)  $, $\left(
N^{i},\Pi_{i}\right)  $, and $\left(  g_{km},\Pi^{km}\right)  $ therefore is
not canonical. One PB is enough to show that the transformation from
$g_{\mu\nu}$ to ADM variables is not canonical, irrespective of the results we
might obtain for the PBs among the other phase-space variables.

We note that equation (\ref{eqn445}) gives zero in the static coordinate
system, but in this case the corresponding components, $g_{0k}$, and their
conjugate momenta have to be dropped out of the formalism from the beginning,
as in the case of the Hamiltonian formulation in the Schwarzschild metric
\cite{Chand} where only four components of the metric tensor are left in the
Lagrangian before passing to the Hamiltonian.

This simple calculation, (\ref{eqn445}), allows us to conclude that
\textit{the two Hamiltonians of Dirac and ADM are not related by a canonical
transformation and the respective failure of the ADM variables and the
Hamiltonian to produce a diffeomorphism transformation is a manifestation of
this non-equivalence}. Moreover, (\ref{eqn445}) shows that the ADM variables
are not the canonical variables of GR. The converse statement is also true and
the metric tensor is not a canonical variable of the ADM formulation. The ADM
Hamiltonian formulation might be considered as a model (geometrodynamics or
ADM gravity) without any reference to Einstein GR, but in this case a
\textquotedblleft correspondence\textquotedblright\ between the two
transformations (\ref{eqn425})-(\ref{eqn428}) is, in fact, meaningless. The
transformations that follow from ADM are given by (\ref{eqn420a}),
(\ref{eqn429}), and (\ref{eqn440a}) and in the absence of canonicity we cannot
return to the transformations of the metric tensor. There is another
characteristic that supports the loss of connection with the original
variables: it is impossible to find any redefinition of $\Pi$ and $\Pi_{k}$ in
terms of Dirac's phase space variables (whether they satisfy (\ref{eqnE}) or
not) that can transform his total Hamiltonian (\ref{eqn401}) into the ADM
total Hamiltonian (\ref{eqn400}).

In general, no algorithm exists for finding a canonical transformation but the
canonicity of a given transformation can be checked. There are no canonical
transformations for Hamiltonians that involve only a change of generalized
coordinates, as this change must be accompanied by transformations of the
momenta that can be found by using the following procedure \cite{Lanczos}. If
the transformations of the generalized coordinates (fields) are given, one can
find the corresponding transformations of the momenta that will guarantee that
the new coordinates and momenta are canonical and satisfy (\ref{eqnE}) using
the relation%

\begin{equation}
p_{i}\delta q_{i}=P_{i}\delta Q_{i}. \label{canon}%
\end{equation}

Let us see what we can obtain\ from this relation for the ADM\ change of variables%

\[
p^{\alpha\beta}\delta g_{\alpha\beta}=\Pi\delta N+\Pi_{k}\delta N^{k}+\Pi
^{km}\delta g_{km}.
\]
By performing the variations $\delta Q=\frac{\delta Q}{\delta g_{\alpha\beta}%
}\delta g_{\alpha\beta}$ using (\ref{eqn407})-(\ref{eqn408n}), we find that%

\[
p^{\alpha\beta}=\Pi\frac{\delta N}{\delta g_{\alpha\beta}}+\Pi_{k}\frac{\delta
N^{k}}{\delta g_{\alpha\beta}}+\Pi^{km}\frac{\delta g_{km}}{\delta
g_{\alpha\beta}}%
\]
which gives%

\begin{equation}
p^{00}=-\Pi\frac{1}{2}\left(  -g^{00}\right)  ^{1/2}, \label{eqn450}%
\end{equation}

\begin{equation}
p^{0m}=\Pi\frac{1}{2}\left(  -g^{00}\right)  ^{-1/2}g^{0m}+\Pi_{k}\frac{1}%
{2}e^{km}, \label{eqn451}%
\end{equation}
and%

\begin{equation}
p^{pq}=-\Pi\frac{1}{2}\left(  -g^{00}\right)  ^{-3/2}g^{0p}g^{0q}+\Pi_{k}%
\frac{1}{2}\left(  \frac{g^{0p}}{g^{00}}e^{kq}+\frac{g^{0q}}{g^{00}}%
e^{kp}\right)  +\Pi^{pq}. \label{eqn452}%
\end{equation}
Now solving for the $\Pi$'s:%

\begin{equation}
\Pi=-2\left(  -g^{00}\right)  ^{-1/2}p^{00}, \label{eqn453}%
\end{equation}

\begin{equation}
\Pi_{n}=2g_{mn}p^{0m}+2g_{0n}p^{00}, \label{eqn454}%
\end{equation}

\begin{equation}
\Pi^{pq}=p^{pq}+\frac{g^{0q}}{g^{00}}\frac{g^{0p}}{g^{00}}p^{00}-\frac{g^{0p}%
}{g^{00}}p^{0q}-\frac{g^{0q}}{g^{00}}p^{0p}. \label{eqn455}%
\end{equation}
Note that, to have $\Pi^{pq}=p^{pq}$, as in the case of the Dirac and ADM
Hamiltonians, we again must impose the condition $g^{0k}=0.$

Only if equations (\ref{eqn453})-(\ref{eqn455}) are taken together with the
relations for the generalized coordinates (\ref{eqn407})-(\ref{eqn408n}), are
the transformations canonical. It is not difficult to check that the canonical
properties of PBs are preserved, and as an example, the PB that we considered
in (\ref{eqn445}) gives%

\[
\left\{  N\left(  x\right)  ,\Pi^{pq}\left(  x^{\prime}\right)  \right\}
=\left\{  \left(  -g^{00}\right)  ^{-1/2},p^{pq}+\frac{g^{0q}}{g^{00}}%
\frac{g^{0p}}{g^{00}}p^{00}-\frac{g^{0p}}{g^{00}}p^{0q}-\frac{g^{0q}}{g^{00}%
}p^{0p}\right\}  _{g,p}=0,
\]
as it should for variables that are connected by canonical transformations.
For \textit{non-singular} Lagrangians and their corresponding Hamiltonians,
the ADM change of variables accompanied by the change of momenta
(\ref{eqn450})-(\ref{eqn452}) would be sufficient to obtain the new set of
canonical variables. If GR were a non-singular theory, these transformations
would guarantee equivalence between the two formulations; but for the
constrained Hamiltonian this is not the case. If the canonical transformation
of (\ref{eqn409}), (\ref{eqn450})-(\ref{eqn452}) are performed in the Dirac
Hamiltonian, we will not obtain the ADM Hamiltonian, and we will not obtain a
consistent result. In particular, for a description of a constrained system,
the total Hamiltonian, $H_{T}$, is important as it includes all the primary
constraints. In the Dirac formulation, these are (\ref{eqn06.5}):%

\begin{equation}
g_{00,0}p^{00}+2g_{0k,0}p^{0k}. \label{eqn06.5a}%
\end{equation}
\qquad\ Substitution of (\ref{eqn409}), (\ref{eqn450})-(\ref{eqn452}) into
this equation gives%

\[
N_{,0}\Pi+N_{,0}^{m}\Pi_{m}+g_{kj,0}\left(  \frac{1}{2}\Pi\frac{N^{k}N^{j}}%
{N}+N^{j}\Pi_{m}e^{mk}\right)
\]
which is nonsensical. The first two terms are equivalent to (\ref{eqn06.5a})
but an extra term appears, which is zero only if $N^{k}=0$. In the course of
the Hamiltonian procedure the space-space velocities have already been
eliminated in favour of their corresponding momenta but now they reappear and
it is not clear what to do with them at this stage. If we treat them on the
same footing as the rest of the terms with time derivatives, we must specify
their coefficients as primary constraints, which would then give a total of
ten primary constraints. The same change of variables in the canonical part of
the Hamiltonian will produce contributions that are quadratic in all momenta.
Without further analysis we see that this will probably lead to some
contradictions, as it is clear that the constraint structure of the
Hamiltonian is changed and one would expect second-class constraints, etc.

The foregoing example, (\ref{eqn407})-(\ref{eqn408n}) and (\ref{eqn453}%
)-(\ref{eqn455}), clearly demonstrates that the condition (\ref{eqnE}), which
is necessary and sufficient for the transformation to be canonical in the case
of non-singular Lagrangians, is not sufficient for singular Lagrangians.

Actually, for the ADM change of variables, the non-canonical nature of the
transformations is immediately clear even from (\ref{eqn445}) and this is not
related to a singular structure of the Lagrangian of GR. The possible
existence of additional restrictions beyond (\ref{eqnE}) is understandable,
and for the Hamiltonian with first-class constraints it can be expected. Gauge
transformations are derived from the first-class constraints and the whole PB
algebra of constraints plays a key role in this derivation. It is not enough
to have the same number of constraints in the two formulations. The ADM change
of variables keeps the same number of constraints as the Dirac formulation;
but the PB algebra of constraints is affected. The simplest example is a PB
among primary and secondary constraints, which are zero in the ADM case and
proportional to the constraint $\chi^{00}$ in the case of the Dirac
Hamiltonian (\ref{eqn90}). From a mathematical point of view the ADM
formulation is just the result of a non-canonical change of variables in
Dirac's Hamiltonian of GR and if Dirac's formulation allows one to derive the
diffeomorphism transformations, then the ADM formulation, because of this
non-canonical change of variables, does not allow one to restore either the
full diffeomorphism invariance for all components of $g_{\mu\nu}$ or even for
its spatial part (as it is usually claimed) without a non-covariant and
field-dependent redefinition of gauge parameters. In addition, we observed
that some equations in the ADM formulation are true only if $g_{0k}=0$. So the
ADM change of variables is somehow also related to a static coordinate system,
but in a mysterious way: $g_{0k}$ is not zero at the outset, but later should
be set equal to zero so that some subsequent relations are made valid.

We conclude that the Dirac Hamiltonian of GR was obtained by following the
\textquotedblleft rule of procedure\textquotedblright\ and, because of this,
it is automatically canonical and the use of the adjective \textquotedblleft
canonical\textquotedblright\ is a tautology. It is not equivalent to the ADM
Hamiltonian, which, as we have demonstrated, is the result of a non-canonical
change of variables. The ADM formulation is obtained by abandoning the
\textquotedblleft rule of procedure\textquotedblright\ and, consequently, only
by a canonization can it be ironically called the \textquotedblleft canonical
formulation of GR\textquotedblright.

\subsection{Dirac and ADM Lagrangians and equations of motion}

One can argue that the ADM Hamiltonian is not obtained from the Dirac
Hamiltonian. There are no references in \cite{ADM} to Dirac's article
\cite{Dirac}, the only reference where a derivation of Dirac's Hamiltonian has
been considered. In their culminating paper \cite{ADM} and in 13 preceding
articles it is mentioned only once in \cite{ADM-7} and in a different context.
However, according to both Dirac and ADM, their respective Hamiltonians are
derived from the same theory - GR. In both cases some modifications of the EH
action were performed. In Dirac's case these modifications are explicitly
stated, in the ADM case it is more difficult to trace what has been done.
There is a statement in \cite{ADM} that ADM formulation is started from
\textquotedblleft the Palatini Lagrangian\textquotedblright\ (see footnote 19
about \textquotedblleft Palatini\textquotedblright\ formulation). But in the
first-order, affine-metric, formulation of GR tertiary constraints appear and
diffeomorphism invariance follows from all the first-class constraints
\cite{Gerry-Ramin-2, Gerry-new, KKaffinemetric}; such constraints, as well as
diffeomorphism invariance, are absent in the ADM Hamiltonian formulation.

Let us try to obtain the ADM Lagrangian from the Dirac Lagrangian. We have
discussed above how, in the course of a Hamiltonian analysis that follows
Dirac, the possibility of additional integrations by parts appears when the
Hamiltonian is to be expressed as a linear combination of secondary
constraints (see our discussion after (\ref{eqn80})). The modified Lagrangian
can be written in the following form (up to total derivatives as in
(\ref{eqn74}))%

\[
L_{Dirac}=\frac{1}{4}\sqrt{\det g_{km}}\left(  \mathbf{-}g^{00}\right)
^{1/2}E^{rsab}\left[  \left(  -g_{ap}\frac{g^{0p}}{g^{00}}\right)
_{,b}+\left(  -g_{bp}\frac{g^{0p}}{g^{00}}\right)  _{,a}-g_{ab,0}+\frac
{g^{0k}}{g^{00}}\left(  g_{ak,b}+g_{bk,a}-g_{ab,k}\right)  \right]
\]

\[
\times\left[  \left(  -g_{rq}\frac{g^{0q}}{g^{00}}\right)  _{,s}+\left(
-g_{sq}\frac{g^{0q}}{g^{00}}\right)  _{,r}-g_{rs,0}+\frac{g^{0m}}{g^{00}%
}\left(  g_{rm,s}+g_{sm,r}-g_{rs,m}\right)  \right]
\]

\begin{equation}
+\sqrt{\det g_{km}}\left(  -g^{00}\right)  ^{-1/2}\left[  g_{mn,kt}%
E^{mnkt}+\frac{1}{4}g_{mn,k}g_{pq,t}\left(  E^{mnpq}e^{kt}-2E^{ktnp}%
e^{mq}-4E^{pqnt}e^{mk}\right)  \right]  .\label{eqn460}%
\end{equation}
Such a Lagrangian can be also constructed if, in addition to Dirac's
modification (\ref{eqn04}), the integrations of (\ref{eqn74}) were performed
at the Lagrangian level \cite{Gitmanbook}. This Lagrangian can be easily
obtained from Dirac's Hamiltonian by performing the inverse Legendre
transformation and eliminating momenta in terms of `velocities'.

By performing the change of variables of (\ref{eqn409}) in (\ref{eqn460}), we
obtain the ADM Lagrangian%

\[
L_{ADM}=\frac{1}{4}\sqrt{\det g_{km}}\frac{1}{N}E^{rsab}\left[  \left(
g_{ap}N^{p}\right)  _{,b}+\left(  g_{bp}N^{p}\right)  _{,a}-g_{ab,0}%
-N^{k}\left(  g_{ak,b}+g_{bk,a}-g_{ab,k}\right)  \right]
\]

\[
\times\left[  \left(  g_{rq}N^{q}\right)  _{,s}+\left(  g_{sq}N^{q}\right)
_{,r}-g_{rs,0}-N^{m}\left(  g_{rm,s}+g_{sm,r}-g_{rs,m}\right)  \right]
\]

\begin{equation}
+\sqrt{\det g_{km}}N\left[  g_{mn,kt}E^{mnkt}+\frac{1}{4}g_{mn,k}%
g_{pq,t}\left(  E^{mnpq}e^{kt}-2E^{ktnp}e^{mq}-4E^{pqnt}e^{mk}\right)
\right]  .\label{eqn461}%
\end{equation}
Or, by using the intrinsic and extrinsic curvatures, $^{3}R_{rs}$ and $K_{rs}%
$, respectively, it can be written in more familiar form \cite{Waldbook,
deWitt}%

\begin{equation}
L_{ADM}=\sqrt{\det g_{km}}N\left(  E^{rsab}K_{rs}K_{ab}+^{3}R\right)
\label{eqn462}%
\end{equation}
with%

\[
K_{rs}=\frac{1}{2N}\left(  N_{r\mid s}+N_{s\mid r}-g_{rs,0}\right)
\]
where, as before, \textquotedblleft$\mid$\textquotedblright\ means covariant
derivative with respect to three dimensional metric ($N_{r\mid s}%
=N_{r,s}-\Gamma_{rs}^{k}N_{k}$).

Now using (\ref{eqn461}) or (\ref{eqn462}) we can easily obtain the ADM
Hamiltonian, which is the same as (\ref{eqn400}). However, following such a
detour we cannot avoid the question of whether the ADM variables are
canonical. The transformations of the metric tensor derived from the
Hamiltonian formulation of the Dirac Lagrangian (\ref{eqn460}) and from the
Hamiltonian formulation of the ADM Lagrangian (\ref{eqn461}) are different, so
the destination is changed. This detour is a wrong turn or perhaps a dead end road.

We combine the results of these two formulations into a compact visual
form\footnote{This `pictorial visualization' is based on results of
calculations, not the other way around.}:%

\begin{equation}%
\begin{array}
[c]{ccc}%
L_{Dirac}\left(  q\right)  & \overset{q=q\left(  Q\right)  }{\Longrightarrow}
& L_{ADM}\left(  Q\right) \\
\Downarrow &  & \Downarrow\\
H_{Dirac}\left(  p,q\right)  & \neq & H_{ADM}\left(  P,Q\right) \\
\Downarrow &  & \Downarrow\\%
\begin{array}
[c]{c}%
\\
\\
\delta_{diff}q
\end{array}
&
\begin{array}
[c]{c}%
\\
\\
\neq
\end{array}
&
\begin{array}
[c]{c}%
\delta_{ADM}Q\\
\Downarrow Q\left(  q\right) \\
\delta_{ADM}q
\end{array}
\end{array}
\label{eqn480}%
\end{equation}
\bigskip

It is reasonable to expect that if by a change of variables
$\overset{q=q\left(  Q\right)  }{\Longrightarrow}$ we can obtain a new
equivalent Lagrangian and find the corresponding Hamiltonian, then the two
Hamiltonians (in both the new and the original variables) should also be
equivalent, be related to each other by a canonical transformation, and
necessarily lead to the same gauge invariance. If the Hamiltonians are not
related by a canonical transformation, then the two Lagrangians are not
equivalent. This is the natural conclusion that one can make. In particular,
application of the Lagrangian methods \cite{Gitmanbook} used by Samanta
\cite{Samanta} to derive the diffeomorphism invariance of GR, when applied to
the ADM Lagrangian, does not give the diffeomorphism transformations
\cite{Banerjee-JHEP} but rather the same transformations as obtained in its
Hamiltonian treatment in \cite{Saha}. This is a consequence of the well-known
equivalence of the Lagrangian and Hamiltonian formulations for any system,
either non-singular or singular \cite{Gitmanbook, Pons4}: the
\textquotedblleft vertical\textquotedblright\ equivalence of (\ref{eqn480}).
If a \textquotedblleft horizontal\textquotedblright\ equivalence is broken,
either for the two Hamiltonians or for the gauge transformations, then it is
broken everywhere, including at the Lagrangian level.

The Hamiltonian formulations of the linearized versions of the Dirac and
gamma-gamma (PSS) Lagrangians lead to the same algebra of constraints and
gauge transformations (although the Hamiltonians themselves and the
constraints are different). But the two Hamiltonians are related by a
canonical transformation \cite{GKK}. Such transformations also exist in the
case of the corresponding full Dirac and the gamma-gamma formulations of GR
\cite{KKX}. The equivalence of the PB algebra of constraints can be easily
seen by comparing \cite{KKRV} to the results of Section II. The explicit
canonical transformation and its effect on constraints, their algebra and
structure functions is given in \cite{KKX}.

The main subject of the present paper is the Hamiltonian formulation of GR in
second-order form and, in particular, a comparison of the formulations related
by a change of variables. However, we think that some comments on similar
changes made to its Lagrangian should be given.

It is often stated that in a Lagrangian formulation of a model any field
redefinition is legitimate provided it is invertible (i.e. it has a non-zero
Jacobian). For singular Lagrangians, and especially gauge invariant ones, this
is obviously not a sufficient condition. One additional restriction in gauge
invariant cases is the preservation of the rank of Hessian as this gives us
the number of gauge parameters (if all constraints are first-class). We cannot
have two equivalent formulations if they have a different number of gauge
parameters, and we cannot, by a change of variables, eliminate some gauge
invariance or create a new gauge invariance. If we were to make such a change,
we can of course, treat the new Lagrangian as some different model, but we
cannot relate it to the original one as any connection with the original
theory is lost. Obviously these two conditions, non-zero Jacobian and the
preservation of the rank of Hessian, are necessary, but not sufficient. The
ADM change of variables satisfies them both, but leads to different gauge
transformations (compare \cite{KKRV} and Section III versus \cite{Saha} and
(\ref{eqn425})-(\ref{eqn428})). A change of variables in singular (in
particular, gauge invariant) Lagrangians is a much more restrictive procedure
if one intends to preserve its equivalence with the initial formulation. One
way is to rely on the Hamiltonian method and check if the new variables are
canonical and the two total Hamiltonians are equivalent, including the primary
constraints. We must also check whether the entire algebra of constraints is
equivalent, as this algebra is responsible for the gauge transformations. This
can be considered as a confirmation of Dirac's statement \cite{Diracbook}
\textquotedblleft I feel that there will always be something missing from them
[non-Hamiltonian methods] which we can only get by working from a Hamiltonian,
or maybe from some generalization of the concept of a
Hamiltonian\textquotedblright.\ However, we think that some criteria for the
equivalence between two sets of variables for singular Lagrangians can be
formulated at the pure Lagrangian level. At the Lagrangian level, a gauge
invariance is related to the existence of gauge identities \cite{Noether,
Gitmanbook, Noether-eng} and an inappropriate change of fields can modify or
even destroy them. This echoes the conclusion of Isham and Kuchar
\cite{Isham-diff-II} that \textquotedblleft... space-time diffeomorphism has
somehow got lost in making the transition from the Hilbert action to the
Dirac-ADM action\textquotedblright\footnote{We have demonstrated the
inequivalence of the Dirac and ADM formulations. In \cite{Isham-diff-II} the
authors discussed geometrodynamics (not the Dirac formulation) which uses ADM
variables and their statements should be applied to the ADM action only.}. The
Dirac Lagrangian gives the same equations of motion as the EH Lagrangian and
consequently the same differential identity (\ref{diff4}) and transformation
of the metric tensor (\ref{diff7}) will be obtained. It is interesting to
investigate what differential identities will follow from the ADM Lagrangian,
but for non-covariant variables such calculation are extremely cumbersome. The
alternative path is to find what differential identities are needed to
describe diffeomorphism invariance of the ADM variables and after that to
check whether these identities are satisfied if the Euler derivatives of the
ADM Lagrangian are substituted (this work is in progress and will be reported
elsewhere). Regge and Teitelboim in \cite{Regge-Teit} stated:
\textquotedblleft...the usual Hamiltonian $H_{0}$ [the ADM Hamiltonian] not
only does not give the correct equations of motion (Einstein's equations) but,
worse than that, it ($H_{0}$) gives no well-defined set of equations of motion
at all.\textquotedblright\ One confirmation of their words comes from
Numerical Relativity.

\subsection{ADM formulation and Numerical Relativity}

There is another indication of the incorrectness of the ADM change of
variables at the Lagrangian level that comes from Numerical Relativity. In
almost all methods of numerical integration of the Einstein equations the
starting point is the ADM 3+1 decomposition. So at the outset the Einstein
equations are replaced by the ADM equations \cite{ADM}. It was shown that, in
contrast to the Einstein equations which are strongly (strictly) hyperbolic
(SH) \cite{FM}, the ADM equations are weakly hyperbolic (WH) (e.g., see
\cite{NR1, NR2, NR3}). \textit{This change in the type of equations }is
related to the different constraint structure and different transformations
derived from a non-equivalent Hamiltonian. There is a fundamental difference
between SH and WH systems of PDEs: the former are well-posed and convergent,
whereas the latter are not well-posed and are divergent \cite{NR4}. There is
\textquotedblleft evidence\textquotedblright\ that ADM-based algorithms are
numerically unstable. As is indicated in \cite{NR5}: \textquotedblleft The
common lore these days is, however, that the standard Arnowitt-Deser-Misner
(ADM) formulation is the one which most easily suffers
instabilities\textquotedblright. Or in \cite{NR4}: \textquotedblleft It took
several years to realize that such instabilities were not associated with the
numerical algorithms but rather with the mathematical structure of the ADMY
[ADM \cite{ADM} and York \cite{York}] system itself\textquotedblright. This is
not just an additional indication of the incorrectness of the ADM change of
variables, but also a demonstration that even as a model, geometrodynamics is
an \textit{ill defined formulation.}

In general, the change in the type of equations, from SH to WH, or a change of
\textquotedblleft level of hyperbolicity\textquotedblright\ \cite{NR6}, is an
indication that in the process of transforming from the Einstein to the ADM
equations, some \textquotedblleft damage\textquotedblright\ was done (as there
is no longer a complete set of eigenvectors associated with the characteristic
matrix \cite{Bona}). The proposed \textquotedblleft cure\textquotedblright%
\ \cite{NR5} of such a \textquotedblleft damage\textquotedblright\ in most
approaches lies in a modification (or \textquotedblleft
adjusting\textquotedblright) of the ADM equations by adding terms involving
constraints \cite{Brown} (trying to restore what has been lost) or by using
different choices of the lapse and shift functions \cite{NR5} to give the ADM
system of equations well-posedness (or \textquotedblleft quasi
well-posedness\textquotedblright\ \cite{NR5}). From our point of view, the
best \textquotedblleft cure\textquotedblright\ of this \textquotedblleft
illness\textquotedblright\ is to return to the original Einstein equations and
constraints (or differential identities) that preserve diffeomorphism invariance.

\subsection{Summary}

The main point of this relatively long Section is not to prove that the ADM
variables are not canonical variables for GR (as is shown by the one simple PB
(\ref{eqn445})), but to demonstrate and discuss the restrictive conditions
that must be made on change of variables in any Hamiltonian formulation of a
singular Lagrangian, using GR as an example. A blind change of variables in
singular systems without performing a thorough analysis and without developing
mathematical criteria for such changes can lead to a wrong result. All new
variables that are introduced in such cases, regardless of their physical or
geometrical meaning, regardless of what new names were given to reflect their
interpretation, or after whom new variables were named, must be carefully
analyzed if one wants to keep all the properties of the original theory intact
or, in other words, if one intends to study the original theory and not a
substitute, which can be ill defined, even as unrelated to the original theory
model. One should under no circumstances attribute to the original theory the
contradictions or problems that arise after such inappropriate changes are
made; and never project any novel result or discovery obtained by abandoning a
\textquotedblleft regular and uniform rule of procedure\textquotedblright%
\ into the original theory or to Nature Herself.

\section{Acknowledgement}

The authors are grateful to A.M. Frolov, T. Hyland, M. Kidd, S.R. Valluri and,
especially, to D.G.C. McKeon and A.V. Zvelindovsky for numerous discussions
and suggestions during the preparation of our paper. We are thankful to P.G.
Komorowski for pointing out, guiding in and providing references on results of
numerical relativity. The partial support of the Huron University College
Faculty of Arts and Social Science Research Grant Fund is greatly acknowledged.

We would also like to express our gratefulness to our Teacher, the late Prof.
A.V. Zatovsky, who long ago taught us the first course in Theoretical Physics,
Analytical Mechanics, in the spirit inherited from Lagrange.


\begin{thebibliography}{999}                                                                                              %


\bibitem {Lagrange}J.L. Lagrange, Mecanique Analytique (Acad\'{e}mie des
Sciences, Paris, 1788).

\bibitem {Lagrange-eng}J.L. Lagrange, Analytical Mechanics (Kluwer, Dordrecht, 1997).

\bibitem {Dirac-1}P.A.M. Dirac, Can. J. Math. 2 (1950) 129.

\bibitem {Bergmann-1}P.G. Bergmann and J.H.M. Brunings, Rev. Mod. Phys. 21
(1949) 480.

\bibitem {Bergmann-2}J.L. Anderson and P.G. Bergmann, Phys. Rev. 83 (1951) 1018.

\bibitem {Rosenfeld}L. Rosenfeld, Annalen der Physik, 397 (1930) 113.

\bibitem {Bergmann-3}P.G. Bergmann and I. Goldberg, Phys. Rev. 98 (1955) 531.

\bibitem {DeWitt-1}B.S. DeWitt, Phys. Rev. 160 (1967) 1113.

\bibitem {DeWitt-2}B.S. DeWitt, Gen. Rel. Grav. 1 (1970) 181.

\bibitem {HRT}A. Hanson, T. Regge and C. Teitelboim, Constrained Hamiltonian
Systems (Accademia Nazionale dei Lincei, Roma, 1976).

\bibitem {Diracbook}P.A.M. Dirac, Lectures on Quantum Mechanics (Belfer
Graduate School of Sciences, Yeshiva University, New York, 1964).

\bibitem {Kurt}K. Sundermeyer, Constrained Dynamics, Lecture Notes in Physics,
vol. 169 (Springer, Berlin, 1982).

\bibitem {Gitmanbook}D.M. Gitman and I.V. Tyutin, Quantization of Fields with
Constraints (Springer, Berlin, 1990).

\bibitem {HTbook}M. Henneaux and C. Teitelboim, Quantization of Gauge Systems
(Princeton University Press, Princeton, New Jersey, 1992).

\bibitem {RRbook}H.J. Rothe and K.D. Rothe, Classical and Quantum Dynamics of
Constrained Hamiltonian Systems, vol. 81 (World Scientific Lecture Notes in
Physics, New Jersey, 2010).

\bibitem {Preprint}D. Salisbury, Preprint 381 (Max Planck Institute for the
History of Science, 2009).

\bibitem {Einstein}A. Einstein, Sitzungsber. preuss. Akad. Wiss., Phys.-Math.
K1 (1925) 414.

\bibitem {Einstein-rus}A. Einstein, In: I.E. Tamm, Ya.A. Smorodinskii, B.G.
Kuznetsov (Eds.), The Complete Collection of Scientific Papers, vol. 2 (Nauka,
Moscow, 1966), p. 171.

\bibitem {Einstein-eng}A. Unzicker, T. Case, arXiv:physics/0503046;
%TCIMACRO{\TEXTsymbol{<}}%
%BeginExpansion
$<$%
%EndExpansion
http:/www.lrz-muenchen.de/`aunzicker/ae1930.html%
%TCIMACRO{\TEXTsymbol{>}}%
%BeginExpansion
$>$%
%EndExpansion
.

\bibitem {Pirani}F.A.E. Pirani, A. Schild and R. Skinner, Phys. Rev. 87 (1952) 452.

\bibitem {Dirac}P.A.M. Dirac, Proc. Roy. Soc. A 246 (1958) 333.

\bibitem {ADM}R. Arnowitt, S. Deser and C. W. Misner, In: L. Witten (Ed.),
Gravitation: An Introduction to Current Research (Wiley, New York, 1962), p.
227; arXiv:gr-qc/0405109.

\bibitem {Castref1}E.C.G. Sudarshan and N. Mukunda, Classical Dynamics - A
Modern Perspective (A Wiley-Interscience Publication, New York, 1974).

\bibitem {Castref2}N. Mukunda, Physica Scripta 21 (1980) 783.

\bibitem {Castellani}L. Castellani, Ann. Phys. 143 (1982) 357.

\bibitem {HTZ}M. Henneaux, C. Teitelboim and J. Zanelli, Nucl. Phys. B 332
(1990) 169.

\bibitem {Novel}R. Banerjee, H.J. Rothe and K.D. Rothe, Phys. Lett. B 479
(2000) 429.

\bibitem {Novel-1}R. Banerjee, H.J. Rothe and K.D. Rothe, Phys. Lett. B 463
(1999) 248.

\bibitem {KKaffinemetric}N. Kiriushcheva and S.V. Kuzmin, Eur. Phys. J. C 70
(2010) 389.

\bibitem {Gravitation}C.W. Misner, K.S. Thorne, J.A. Wheeler, Gravitation
(W.H. Freeman and Company, San Francisco, 1973).

\bibitem {Waldbook}R.M. Wald, General Relativity (The University of Chicago
Press, Chicago, 1984).

\bibitem {Saha}P. Mukherjee and A. Saha, Int. J. Mod. Phys. A 24 (2009) 4305.

\bibitem {Landau}L.D. Landau and E.M. Lifshitz, The Classical Theory of
Fields, fourth ed. (Pergamon Press, Oxford, 1975).

\bibitem {Bergmann}P.G. Bergmann and A. Komar, Int. J. Theor. Phys. 5\textbf{
}(1972) 15.

\bibitem {SalSund}D.C. Salisbury and K. Sundermeyer, Phys. Rev. D 27 (1983) 740.

\bibitem {PonsSS}J.M. Pons, D.C. Salisbury and L.C. Shepley, Phys. Rev. D 62
(2000) 064026.

\bibitem {Pons}J.M. Pons, Class. Quant. Grav. 20 (2003) 3279.

\bibitem {PonsS}J.M. Pons and D.C. Salisbury, Phys. Rev. D 71 (2005) 124012.

\bibitem {SalisburyPS}J.M. Pons, D.C. Salisbury and L.C. Shepley, Phys. Rev. D
55 (1997) 658.

\bibitem {Salisbury-hist}D.C. Salisbury, The Eleventh Marcel Grossmann
Meeting, On Recent Developments in Theoretical and Experimental General
Relativity, Gravitation and Relativistic Field Theories (In 3 Volumes),
Proceedings of the MG11 Meeting on General Relativity Berlin, Germany, 23 --
29 July 2006; arXiv:physics/0701299 [physics.hist-ph].

\bibitem {Samanta}S. Samanta, Int. J. Theor. Phys. 48 (2009) 1436.

\bibitem {Hawking}S.W. Hawking, In: S. W. Hawking and W. Israel (Eds.),
General Relativity. An Einstein Centenary Survey (Cambridge University Press,
Cambridge, 1979), p. 746.

\bibitem {Rovelli}C. Rovelli, In: H. Garcia-Compe\`{a}n (Ed.), Topics in
Mathematical Physics. General Relativity and Cosmology, in honor of Jerzy
Plebanski, Proceedings of 2002 International Conference (Cinvestav, Mexico
City, 17-20 September, 2002); arXiv:gr-qc/0202079.

\bibitem {Pulin}J. Pullin, In: Cosmology and Gravitation: Xth Brazilian School
of Cosmology and Gravitation; 25th Anniversary (1977-2002), vol 668 (AIP
conference proceedings, 2003), p. 141; arXiv:gr-qc/0209008.

\bibitem {KKRV}N. Kiriushcheva, S.V. Kuzmin, C. Racknor and S.R. Valluri,
Phys. Lett. A 372 (2008) 5101.

\bibitem {HKT}S.A. Hojman, K. Kuchar, and C. Teitelboim, Ann. Phys. 96 (1976) 88.

\bibitem {Kuchar}K. Kuchar, In: W. Israel (Ed.), Relativity, Astrophysics and
Cosmology (D. Reidel Publishing Company, Dordrecht, Holland, 1973), p. 237.

\bibitem {Ann33}S.W. Hawking and R. Penrose, Proc. Roy. Soc. A 246 (1970) 529.

\bibitem {Ann34}F. Markopoulou, Commun. Math. Phys. 211 (2000) 559.

\bibitem {Thiemann}T. Thiemann, arXiv:gr-qc/0110034.

\bibitem {Smolin}L. Smolin, Three roads to Quantum Gravity (Basic Books, New
York, 2001).

\bibitem {GKK}K.R. Green, N. Kiriushcheva and S.V. Kuzmin, Eur. Phys. J. C 71
(2011) 1678.

\bibitem {Carmeli}M. Carmeli, Classical Fields: General Relativity and Gauge
Theory (World Scientific, New Jersey, 2001).

\bibitem {Chand}S. Chandrasekhar, The Mathematical Theory of Black Holes
(Clarendon Press, Oxford, 2005).

\bibitem {Isham}F. Antonsen, F. Markopoulou, arXiv:gr-qc/9702046.

\bibitem {Teit-PRL}C. Teitelboim, Phys. Rev. Lett. \ 38 (1977) 1106.

\bibitem {ADM-1}R. Arnowitt, S. Deser and C.W. Misner, Phys. Rev. 113 (1959) 745.

\bibitem {Teit-in-Held}C. Teitelboim, In: A. Held (Ed.) General Relativity and
Gravitation, One Hundred Years After the Birth of Albert Einstein, vol.1
(Plenum Press, New York and London, 1980).

\bibitem {Teit-AOP}C. Teitelboim, Ann. Phys. 79 (1973) 542.

\bibitem {Relativity}J.A. Wheeler, In: B.S. DeWitt and C.M. DeWitt (Eds.)
Relativity, Groups and Topology (Gordon and Breach, New York - London, 1064),
p. 346.

\bibitem {Faddeev}L.P. Faddeev, Usp. Fiz. Nauk 136 (1982) 435; Sov. Phys. Usp.
25 (1982) 130.

\bibitem {Teit-PRD}C. Teitelboim, Phys. Rev. D \ 28 (1983) 297.

\bibitem {Gerry-Ramin-2}R.N. Ghalati and D.G.C. McKeon, arXiv:0712.2861v3 [gr-qc].

\bibitem {Gerry-new}D.G.C. McKeon, Int. J. Mod. Phys. A25 (2010) 3453;
arXiv:1005.3001 [gr-qc].

\bibitem {KKX}A.M. Frolov, N. Kiriushcheva and S.V. Kuzmin, arXiv:0809.1198 [gr-qc].

\bibitem {KKAnn}N. Kiriushcheva and S.V. Kuzmin, Ann. Phys. 321 (2006) 958.

\bibitem {Lovelock}D. Lovelock, J. Math. Phys. 12 (1971) 498.

\bibitem {Torre}C.G. Torre and I.M. Anderson, Phys. Rev. Lett. 70 (1993) 3525.

\bibitem {Grishchuk}L.P. Grishchuk, A.N. Petrov and A.D. Popova, Commun. Math.
Phys. 94 (1984) 379.

\bibitem {Franc}M. Ferraris, M. Francaviglia and C. Reina, Gen. Rel. Grav. 14
(1982) 243.

\bibitem {Palatini}A. Palatini, Rend. Circolo Math. Palermo 43 (1919) 203.

\bibitem {Palatini-eng}A. Palatini, In: P.G. Bergmann and V. De Sabbata (Eds.)
(Cosmology and Gravitation, Plenum Press, 1979), p. 477.

\bibitem {Kummer}W. Kummer and H. Schuetz, Eur. Phys. J. C 42 (2005) 277.

\bibitem {KKM}N. Kiriushcheva, S.V. Kuzmin and D.G.C. McKeon, Int. J. Mod.
Phys. A 21 (2006) 3401.

\bibitem {GM}R.N. Ghalati and D.G.C. McKeon, arXiv:0712.2861 [gr-qc].

\bibitem {ADM-2}R. Arnowitt, S. Deser and C.W. Misner, Phys. Rev. 116 (1959) 1322.

\bibitem {Eisenhart}L.P. Eisenhart, Riemannian Geometry (Princeton University
Press, Princeton, New Jersey, 1997).

\bibitem {Muller}C. M\o ller, The theory of Relativity (Clarendon Press,
Oxford, 1952).

\bibitem {Weinberg}S. Weinberg, Gravitation and Cosmology: Principles and
Applications of the General Relativity (Wiley, New York, 1972).

\bibitem {Notexact}S. Deser, R. Jackiw and S. Templeton, Ann. Phys. 140 (1982) 372.

\bibitem {Notexact1}S. Deser, Class. Quant. Grav. 23 (2006) 5773.

\bibitem {Notexact2}N. Kiriushcheva and S.V. Kuzmin, Class. Quant. Grav. 24
(2007) 1371.

\bibitem {Notexact3}R.N. Ghalati, N. Kiriushcheva, S.V. Kuzmin and D.G.C.
McKeon, arXiv:hep-th/0609220.

\bibitem {Noether}E. Noether, Nachr. d. K\"{o}nig. Gesellsch. d. Wiss. zu
G\"{o}ttingen, Math-phys. Klasse (1918) 235.

\bibitem {Noether-eng}E. Noether (M.A. Tavel's English translation), arXiv:physics/0503066.

\bibitem {trans}N. Kiriushcheva and S.V. Kuzmin, Gen. Rel. Grav. 42 (2010) 2613.

\bibitem {Pictorial}K. Kuchar, arXiv:gr-qc/9304012.

\bibitem {EinsteinGR}A. Einstein, Sitzungsber. preuss. Akad. Wiss. 44, 2
(1915) 778. 

\bibitem {EinsteinGR-rus}A. Einstein, In: I.E. Tamm, Ya.A. Smorodinskii, B.G.
Kuznetsov (Eds.), The Complete Collection of Scientific Papers, vol. 1 (Nauka,
Moscow, 1965), p. 425.

\bibitem {Shestakova}T.P. Shestakova, Proceedings of Russian summer school -
seminar on Gravitation and Cosmology GRACOS - 2007, Kazan (2007) p. 179;
arXiv:0801.4854 [gr-qc].

\bibitem {AndBar}E. Anderson, J. Barbour, B.Z. Foster, B. Kelleher, N.O.
Murchadha, Class. Quant. Grav. 22 (2005) 1795.

\bibitem {Dirac-2}P.A.M. Dirac, Phys. Rev. 114 (1959) 924.

\bibitem {ADM-6}R. Arnowitt, S. Deser and C.W. Misner, Phys. Rev. 118 (1960) 1100.

\bibitem {ADM-12}R. Arnowitt, S. Deser and C.W. Misner, Phys. Rev. 121 (1961) 1556.

\bibitem {Franke}V.A. Franke, Theor. Math. Phys. 148 (2006) 995;
arXiv:0710.4953 [gr-qc].

\bibitem {Franke-rus}V.A. Franke, Teor. Mat. Fiz. 148 (2006) 143. 

\bibitem {ADM-7}R. Arnowitt, S. Deser and C.W. Misner, J. Math. Phys. 1 (1960) 434.

\bibitem {Nicolai}H. Nicolai, K. Peeters and M. Zamaklar, Class. Quant. Grav.
22 (2005) R193.

\bibitem {deWitt}B.S. DeWitt, Phys. Rev. 160 (1967) 1113.

\bibitem {Isham-diff-II}C.J. Isham and K.V. Kuchar, Ann. Phys. 164 (1985) 316.

\bibitem {Lanczos}C. Lanczos, The variational princiles of mechanics, fourth
ed. (Dover Publications, New York, 1970).

\bibitem {Whittaker}E.T. Whittaker, A Treatise on the Analytical Dynamics of
Particles and Rigid Bodies, fourth ed. (Cambridge University Press, 1999).

\bibitem {Gomis}D. Dominici and J. Gomis, J. Math. Phys. 21 (1980) 2124; ibid
23 (1982) 256.

\bibitem {Gomis2}J. Gomis, J. Liosa and N. Roman, J. Math. Phys. 25 (1984) 1348.

\bibitem {DeserIsham}S. Deser and C.J. Isham, Phys. Rev. D 14 (1976) 2505.

\bibitem {Stefano}R. Di Stefano and R.T. Rauch, Phys. Rev. D 26 (1982) 1242.

\bibitem {CastNieu}L. Castellani, P. van Nieuwenhuizen, M. Pilati, Phys. Rev.
D 26 (1982) 352.

\bibitem {3D}A.M. Frolov, N. Kiriushcheva and S.V. Kuzmin,  Grav. \& Cosm. 16
(2010) 181.

\bibitem {Report}N. Kiriushcheva and S.V. Kuzmin, arXiv:0907.1553 [gr-qc].

\bibitem {ADM-footnote}R. Arnowitt, S. Deser and C.W. Misner, Phys. Rev. 116
(1959) 1322.

\bibitem {Banerjee-JHEP}R. Banerjee, S. Gangopadhyay, P. Mukherjee and D. Roy,
JHEP 1002 (2010) 075.

\bibitem {Pons4}C. Battle, J. Gomis, J.M. Pons, N. Roman-Roy, J. Math. Phys.
27 (1986) 2953.

\bibitem {Regge-Teit}T. Regge and C. Teitelboim, Ann. Phys. 88 (1974) 286.

\bibitem {FM}A.E. Fischer and J.E. Marsden, In: S.W. Hawking and W. Israel
(Eds.), General Relativity. An Einstein Centenary Survey (Cambridge University
Press, Cambridge, 1979), p. 138.

\bibitem {NR1}L.E. Kidder, M.A. Scheel, and S.A. Teukolsky, Phys. Rev. D 64
(2001) 064017.

\bibitem {NR2}O. Sarbach, G. Calabrese, J. Pullin, and M. Tiglio, Phys. Rev. D
66 (2002) 064002.

\bibitem {NR3}V. Paschalidis, A. Khokhlov, and I. Novikov, Phys. Rev. D 75
(2007) 024026.

\bibitem {NR4}M. Salgado, D. Mart\'{\i}nez-del R\'{\i}o, M. Alcubierre, and D.
N\'{u}\~{n}ez, Phys. Rev. D 77 (2008) 104010.

\bibitem {NR5}B. Kelly, P. Laguna, K. Lockitch, J. Pullin, E. Schnetter, D.
Shoemaker and M. Tiglio, Phys. Rev. D 64 (2001) 084013.

\bibitem {York}J. York, In: L. Smarr (Ed.), Sources of Gravitational Radiation
(Cambridge University Press, Cambridge, England, 1979).

\bibitem {NR6}G. Calabrese, J. Pullin, O. Sarbach, and M. Tiglio, Phys. Rev. D
\textbf{66} (2002) 064011.

\bibitem {Bona}C. Bona and C. Palenzuela-Luque, Elements of Numerical
Relativity, Lecture Notes in Physics, vol. 673 (Springer Verlag, Berlin, 2005).

\bibitem {Brown}J. D. Brown, arXiv:0803.0334 [gr-qc].
\end{thebibliography}
\end{document}